\documentclass[letterpaper,11pt,fleqn]{article}
\pdfoutput=1
\usepackage{jheppub}

\usepackage{epstopdf}
\usepackage[T1]{fontenc}
\usepackage{extarrows}
\usepackage{bm,amsmath,amssymb}
\usepackage[mathscr]{eucal}
\usepackage{graphicx}
\usepackage{cancel}

\long\def\comment#1{ }

\newcommand{\nlo}{{\rm \scriptscriptstyle NLO}}
\newcommand{\rnlo}{{\rm \scriptscriptstyle rNLO}}

\newcommand{\lo}{{\rm \scriptscriptstyle LO}}

\newcommand{\eqn}[1]{Eq.~\eqref{#1}}
\newcommand{\beq}{\begin{equation}}
\newcommand{\eeq}{\end{equation}}

\newcommand{\order}[1]{\mathcal{O}{(#1)}}
\newcommand{\abar}{\bar{\alpha}_s}
\newcommand{\nn}{\nonumber\\}

\newcommand{\rme}{{\rm e}}

\title{Forward dijets in proton-nucleus collisions at next-to-leading order: the real corrections}

\author[a]{Edmond Iancu}
\author[b,c]{Yair Mulian}
\affiliation[a]{Institut de Physique Th\'eorique, Universite Paris Saclay, CNRS, CEA, F-91191, Gif-sur-Yvette, France}
\emailAdd{edmond.iancu@ipht.fr, yair25m@gmail.com}

\affiliation[b]{
Department of Physics, University of Jyv\"askyl\"a, \\
 P.O. Box 35, 40014 University of Jyv\"askyl\"a, Finland
}

\affiliation[c]{
Helsinki Institute of Physics, \\ P.O. Box 64, 00014 University of Helsinki, Finland
}

\abstract{Using the CGC effective theory together with the hybrid factorisation, we study 
forward dijet production in proton-nucleus collisions beyond leading order. In this paper, we 
compute the ``real'' next-to-leading order (NLO) corrections, i.e. the radiative corrections associated
with a three-parton final state, out of which only two are being measured. To that aim, we start
by revisiting our previous results for the three-parton cross-section presented in 
\cite{Iancu:2018hwa}. After some reshuffling of terms, we deduce new expressions for
these results, which not only look considerably simpler, but are also physically more transparent.
We also correct several errors in this process. The real NLO corrections to inclusive 
dijet production are then
obtained by integrating out the kinematics of any of the three final partons. We explicitly
work out the interesting limits where the unmeasured parton is either a soft gluon, or the
product of a collinear splitting. We find the expected results in both limits: the B-JIMWLK
evolution of the leading-order dijet cross-section in the first case (soft gluon) and,
respectively, the DGLAP evolution of the initial and final states in the second case (collinear
splitting). The ``virtual'' NLO corrections to dijet production will be presented in a subsequent publication.}

\begin{document}
\maketitle

\flushbottom

\section{Introduction}

In this paper we shall study inclusive dijet production in proton-nucleus ($pA$) collisions 
at {\it forward rapidities}, i.e. at very small angles with respect to the collision axis in the
fragmentation region of the proton. This kinematics is interesting
in that it gives us access to the small-$x$ part of the gluon distribution in the nuclear target,
where high-density  phenomena like gluon saturation are expected to be important.
For this particular set-up, it is a good
approximation to assume that the final state --- the two measured jets 
possibly accompanied  by unmeasured particles --- is produced via radiation from
a single parton from the incoming proton.
The role of the scattering is to put on-shell the otherwise virtual quanta from the 
parton (light-cone) wavefunction  
and also to give these quanta a non-trivial distribution in energies and momenta.
In particular, their final distribution in transverse momenta (or azimuthal angles) should reflect
properties of the nuclear gluon distribution, that we are ultimately interested in.
To avoid a proliferation of cases, we shall restrict ourselves to the {\it quark channel}, that is,
we shall consider only the case where the parton from the proton which participates in
the scattering is a quark. The addition of the gluon channel is  in 
principle straightforward and will be addressed in a subsequent publication.

As implicit in the previous discussion, we shall use the framework of the so-called ``hybrid factorisation''
\cite{Dumitru:2005gt,Albacete:2013tpa,Albacete:2014fwa}, 
where the leading quark is assumed to be collinear with the proton (and described by the standard 
quark distribution), whereas its interactions with the nuclear target are described within the
Colour Glass Condensate (CGC) effective theory  \cite{Iancu:2003xm,Gelis:2010nm}. The nucleus is viewed as a source of strong, 
random, colour fields representing the small-$x$ gluons and their correlations. The
high-energy scattering between a parton from the proton --- the leading quark and its radiation ---
and these strong colour fields will be computed in the {\it eikonal approximation}, that is, by associating
a Wilson line (describing colour precession) to each parton projectile.  As a result, the cross-section
corresponding to a given (partonic) final state can be related to 
gauge-invariant products of Wilson lines, whose high-energy evolution (via soft gluon emissions) 
is described by the Balitsky-JIMWLK equations  \cite{Balitsky:1995ub,Kovchegov:1999yj,JalilianMarian:1997jx,JalilianMarian:1997gr,Kovner:2000pt,Iancu:2000hn,Iancu:2001ad,Ferreiro:2001qy}  
--- actually, an infinite hierarchy of coupled equations for multi-point correlations. This evolution
becomes considerably simpler in the limit of a large number $N_c\gg 1$ of colours, where the
hierarchy acquires a ``triangular'' structure. In particular, the first equation in this hierarchy reduces
to a closed, non-linear, equation --- the Balitsky-Kovchegov equation  \cite{Balitsky:1995ub,Kovchegov:1999yj} --- for the elastic $S$-matrix of a colour dipole.

The formalism as a whole --- meaning the non-linear evolution equations with increasing energy and
the hybrid factorisation --- has been originally proposed to leading order (LO) in perturbative QCD
\cite{Iancu:2003xm,Gelis:2010nm,Albacete:2014fwa}.
But for realistic applications to the phenomenology, one needs next-to-leading order (NLO) accuracy,
at least. Note that, unlike for the collinear factorisation (and the associated DGLAP evolution),
the compatibility between the high-energy factorisation and the weak coupling expansion is
a {\it a priori} unclear, at both conceptual and practical level. Yet, via explicit calculations and tenuous
efforts, one was able to build the NLO versions of the BK  \cite{Balitsky:2008zza} and B-JIMWLK
\cite{Balitsky:2008zza,Balitsky:2013fea,Kovner:2013ona,Kovner:2014lca,Lublinsky:2016meo} 
equations, and also
of the ``'impact factors'' (the scattering matrix elements void of evolution) for a few,
relatively simple, processes, like single inclusive hadron production in $pA$ collisions
 \cite{Chirilli:2011km,Chirilli:2012jd,Altinoluk:2014eka}, the structure functions for electron-proton
  ($ep$) or electron-nucleus ($eA$) deep 
 inelastic scattering  (DIS) at small Bjorken  $x$ \cite{Balitsky:2012bs,Beuf:2016wdz,Beuf:2017bpd},
exclusive diffractive vector-meson and dijet production in DIS at  small $x$  \cite{Boussarie:2014lxa,Boussarie:2016ogo,Boussarie:2016bkq}, and inclusive photon + dijet production in  $eA$ collisions at small $x$ \cite{Roy:2019hwr}.

The original NLO results for both the BK equation and (some of) the impact factors turned out to be problematic, in the sense of generating instabilities in the high energy evolution \cite{Lappi:2015fma}, 
or negative cross-sections for particle production
\cite{Stasto:2013cha,Stasto:2014sea,Watanabe:2015tja,Ducloue:2016shw}. 
These difficulties have been eventually understood and cured.
In the case of the BK equation, the solution  \cite{Beuf:2014uia,Iancu:2015vea,Iancu:2015joa,Lappi:2016fmu,Hatta:2016ujq,Ducloue:2019ezk,Ducloue:2019jmy,Beuf:2020dxl} involves all-order resummations which enforce the
proper time-ordering for the lifetimes of the soft gluon fluctuations of the dilute projectile (a parton
from the proton, or a colour dipole in the case of DIS). The scheme-dependence of such
resummations is considerably reduced when using the rapidity of 
the dense target (the large nucleus) as the ``evolution time'' \cite{Ducloue:2019ezk}. 
For the NLO impact factors, the 
negativity problem arises when enforcing a separation between the leading-order BK evolution
and the NLO impact factor which is local in rapidity \cite{Iancu:2016vyg} 
(thus following the traditional prescription
of the  $k_T$-factorisation \cite{Catani:1994sq,Catani:1990eg}). So, the simplest way
to avoid the problem --- and obtain a positive-definite cross-section --- is to give up this separation,
that is, to keep the high-energy evolution and the NLO corrections together, in a generalised
impact factor \cite{Iancu:2016vyg,Ducloue:2017mpb,Ducloue:2017dit,Ducloue:2017ftk}. 
In this ``unsubtracted'' scheme, the only evolution which needs to be factorised
from the ``hard'' impact factor is the DGLAP evolution on the external lines --- that is, the
evolution of the parton distribution for the parton from the proton which initiates the scattering,
and that of the fragmentation functions for the produced hadrons.

One process that has attracted much interest over the last years is the production of a pair of jets
or hadrons in ``dilute-dense''  --- proton(deuteron)-nucleus ($pA$) \cite{Marquet:2007vb,Albacete:2010pg,Dominguez:2011wm,Stasto:2011ru,Iancu:2013dta,Lappi:2012nh,Kotko:2015ura,Marquet:2016cgx,vanHameren:2016ftb,Albacete:2018ruq} or electron-nucleus ($eA$) \cite{Dominguez:2011wm,Metz:2011wb,Dumitru:2015gaa,Altinoluk:2015dpi,Hatta:2016dxp,Dumitru:2018kuw,Salazar:2019ncp,Mantysaari:2019hkq} ---
collisions at forward rapidities.
 This process could be used to probe saturation even in the (experimentally more accessible)
 set-up where the final jets/hadrons have transverse momenta significantly larger than
 the saturation momentum in the dense target. Indeed, the multiple scattering off the saturated
 gluons in the target is one of the mechanisms responsible for the transverse momentum imbalance
 between the final jets, hence for their distribution in the relative azimuthal angle.
 (The other important such a mechanism is the final state radiation, leading
to the so-called ``Sudakov factor'' \cite{Mueller:2013wwa}.) 
In turn, this imbalance leads to a broadening of the final
particles distribution in the relative azimuthal angle $\Delta\phi$, around the back-to-back peak
at $\Delta\phi=\pi$. Such a broadening has indeed been observed in d+Au collisions
at RHIC \cite{Braidot:2011zj,Adare:2011sc}, although at this level it looks difficult to distinguish
the effects of saturation from those of the final state radiation \cite{Zheng:2014vka}.

That said, it would be important to have a more precise calculation of the cross-section for dijets 
(or dihadrons) production within the CGC formalism. This is our main purpose in this and a 
subsequent paper, where we will compute the respective NLO corrections to the impact factor
for the case of $pA$ collisions.
Specifically, in this paper we shall present the ``real'' NLO corrections  --- those associated with a
final state which involves three partons, out of which only two are measured. By integrating
out the kinematics of the unmeasured parton, one generates a NLO correction to the cross-section
for dijet production. In the companion paper, we will compute the corresponding
 ``virtual'' corrections --- those associated with one-loop corrections to the amplitude.
 
 For simplicity and to avoid a proliferation of cases, we shall restrict ourselves to the quark channel,
 that is, we shall only consider processes in which the final state is obtained via radiation
 from an initial quark from the wavefunction of the incoming proton. The quark channel is expected
 to dominate the forward particle production in the kinematics at RHIC, but the gluon channel
 should be added too in view of realistic applications to the LHC. We plan to do so in a later paper.
 
The starting point for computing the ``real'' NLO corrections to the dijet cross-section is
the leading-order (tree-level) cross-section for ``trijet'' (three partons) production, that we 
obtained in a previous paper \cite{Iancu:2018hwa}, by using the formalism of the light-cone
wavefunction (LCWF). So, for our present purposes, it should
 be enough to ``integrate out'' one (any) of the three final partons in our results in~\cite{Iancu:2018hwa}.
However, this is not what we shall do in practice. Indeed, the LO trijet cross-section is by itself a very
complex quantity and our respective results in \cite{Iancu:2018hwa} were presented in a 
cumbersome way, which is not convenient for the present purposes. (We have realised that when 
trying to match our results for ``real'' and  ``virtual'' corrections to dijet production, 
e.g. in order to check the cancellation of the infrared divergences.) So, our first step in this
paper will be to re-derive the results for 3-parton production in a streamlined way and with a
better strategy for organising the final expressions. In this process, we shall also correct several
errors which occurred in our original results \cite{Iancu:2018hwa}, that we were able
to identify thanks to various tests to be discussed below. 

In view of the above, the results
for the LO trijet cross-section to be presented in this paper should be viewed as our final respective
results, in replacement of the previous ones in \cite{Iancu:2018hwa}.  These results are still formal, 
as they involve undone Fourier transforms in the transverse plane: the calculation of the LCWF is performed using the transverse coordinate representation, to take profit of the eikonal approximation. 
Hence, our expressions for the cross-section involve Fourier transforms
relating the transverse coordinates of the three final partons (in both the amplitude and the
complex conjugate amplitude) to their respective momenta, as measured in the final state.
To actually compute these Fourier transforms --- say, in view of applications to the phenomenology
--- one would also need explicit results for the partonic $S$-matrices probed by 
these particular processes. 
The techniques for computing such $S$-matrices --- from numerical solutions to the BK and JIMWLK
equations, or via mean field approximations to the latter --- are well documented in the literature, 
but their discussion goes beyond our present purposes. In agreement with previous
studies \cite{Dominguez:2011wm,Dominguez:2012ad}, we shall find that in the multicolour limit
$N_c\to\infty$, the various $S$-matrices that we shall encounter are built with just two non-trivial
colour structures: the dipole and the quadrupole.

Given these results for trijet production, the ``real'' NLO corrections to dijets are easily
obtained (at least, formally) by integrating over the longitudinal and the transverse momentum of one
of the three final partons and then convoluting with the proton distribution function for the incoming
quark. If one measures two hadrons in the final state, convolutions with the appropriate 
parton-to-hadron fragmentation functions are also needed.

As a check of our calculations, we shall study two special limits for which the results are {\it a priori}
known: the case where the unmeasured parton is a soft gluon and that where this parton is
produced via a collinear splitting. In the first case, we expect to recover (one step in) the B-JIMWLK evolution\footnote{More precisely, one must recover only the ``real'' part of the B-JIMWLK 
equation and similarly for the DGLAP equation; the respective ``virtual'' parts will be generated by the ``virtual'' NLO corrections to the dijet cross-section.}
of the LO cross-section for producing a quark-gluon pair. In the second case, one should
find the DGLAP evolution for the incoming quark distribution and for the fragmentation functions
of the produced quark and gluon. We indeed recover these expected results, but only after rather
non-trivial manipulations, which in particular involve cancellations between many terms. Thus the
good results that we find in these limits provides a rather stringent test on our results. 

We shall not insist in separating out the LO 
evolution from the NLO impact factor. In the case of the B-JIMWLK evolution, this is in line
with our preference for the ``unsubtracted'' scheme  \cite{Iancu:2016vyg} 
alluded to above, which avoids potential
problems associated with a local subtraction in rapidity. In the case of the DGLAP evolution,
the subtraction of the collinear divergences is indeed mandatory, but it requires more refined
techniques like the dimensional regularisation, that we shall develop in the companion paper
devoted to ``virtual'' NLO corrections.

This paper is structured as follows. In  Sect.~\ref{sec:LO} we shall briefly recall the result
for the LO dijet cross-section in the quark channel, i.e. the Born-level
cross-section for producing a quark-gluon ($qg$) pair at forward rapidities. This result
is well known \cite{Marquet:2007vb,Dominguez:2011wm,Iancu:2018hwa} 
but we shall often need it for comparisons with the NLO results to be obtained later.
In Sect.~\ref{sec:trijet}, we present the LO (tree-level) results for the production of three partons:
$qq\bar q$ and $qgg$. We first show the three-parton components of the quark LCWF 
in the final state and then use them to compute the relevant cross-sections. As already mentioned,
in doing that we shall both reorganise and correct our original results in \cite{Iancu:2018hwa}.
In Sect.~\ref{sec:NLOreal} we  compute the "real" NLO corrections to dijet production in
the quark-initiated channel by (formally) ``integrating out'' one of the three partons in the
trijet results. We also explain the simplifications which occur in the colour structure (i.e.
in the partonic $S$-matrices) due to the fact that one of the partons is not measured.
In Sect.~\ref{sec:JIMWLK} we show that in the limit where the unmeasured parton is
a soft gluon, our NLO results reproduce, as expected, the (real part of the) JIMWLK
evolution for the $qg$ dijet cross-section. Finally, in Sect.~\ref{sec:DGLAP}, we consider
the limit where the unmeasured parton is produced by a collinear splitting. We show that,
in this limit, our NLO results develop collinear singularities, that we isolate to leading
logarithmic accuracy and verify that they can be interpreted as one-step in the DGLAP
evolution of the initial ($q$) and final ($qg$) states.

%
%
%

 \section{ Dijet cross-section at leading-order}
\label{sec:LO}

As a warm-up, let us briefly recall some steps in the derivation of the leading-order (LO) 
result for the cross-section for dijet production in $pA$ collisions at forward rapidity (see
e.g.  \cite{Marquet:2007vb,Iancu:2018hwa} for more details).
As mentioned in the Introduction, we consider only the quark channel:
the parton collinear with
the incoming proton and which initiates the process is taken to be a quark.
(The corresponding result for the gluon channel 
can be found in Refs.~\cite{Iancu:2013dta,Iancu:2018hwa}.)

\subsection{Kinematics}

We start with some generalities on the kinematics and use this opportunity to introduce some of the
notations. We chose the $z$ direction along the collision axis and work in a frame
in which the proton is an energetic right-mover with (light-cone) longitudinal momentum
$Q^+$, whereas the nucleus is an ultrarelativistic left mover, with longitudinal momentum $P^-$
per nucleon. (We shall neglect the nucleon masses in what follows.) We more precisely assume
that the nucleus target carries most of the total energy, so the high-energy evolution via the
successive emissions of soft gluons is fully encoded in the nuclear gluon distribution. 

To leading order, dijet production at forward rapidities and in the quark channel proceeds as 
follows (see also Fig.~\ref{emiglu}): a quark initially collinear with the proton, 
with longitudinal momentum $q^+=x_pQ^+$,
scatters off the dense gluon system in the nuclear target and emits a gluon in the process.
The two ``jets'' are the final quark, with longitudinal momentum $p^+=x_1Q^+$
and transverse momentum $\bm{p}$, and the emitted gluon, with $k^+=x_2Q^+$
and final transverse momentum $\bm{k}$. The target has zero ``plus'' momentum, hence
the respective component is preserved by the scattering: $x_p=x_1+x_2$; accordingly, we shall
also write $k^+=\vartheta q^+$ and hence $p^+=(1-\vartheta)q^+$, 
with $\vartheta =x_2/x_p\le 1$ the gluon splitting fraction. On the other hand, the collision can
transmit a transverse momentum of the order of the saturation momentum $Q_s$ ---
the typical momentum of a gluon at saturation ---, hence we expect an imbalance
$|\bm{p}+\bm{k}|\sim Q_s$ between the final jets. 

The two jets are put on-shell by the collision, hence they must receive from the
nucleus a ``minus'' component equal to their total light-cone energy; writing this as a fraction $x_g$
of $P^-$, we deduce
\beq\label{xg}
x_gP^-=\,\frac{\bm{p}^2}{2p^+}\,+\,\frac{\bm{k}^2}{2k^+}\quad\Longrightarrow\quad
x_g=\,\frac{\bm{p}^2}{x_1s}\,+\,\frac{\bm{k}^2}{x_2s}\,,
\eeq
with $s=2Q^+P^-$ (the center-of-mass energy squared of the collision). It is customary to 
express the longitudinal fractions  $x_p$ and $x_g$ in terms of the 
(pseudo)rapidities $\eta_1\equiv (1/2)\ln(p^+/p^-)$ and $\eta_2\equiv (1/2)\ln(k^+/k^-)$ of the 
produced jets in the center-of-mass frame, where $Q^+=P^-= \sqrt{s/2}$. Using $p^+=(p_\perp/\sqrt{2})\rme^{\eta_1}$ and similarly  $k^+=(k_\perp/\sqrt{2})\rme^{\eta_2}$ (with $p_\perp\equiv |\bm{p}|$ etc), 
one finds
\beq
\label{COM}
x_p=\frac{p_\perp}{\sqrt{s}}\,\rme^{\eta_1}\,+\,\frac{k_\perp}{\sqrt{s}}\,\rme^{\eta_2}
\,,\qquad x_g=\frac{p_\perp}{\sqrt{s}}\,\rme^{-\eta_1}\,+\,\frac{k_\perp}{\sqrt{s}}\,\rme^{-\eta_2}
\,.
\eeq
The {\em forward dijet kinematics} corresponds to the situation where $\eta_1$ and $\eta_2$
are both positive and larger than 1.  In this regime,  one has $x_g\ll x_p < 1$, showing that
the forward particle production explores the small-$x_g$ part of the nuclear wavefunction.
The relevant value of the nuclear saturation momentum increases
with decreasing $x_g$, due to the rise in the gluon density in the transverse plane,
via soft gluon emissions. This is governed by the non-linear evolution equations (BK or B-JIMWLK),
which imply $Q_s^2(x_g) \sim x_g^{-\lambda_s}$, with 
$\lambda_s\simeq 0.20$ (see e.g. \cite{Ducloue:2019jmy} for a recent study).

Note that in deriving \eqn{COM} we did not necessarily assume $2\to 2$ kinematics (i.e. $qg\to qg$).
In fact, the most interesting case for us here is that of multiple scattering, which involves the exchange
of arbitrarily many soft gluons between the quark-gluon fluctuation of the
proton and the nucleus. In such a case, $x_gP^-$ is the total LC energy transmitted from the
target to the quark-gluon system and $\bm{p}+\bm{k}$ is similarly the total transverse momentum.

Under the present assumptions, multiple scattering can be resummed to all orders within the eikonal approximation: the nuclear target appears to the proton as a Lorentz-contracted shockwave 
(say, localised at $x^+=0$) and the transverse coordinates of a projectile parton (quark or gluon)
is not modified by the collision. It is then convenient to work in the transverse coordinate
representation and Fourier transform to transverse momenta only at the end of the calculation.
In this representation,  the only effect of the scattering is a colour precession of the parton 
wavefunction, represented by a Wilson line in the appropriate representation of the colour
group SU($N_c$).

\subsection{The quark-gluon component of the quark light-cone wavefunction}

We use the light-cone wavefunction (LCWF) formalism and the projectile LC gauge $A^+=0$.
The initial state at (LC) time $x^+\to -\infty$ is taken to be a bare quark:
$\left|q\right\rangle^{in}= \left|q_{\lambda}^{\alpha}(q^{+},\,\bm{q})\right\rangle$,
with $q^{+},\,\bm{q}, \,{\lambda}$ and $\alpha$ denoting the quark longitudinal and transverse
momenta, its spin, and its colour state, respectively. (Our conventions for the bare Fock
states and for the action of the associated creation and annihilation operators are presented
in Appendix~\ref{fieldef}.) The outgoing state at $x^+\to \infty$ is computed
as $\left|q\right\rangle^{out}=U_I(\infty,0) \,\hat S\, U_I(0,-\infty)\left|q\right\rangle^{in}$,
where the QCD evolution operators $U_I(0,-\infty)$ and $U_I(\infty,0)$ describe QCD radiation
prior and after the scattering, respectively, and  $\hat S$ is the $S$-matrix operator for the
scattering between the parton system which exists at time $x^+=0$ and the shockwave.

\begin{figure}[t]\center
  \includegraphics[scale=0.85]{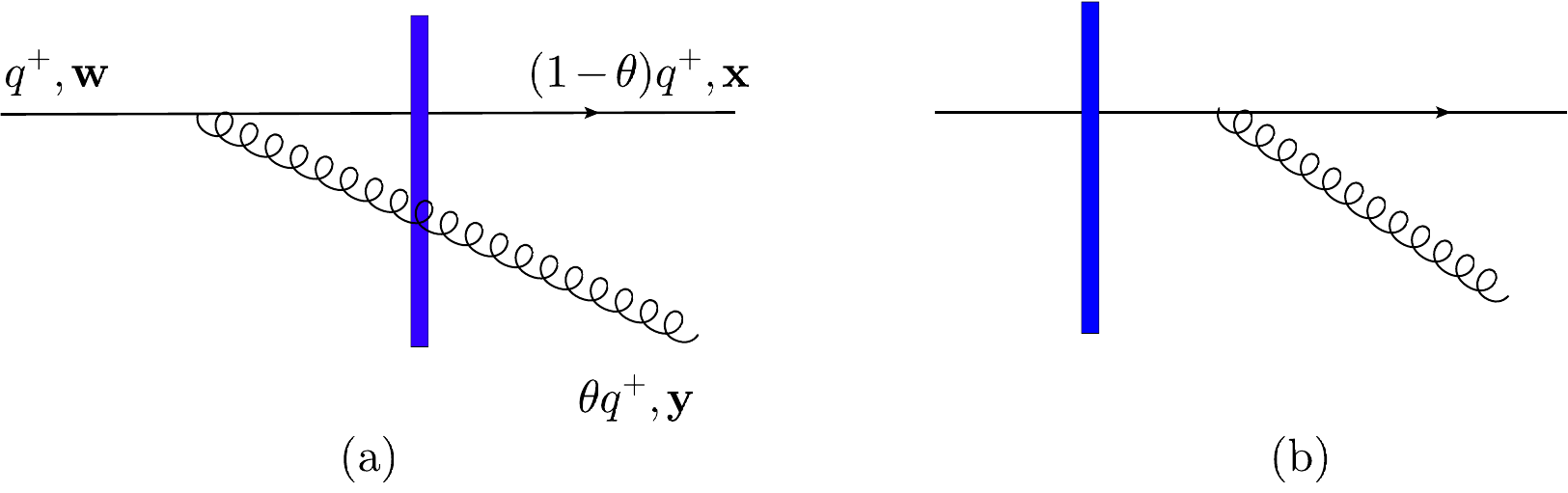}
  \caption{The quark-gluon component of the outgoing wave-function for an incoming quark. There are two ways to insert the shockwave, before and after the gluons emission, represented in Fig.~a and b, respectively.   \label{emiglu}}
\end{figure}

To compute quark-gluon production to leading order (LO), it is enough to expand the action of the evolution operators to $\order{g}$, e.g. 
\beq\label{pertUin}
U_I(0,-\infty)\,|i\rangle=\,|i\rangle - \sum_j\frac{\langle j| H_{\rm int} |i\rangle}{E_j-E_i} |j\rangle+\,\cdots  ,
\eeq
where $H_{\rm int}$ is the interaction part of the Hamiltonian,  $\left|i\right\rangle$, $\left|j\right\rangle$ are energy-momentum eigenstates of the free QCD Hamiltonian $H_0$ (i.e. the Fock states built with bare partons) and $E_i$, $E_j$ are the corresponding 
LC energies (the sum of the minus components of the partonic 4-momenta for all the partons
composing a Fock state, assumed to be on-shell). Taking $|i\rangle=\left|q_{\lambda}^{\alpha}(q^{+},\,\bm{q})\right\rangle$, the outgoing state is a superposition of a single-quark state (whose colour
has been rotated by the scattering) and the quark-gluon state, in which we are primarily interested.
The $qg$ Fock-space component is most convenient written in the transverse coordinate representation, where the matrix elements of $\hat S$ are diagonal; it then reads \cite{Iancu:2018hwa}:
 \begin{equation}\begin{split}\label{asilo}
&\left|q_{\lambda}^{\alpha}(q^{+},\,\bm{w})\right\rangle _{qg}^{out}\,=\,-\int d^{2}\bm{x}\,d^{2}\bm{y}
\int_{0}^{1}d\vartheta\,\frac{ig\phi_{\lambda_{1}\lambda}^{ij}(\vartheta)\sqrt{q^{+}}}
{4\sqrt{2\vartheta}\pi^{2}}\,
\frac{\bm{x}^{j}-\bm{y}^{j}}{(\bm{x}-\bm{y})^{2}}
\,\delta^{(2)}(\bm{w}-(1-\vartheta)\bm{x}-\vartheta\bm{y})\\
&\times\left[V^{\sigma\beta}(\bm{x})\,U^{ba}(\bm{y})\,t_{\beta\alpha}^{a}\,-\,t_{\sigma\beta}^{b}\,V^{\beta\alpha}(\bm{w})\right]\,\left|q_{\lambda_{1}}^{\sigma}((1-\vartheta)q^{+},\,\bm{x})\,g_{i}^{b}(\vartheta q^{+},\,\bm{y})\right\rangle,
\end{split}\end{equation}
where $\bm{w}$ is the transverse coordinate of the incoming quark (conjugated to its 
transverse momentum $\bm{q}$), $\bm{x}$ and $\bm{y}$ are the transverse coordinates of the
final quark and gluon, and the constraint $\bm{w}=(1-\vartheta)\bm{x}+\vartheta\bm{y}$ follows
from the conservation of the transverse momentum. Furthermore, $g_{i}^{b}(\vartheta q^{+},\,\bm{y})$
is a gluon with transverse helicity $i$, colour state $b$, longitudinal momentum $\vartheta q^{+}$
and transverse position $\bm{y}$. We also used the following notation for the spinor/helicity structure: 
  \begin{equation}
  \label{phidef}
\phi_{\lambda_{1}\lambda}^{ij}(\vartheta)\,\equiv\,\chi_{\lambda_{1}}^{\dagger}\left[(2-\vartheta)\delta^{ij}-i\vartheta\varepsilon^{ij}\sigma^{3}\right]\chi_{\lambda}\,=\,\delta_{\lambda\lambda_{1}}\left[(2-\vartheta)\delta^{ij}-2i\vartheta\varepsilon^{ij}\lambda\right].
\end{equation}
The two terms inside the square bracket in \eqn{asilo} come from the action of the $S$-matrix
 $\hat S$ and  refer to the situation when the scattering with the shockwave occurs before and respectively after the gluon emission (see Fig.~\ref{emiglu}).
$U_{ba}(\bm{x})$ and $V_{\beta\alpha}(\bm{x})$ are Wilson lines in the adjoint (for the gluon)
 and respectively fundamental (for the quark) representation. In matrix notations,
\beq\label{SA}
U(\bm{x})\,=\,{\rm T}\exp\left\{ ig\int dx^{+}\,T^{a}A_{a}^{-}(x^{+},\,\bm{x})\right\} ,
\qquad   
  V(\bm{x})\,=\,{\rm T}\exp\left\{ ig\int dx^{+}\, t^{a}A^{-}_a(x^{+},\, \bm{x})\right\}.
   \eeq
with the colour field  $A^{-}_a$ representing the small-$x$ gluons in the target. In the CGC effective theory, this field is random and must be averaged out at the level of the cross-section.

The outgoing state  in \eqn{asilo} is clearly vanishing in the absence of scattering,
i.e. in the limit where $U\to 1$ and $V\to 1$. Indeed, in the absence of any interaction
(like the scattering off a nuclear target), the quark-gluon pair cannot be produced in the
final state (since an on-shell quark cannot emit a gluon). Furthermore, using the identity
 \begin{equation}\label{VUid}
\left[V^{\dagger}(\bm{y})\,t^{b}\,V(\bm{y})\right]_{\beta\alpha}\,=\,
U^{ba}(\bm{y})\,t_{\beta\alpha}^{a}\,,
\end{equation}
one sees that the difference of Wilson lines in \eqn{asilo} also vanishes when $\bm{y}
\to \bm{x}$. 
This is the expression of colour conservation: in the limit where the transverse separation
between the daughter partons shrinks to zero, 
the quark-gluon pair scatters off the nuclear target
in the same way as its parent quark, hence the bare quark and its $qg$  fluctuation 
cannot be disentangled by the collision.

\subsection{The quark-gluon cross-section at leading order}

The inclusive cross section for forward quark-gluon production in quark-nucleus ($qA$) scattering can be computed as the average of the respective product of number density operators 
over the momentum-space version of the outgoing state, as obtained from
\eqn{asilo}  via a Fourier transform:
\beq\label{backF}
\left|q_{\lambda}^{\alpha}(q^{+},\,\bm{q})\right\rangle _{qg}^{out}\,=\int_{\bm{w}}\,\,\rme^{i\bm{w}\cdot\bm{q}}
\left|q_{\lambda}^{\alpha}(q^{+},\,\bm{w})\right\rangle _{qg}^{out}\,.\eeq
Specifically, with our present conventions one can write 
\begin{equation}\begin{split}\label{locrosdefin}
\frac{d\sigma_{\lo}^{qA\rightarrow qg+X}}{d^{3}p\,d^{3}k}\,(2\pi)\delta(k^++p^+-q^+)
\equiv\,\frac{1}{2N_{c}}\!\!{}_{\ \ \ qg}^{\ \ out}\!\left\langle q_{\lambda}^{\alpha}(q^{+},\,\bm{q})\right|\,\hat{\mathcal{N}}_{q}(p)\,\hat{\mathcal{N}}_{g}(k)\,\left|q_{\lambda}^{\alpha}(q^{+},\,\bm{q})\right\rangle _{qg}^{out}.
\end{split}\end{equation}

\begin{figure}[t]\center
  \includegraphics[scale=0.75]{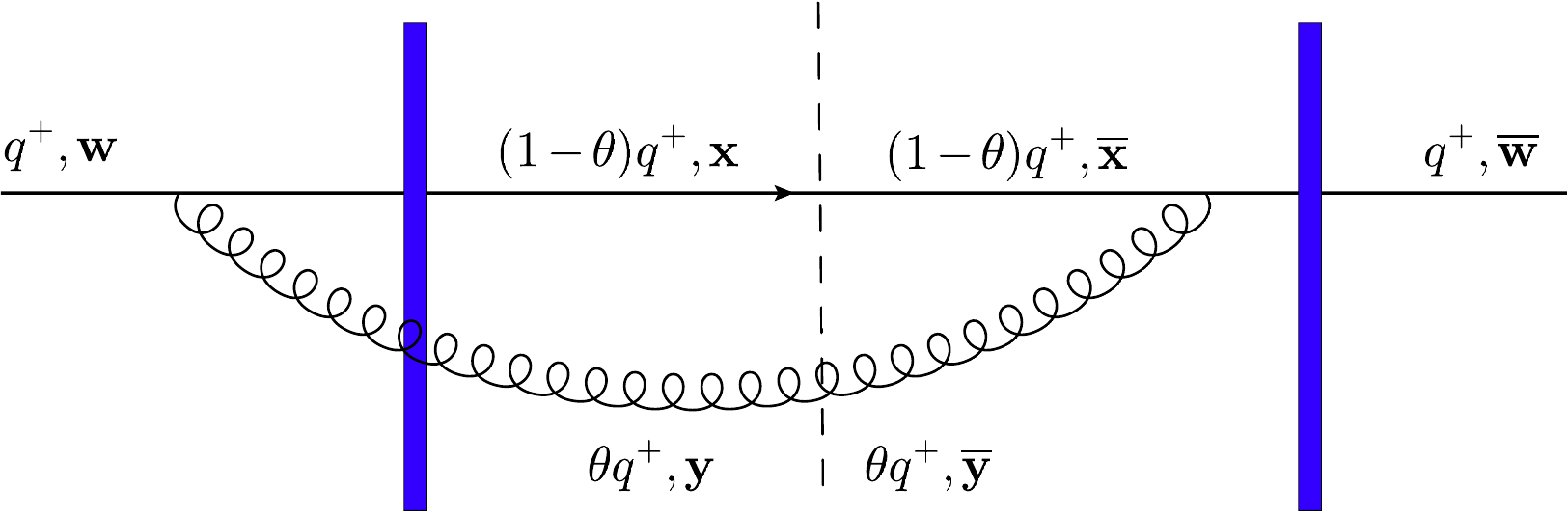}
  \caption{One of the contributions (the one proportional to $S_{qg\bar{q}}\left(\bm{x},\,\bm{y},\,\overline{\bm{w}}\right)$) to the quark-gluon production cross section in Eq.~(\ref{LOfinal}). There are four such contributions, in which the shockwave can be located either before or after the gluon emission, in both the direct amplitude and the complex conjugate amplitude. The dashed line represents the final state at $x^{+}\rightarrow\infty$. \label{loopfig} }
\end{figure}

As already mentioned, we can choose $\bm{q}=0$ (the longitudinal axis is taken to be parallel to
the collision axis). After some simple algebra, one finds the
following result for the forward quark-gluon production cross section at LO \cite{Iancu:2018hwa,Marquet:2007vb}:
\begin{equation}\begin{split}\label{LOfinal}
\frac{d\sigma_{\lo}^{qA\rightarrow qg+X}}{dp^{+}\,d^{2}\bm{p}\,dk^{+}\,d^{2}\bm{k}}\,&=\,\frac{2\alpha_{s}C_{F}\left[1+(1-\vartheta)^{2}\right]}{(2\pi)^{6}\vartheta q^{+}}\,\delta(q^{+}-k^{+}-p^{+})
\\*[0.2cm]
&\times\,\int_{\bm{x},\,\overline{\bm{x}},\,\bm{y},\,\overline{\bm{y}}}\,\frac{(\bm{x}-\bm{y})\cdot
(\overline{\bm{x}}-\overline{\bm{y}})}{(\bm{x}-\bm{y})^{2}\,(\overline{\bm{x}}-\overline{\bm{y}})^{2}}\,\rme^{-i\bm{p}\cdot(\bm{x}-\overline{\bm{x}})-i\bm{k}\cdot(\bm{y}-\bar{\bm{y}})}
\\*[0.2cm]
&\times\left[S_{qg\bar{q}g}\left(\bm{x},\,\bm{y},\,\overline{\bm{x}},\,\overline{\bm{y}}\right)\,-\,S_{qg\bar{q}}\left(\bm{x},\,\bm{y},\,\overline{\bm{w}}\right))\,-\,S_{q\bar{q}g}\left(\bm{w},\,\overline{\bm{x}},\,\overline{\bm{y}}\right)\,+\,\mathcal{S}\left(\bm{w},\,\overline{\bm{w}}\right)\right],
\end{split}\end{equation}
with $C_{F}\equiv t^{a}t^{a}=\frac{N_{c}^{2}-1}{2N_{c}}$ and a compact notation for the transverse integrations: $\int_{\bm{x}}\,\equiv\,\int d^{2}\bm{x}$.  It is understood that $\vartheta = k^+/q^+$,
 $\bm{w}=(1-\vartheta)\bm{x}+\vartheta\bm{y}$ and $\overline{\bm{w}}=(1-\vartheta)\overline{\bm{x}}+\vartheta\overline{\bm{y}}$. To obtain this result, we have also used the identity $\varepsilon^{il}\varepsilon^{jk}=\delta^{ij}\delta^{kl}-\delta^{ik}\delta^{jl}$ to perform the sums over the quark helicities and the gluon polarisations (recall \eqn{phidef}):
\begin{equation}\label{simptheta}
\phi_{\lambda_{1}\lambda}^{ij\dagger}(\vartheta)\,\phi_{\lambda_{1}\lambda}^{ik}(\vartheta)\,
=\,4\delta^{jk}\,\left(1+(1-\vartheta)^{2}\right).
\end{equation}

In writing \eqn{LOfinal}, we have also performed the average over the random colour fields $A^-$
describing the soft gluons from the target and we have introduced the following $S$-matrices describing the forward scattering of colourless systems made with up to four partons: a quark-antiquark dipole,
\begin{equation}\label{lowils2}
\mathcal{S}\left(\bm{w},\overline{\bm{w}}\right)
\,\equiv\,\frac{1}{N_{c}}\,\left\langle \mathrm{tr}\big(
V(\bm{w})V^{\dagger}(\overline{\bm{w}})\big)\right\rangle,
\end{equation}
a quark-antiquark-gluon triplet (this is illustrated in Fig.~\ref{loopfig}),
\begin{equation}\begin{split}\label{lowils3}
&S_{qg\bar{q}}\left(\bm{x},\,\bm{y},\,\overline{\bm{w}}\right)\,\equiv\,\frac{1}{C_{F}N_{c}}\,\left\langle \mathrm{tr}\left(V^{\dagger}(\overline{\bm{w}})\,t^{b}\,V(\bm{x})\,t^{a}\right)\,U^{ba}(\bm{y})\right\rangle \\
&=\,\frac{1}{2C_{F}N_{c}}\,\left(N_{c}^{2}\,
\mathcal{S}(\bm{x},\,\bm{y})\,
\mathcal{S}(\bm{y},\,\overline{\bm{w}})
\,-\,\mathcal{S}(\bm{x},\,\overline{\bm{w}})\right)\,\simeq\,\mathcal{S}(\bm{x},\,\bm{y})\,
\mathcal{S}(\bm{y},\,\overline{\bm{w}})\,,
\end{split}\end{equation}
and finally a quark-antiquark pair accompanied by two gluons:
\begin{equation}\begin{split}\label{lowils1}
&S_{qg\bar{q}g}\left(\bm{x},\,\bm{y},\,\overline{\bm{x}},\,\overline{\bm{y}}\right)\,\equiv\,\frac{1}{C_{F}N_{c}}\,\left\langle \mathrm{tr}\left(V^{\dagger}(\overline{\bm{x}})\,V(\bm{x})\,t^{a}\,t^{c}\right)\,\left[U^{\dagger}(\overline{\bm{y}})\,U(\bm{y})\right]^{ca}\right\rangle \\
&=\,\frac{1}{2C_{F}N_{c}}\,\left(N_{c}^{2}\,\mathcal{Q}(\bm{x},\,\bm{y},\,\overline{\bm{y}},\,\overline{\bm{x}})\,\mathcal{S}(\bm{y},\,\overline{\bm{y}})\,-\,\mathcal{S}(\bm{x},\,\overline{\bm{x}})\right)\,\simeq\,\mathcal{Q}(\bm{x},\,\bm{y},\,\overline{\bm{y}},\,\overline{\bm{x}})\,\mathcal{S}(\bm{y},\,\overline{\bm{y}})\,.
\end{split}\end{equation}
The second equalities in the r.h.s. of Eqs.~(\ref{lowils1}) and (\ref{lowils3}) are obtained after using \eqn{VUid} together with standard Fierz identities. Besides  the colour dipole already introduced in \eqn{lowils2} they also involve the quadrupole,
\begin{equation}
\label{quadrupole}
\mathcal{Q}\,(\bm{x},\,\bm{y},\,\overline{\bm{y}},\,\overline{\bm{x}})\,\equiv\,\frac{1}{N_{c}}\,\left\langle
\mathrm{tr}\big(V(\bm{x})\,V^{\dagger}(\bm{y})\,V(\overline{\bm{y}})\,
V^{\dagger}(\overline{\bm{x}})\big)\right\rangle.
\end{equation}
The final, approximate, equalities in Eqs.~(\ref{lowils1}) and (\ref{lowils3}) hold in the multi-colour limit $N_c\to\infty$, which allows for important simplifications, as already visible in the above results.

It is understood that the target averages occurring in the above equations \eqref{lowils2}--\eqref{quadrupole} are computed over the nuclear gluon distribution evolved down to a longitudinal momentum fraction $x_g$, cf. \eqn{xg}. So, in principle, these $S$-matrices must be obtained from
solutions to the non-linear  B-JIMWLK equations, which in general form an infinite hierarchy. At large $N_c$, one can use simpler equations, which are closed:  the BK equation for the dipole $S$-matrix 
together with the evolution equation for the quadrupole obtained in  \cite{Dominguez:2011gc,Iancu:2011ns,Iancu:2011nj}. The latter is still cumbersome to use in practice, so it is useful to notice that a
mean field approximation relating the quadrupole to the dipole
\cite{Blaizot:2004wv,Dominguez:2011wm,Iancu:2011ns,Iancu:2011nj,Lappi:2012nh}
appears to work quite well, even for finite $N_c$ \cite{Dumitru:2011vk}. 

%

Given the partonic cross-section \eqref{LOfinal}, the contribution of the quark channel to the leading-order 
cross-section for dijet production in $pA$ collisions at forward rapidity is obtained as
\begin{equation}\label{pALO}
\frac{d\sigma_{\lo}^{pA\rightarrow 2jet+X}}{d^{3}p\,d^{3}k} \bigg |_{q-channel}
\,=\,\int dx_{p}\,q_f(x_{p},\mu^{2})
\frac{d\sigma_{\lo}^{qA\rightarrow qg+X}}{d^{3}p\,d^{3}k}\,,
 \end{equation}
where $q_f(x_p,\mu^2)$ is the quark distribution function of the proton, for a longitudinal momentum fraction $x_p=q^+/Q^+$ and a resolution scale $\mu^2$. At LO,
the value of  $x_p$ is fixed by the  $\delta$-function in \eqn{LOfinal}. Furthermore,  $\mu^2$ should be chosen of the order of the hardest among the transverse momenta,
$\bm{k}^2$ or  $\bm{p}^2$, of the produced jets.
The final result at LO and for large $N_c$ reads 
 \begin{align}
 \label{pALOqchannel}
\frac{d\sigma_{\lo}^{pA\rightarrow qg+X}}{d^{3}p\,d^{3}k} 
&\,=\,x_{p}\,q_f(x_{p},\mu^{2})\,\frac{\abar}{(2\pi)^5}\,\frac{1+(1-\vartheta)^2}{2\vartheta (q^{+})^{2}}
\nonumber
\\*[0.2cm]
&\times\,\int_{\bm{x},\,\overline{\bm{x}},\,\bm{y},\,\overline{\bm{y}}}\,\frac{(\bm{x}-\bm{y})\cdot
(\overline{\bm{x}}-\overline{\bm{y}})}{(\bm{x}-\bm{y})^{2}\,(\overline{\bm{x}}-\overline{\bm{y}})^{2}}\,\rme^{-i\bm{p}\cdot(\bm{x}-\overline{\bm{x}})-i\bm{k}\cdot(\bm{y}-\bar{\bm{y}})}\nonumber
\\*[0.2cm]
&\times\,\big[\mathcal{Q}(\bm{x},\,\bm{y},\,\overline{\bm{y}},\,\overline{\bm{x}})\,\mathcal{S}(\bm{y},\,\overline{\bm{y}})-\mathcal{S}(\bm{x},\,\bm{y})\,
\mathcal{S}(\bm{y},\,\overline{\bm{w}})-\mathcal{S}(\bm{w},\,\overline{\bm{y}})\,
\mathcal{S}(\overline{\bm{y}},\,\overline{\bm{x}})
+\mathcal{S}\left(\bm{w},\,\overline{\bm{w}}\right)\big],
\end{align}
where  $q^+=p^++k^{+}$, $\vartheta=k^+/q^+$, and $x_p=q^+/Q^+$.

We shall later need also the expression of the LO cross-section for the case where one is measuring
a pair of {\it hadrons} (instead of jets) in the final state.  
Under the assumption that the parton $\rightarrow$
hadron splitting is collinear, this cross-section is obtained by convoluting \eqn{pALOqchannel} with
the quark $\to$ hadron and gluon $\to$ hadron fragmentation functions:
\begin{equation}\label{pALOfrag}
\frac{d\sigma_{\lo}^{pA\rightarrow h_1h_2+X}}{d^{3}p\,d^{3}k} \bigg |_{q-channel}
\,=\,\int \frac{dz_1}{z_1^3} \int \frac{dz_2}{z_2^3}\,\int dx_{p}\,q_f(x_{p},\mu^{2}) \,
\frac{d\sigma_{\lo}^{qA\rightarrow qg+X}}{d^{3}p_1\,d^{3}k_1}\,D_{h_1/q}(z_1)\,D_{h_2/g}(z_2)\,,
 \end{equation}
where it is understood that $p_1=p/z_1$, $k_1=k/z_2$ (for both longitudinal and transverse components), $x_p=(p_1^++k_1^+)/Q^+$, and $\vartheta=k^+_1/(p_1^++k_1^+)$. This expression \eqref{pALOfrag} for the dijet cross-section is a particular example of the hybrid factorisation, originally introduced
in the context of single hadron production in $pA$ collisions \cite{Dumitru:2005gt,Albacete:2013tpa,Albacete:2014fwa}. Thus is ``hybrid'' in the sense that the initial and final states are treated 
in the spirit of the collinear factorisation, whereas the hard process is rather computed using
the CGC effective theory for QCD at high energy (which in turn can be viewed as 
a generalisation of the $k_T$-factorisation
originally developed in the context of the BFKL evolution \cite{Catani:1994sq,Catani:1990eg}).

 \section{Trijet cross-section at  leading order}
 \label{sec:trijet}
 
 In preparation for the calculation of the real NLO corrections to dijet production,
let us first revisit our results for the
 LO trijet ($qq\bar q$ and $qgg$) cross-sections, originally presented in 
 Ref.~\cite{Iancu:2018hwa}. As compared to \cite{Iancu:2018hwa}, we shall improve these results
 at two levels: \texttt{(i)} we shall rewrite them in a different way, which is not only more compact and physically more transparent, but also better suited for the purposes of the NLO calculation,
  and  \texttt{(ii)} we shall correct several errors and misprints, that we failed to 
  identify in Ref.~\cite{Iancu:2018hwa}, by lack of appropriate tests.  Hence, the
 results for the trijet cross-section  to be presented in this section
 should be viewed as our final respective results, in replacement
 of those in  Ref.~\cite{Iancu:2018hwa}.  And as matter of facts, we have obtained
 these new results via an independent calculation, and not by just reshuffling terms in
 the original results from Ref.~\cite{Iancu:2018hwa}. 
  
 \subsection{The tri-parton components of the quark outgoing state}
 \label{sec:LCWF}

We first present the expressions of the tri-parton Fock-space components of
the quark outgoing state. These are obtained by expanding the QCD evolution operators
in $\left|q\right\rangle^{out}=U_I(\infty,0) \,\hat S\, U_I(0,-\infty)\left|q\right\rangle^{in}$ to second
order in the interaction Hamiltonian (see e.g. Ref.~\cite{Iancu:2018hwa} for details). In this
expansion, we shall ignore the virtual corrections for the time being (they will be computed
in a subsequent paper); that is, we shall only keep those second-order terms which contribute
to a tri-parton ($qq\bar q$ and $qgg$) final state. We shall not present the details of the calculations
--- they would be very similar to those described at length in Ref.~\cite{Iancu:2018hwa} --- but only
emphasise the differences w.r.t. \cite{Iancu:2018hwa}, which mainly refer to a reshuffling of terms.
For the convenience of the reader, our conventions for the light-cone wavefunction formalism are summarised in Appendix~\ref{fieldef}, whereas Appendix~\ref{mateleapp} exhibits all the
matrix elements of the QCD interaction Hamiltonian which are relevant for this calculation.

%
%

\subsubsection{The tri-quark final state}

\begin{figure}[!t]\center
 \includegraphics[scale=0.98]{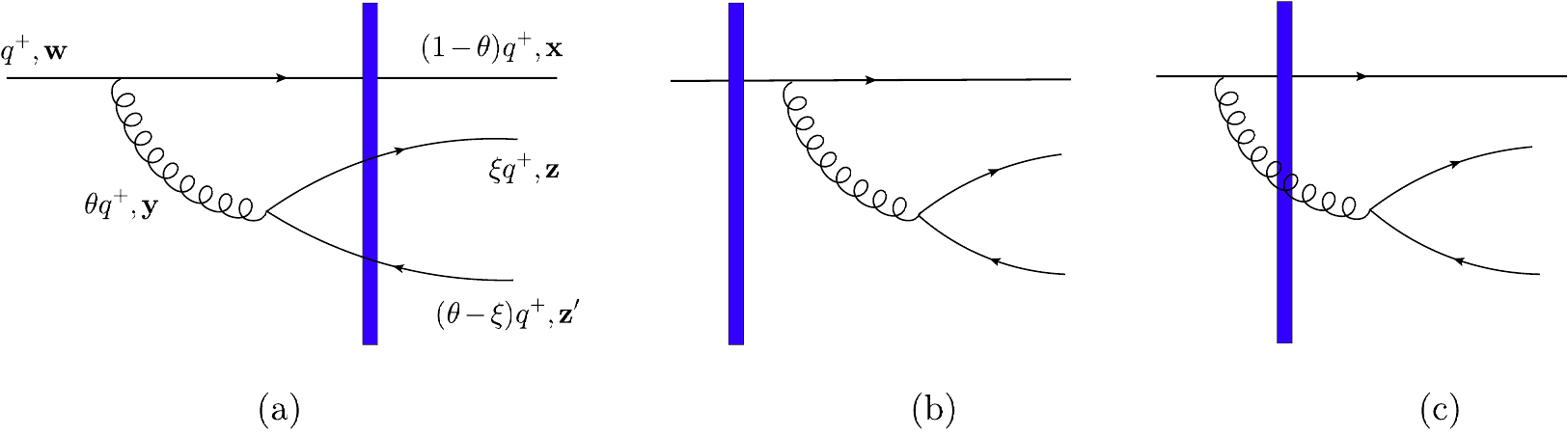}
\caption{The three possible configurations for the interplay between parton branching and scattering, for a final partonic state built with three quarks: (a) initial-state evolution (the first term in the r.h.s. of Eq.~(\ref{out.qqq1})), (b) mixed evolution: the gluon is emitted prior to the scattering, but it splits after the scattering  (the second term in Eq.~(\ref{out.qqq1})), (c) final-state evolution (the third term in Eq.~(\ref{out.qqq1})).}
\label{quarinsert}
\end{figure}

To explain this reshuffling, we shall focus on the three-quark ($qq\bar q$) final state. As explained
in  \cite{Iancu:2018hwa}, this state receives two types of contributions,
\begin{equation}\begin{split}\label{nloasi}
\left|q_{\lambda}^{\alpha}\right\rangle^{out}_{qq\overline{q}}\,=\,\left|q_{\lambda}^{\alpha}\right\rangle _{qq\overline{q}}^{reg}\,+\,\left|q_{\lambda}^{\alpha}\right\rangle _{qq\overline{q}}^{inst},
\end{split}\end{equation}
where the second term, with upper-script {\it inst}, refers to the instantaneous piece of the intermediate
gluon propagator in the LC gauge, where the first term with upper-script {\it reg} 
refers to its regular piece, which is non-local in LC time (see also Figs.~\ref{quarinsert} and
\ref{quarkinst}).

The general expression for the regular piece,
 as obtained to second order in light-cone perturbation theory, reads (see Eq.~(4.2) in \cite{Iancu:2018hwa})
\begin{equation}\begin{split}\label{out.qqq1}
\left|q_{\lambda}^{\alpha}\right\rangle _{qq\overline{q}}^{reg}\,\equiv \,\frac{1}{2}\left|\overline{q}_{\lambda_{3}}^{\rho}\,q_{\lambda_{2}}^{\varrho}\,q_{\lambda_{1}}^{\sigma}\right\rangle & \,\left\{ \frac{\left\langle \overline{q}_{\lambda_{3}}^{\rho}\,q_{\lambda_{2}}^{\varrho}\,q_{\lambda_{1}}^{\sigma}\left|\hat{S}\right|\overline{q}_{\lambda_{7}}^{\delta}\,q_{\lambda_{6}}^{\epsilon}\,q_{\lambda_{5}}^{\kappa}\right\rangle \left\langle \overline{q}_{\lambda_{7}}^{\delta}\,q_{\lambda_{6}}^{\epsilon}\,q_{\lambda_{5}}^{\kappa}\left|\mathsf{H}_{g\rightarrow q\overline{q}}\right|q_{\lambda_{4}}^{\beta}\,g_{i}^{a}\right\rangle \left\langle q_{\lambda_{4}}^{\beta}\,g_{i}^{a}\left|\mathsf{H}_{q\rightarrow qg}\right|q_{\lambda}^{\alpha}\right\rangle }{(E_{qq\overline{q}}-E_{q})\,(E_{qg}-E_{q})}\right.\\
&\,+\,\frac{\left\langle \overline{q}_{\lambda_{3}}^{\rho}\,q_{\lambda_{2}}^{\varrho}\,q_{\lambda_{1}}^{\sigma}\right|\mathsf{H}_{g\rightarrow q\overline{q}}\left|q_{\lambda_{5}}^{\gamma}\,g_{i}^{a}\right\rangle \left\langle q_{\lambda_{5}}^{\gamma}\,g_{i}^{a}\right|\mathsf{H}_{q\rightarrow qg}\left|q_{\lambda_{4}}^{\beta}\right\rangle \left\langle q_{\lambda_{4}}^{\beta}\left|\hat{S}\right|q_{\lambda}^{\alpha}\right\rangle }{(E_{qg}-E_{qq\overline{q}})(E_{q}-E_{qq\overline{q}})}\\
&\left.\,+\,\frac{\left\langle \overline{q}_{\lambda_{3}}^{\rho}\,q_{\lambda_{2}}^{\varrho}\,q_{\lambda_{1}}^{\sigma}\right|\mathsf{H}_{g\rightarrow q\overline{q}}\left|q_{\lambda_{5}}^{\gamma}\,g^{j}\right\rangle \left\langle q_{\lambda_{5}}^{\gamma}\,g^{j}\right|\hat{S}\left|q_{\lambda_{4}}^{\beta}\,g_{i}^{a}\right\rangle \left\langle q_{\lambda_{4}}^{\beta}\,g_{i}^{a}\left|\mathsf{H}_{q\rightarrow qg}\right|q_{\lambda}^{\alpha}\right\rangle }{(E_{qg}-E_{qq\overline{q}})(E_{qg}-E_{q})}\right\} ,
\end{split}\end{equation}
The three terms in the r.h.s. of this equation correspond to the different possible insertions of the
shockwave relative to the two parton branchings, 
as  illustrated by the Feynman graphs in Fig.~\ref{quarinsert}. In the first term, cf. 
Fig.~\ref{quarinsert}.a,  both parton branchings occur prior to the scattering off the shockwave, hence
they are generated by the second-order expansion of the evolution operator $U_I(0,-\infty)$. Accordingly,
both energy denominators involve differences w.r.t. the energy $E_q$ of the initial state. 
Similarly, the second  term, cf. Fig.~\ref{quarinsert}.b, describes
the process where both branchings occur in the final state (after the scattering), hence they are generated
by $U_I(\infty,0)$. The corresponding energy denominators involve differences w.r.t. 
the energy $E_{qq\overline{q}}$ of the final state. Finally, the third term, cf. Fig.~\ref{quarinsert}.c, 
describes one emission prior to the scattering, as generated by the first order term in $U_I(0,-\infty)$, and
a second one after the scattering --- the first order term in $U_I(\infty,0)$.

The reshuffling of terms is based on the following identity:
\begin{equation}\label{iden}
\frac{1}{(E_{qg}-E_{qq\overline{q}})(E_{q}-E_{qq\overline{q}})}\,=\,-\frac{1}{(E_{qq\overline{q}}-E_{q})\,(E_{qg}-E_{q})}\,-\,\frac{1}{(E_{qg}-E_{qq\overline{q}})(E_{qg}-E_{q})}\,,
\end{equation}
which has a simple physical meaning: it shows that in the absence of scattering (i.e. in the limit $\hat S\to 1$),
the three second-order corrections in \eqn{out.qqq1} exactly cancel each other.  
In turn, this is a consequence of the fact
that an on-shell quark  cannot radiate gluons in the absence of any scattering
(see the discussion in \cite{Iancu:2018hwa}). In what follows, we shall rely on this identity to re-express 
the middle term in \eqn{out.qqq1}, cf. Fig.~\ref{quarinsert}.b, as the sum of two negative contributions
which exhibit the same energy denominators as the two other terms there (see Fig.~\ref{reshuffle}
for a graphical illustration of this manipulation). After this rewriting, the contribution of
the initial-state scattering is treated as a subtraction term for the two other contributions (those
shown in Figs.~\ref{quarinsert}.a and \ref{quarinsert}.c).

 \begin{figure}[!t]\center 
\includegraphics[scale=0.85]{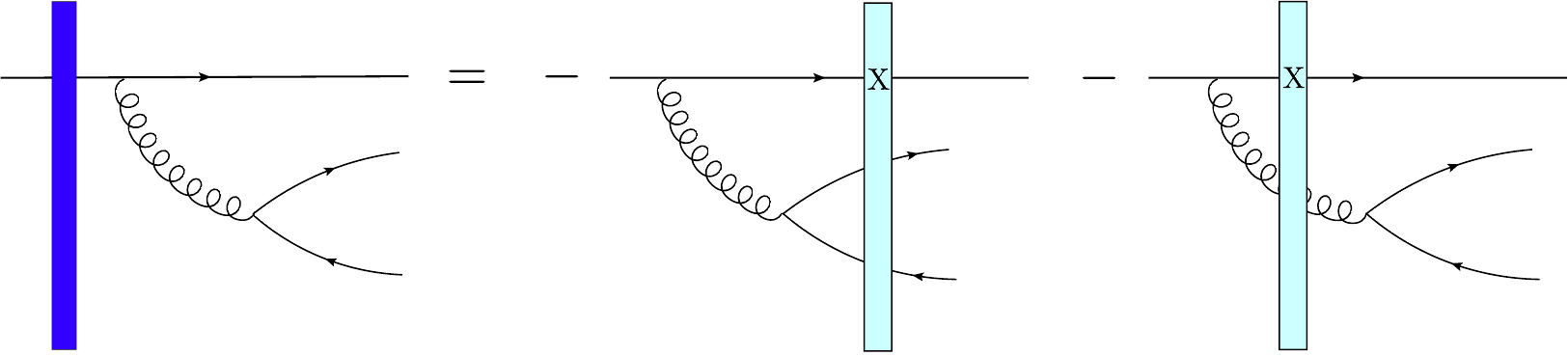}
\caption{The new way in which we express the contribution of the initial-state interaction
(the middle term in \eqn{out.qqq1}) after using the identity \eqn{iden}. 
The light-colour shockwave in the r.h.s. represent the $S$-matrix (a Wilson line) for the scattering of only
one parton: the original quark, denoted with a cross.}
\label{reshuffle}
\end{figure}

\comment{ \begin{figure}[!t]\center 
\includegraphics[scale=0.85]{3quark-subtracted}
\caption{The new way in which we express the contribution of the initial-state interaction
(the middle term in \eqn{out.qqq1}) after using the identity \eqn{iden}. 
The light-colour shockwave in the r.h.s. represent the $S$-matrix (a Wilson line) for the scattering of only
one parton: the original quark, denoted with a cross.}
\label{reshuffle}
\end{figure}
}

It is in principle straightforward to implement this reshuffling of terms via manipulations of the original
results for the trijet final state, as presented in \cite{Iancu:2018hwa}. However, in presenting
our respective results in what follows, we shall often use different integration variables compared to \cite{Iancu:2018hwa}. Moreover, as stated at the beginning of this section, our original results in \cite{Iancu:2018hwa} contained some misprints and errors, that we have  identified (and corrected) 
after explicitly working out special limits. In view of that, it is perhaps not so useful to compare with
Ref.~\cite{Iancu:2018hwa} anymore, but simply rely on the new results to be presented here.

This is in particular true for the $qq\bar q$ component of the outgoing quark state. 
In Ref.~\cite{Iancu:2018hwa}, this has been obtained by separately evaluating the three terms in
the r.h.s. of  \eqn{out.qqq1}, with the result shown in Eq.~(4.9) of that paper. After treating the 
second term there (which corresponds to  Fig.~\ref{quarinsert}.b) as a subtraction term,
performing some changes of variables, and correcting an error, one finds
\begin{equation}\label{qqq_out}
\begin{split}
&\left|q_{\lambda}^{\alpha}(q^{+},\,\bm{w})\right\rangle _{qq\bar{q}}^{reg}\,=\,\frac{g^{2}\,q^{+}}{2(2\pi)^{4}}\int_{\bm{x},\,\bm{z},\,\bm{z}^{\prime}}\,\int_{0}^{1}d\vartheta\,\int_{0}^{\vartheta}d\xi\:\varphi_{\lambda_{2}\lambda_{3}}^{il}\left(\frac{\xi}{\vartheta}\right)\,\phi_{\lambda_{1}\lambda}^{ij}(\vartheta)\,\frac{\vartheta^{2}(1-\vartheta)\,\bm{Z}^{l}\,\bm{R}^{j}}{\bm{Z}^{2}}\\
&\times\left[\frac{V^{\varrho\delta}(\bm{z})\,t_{\delta\epsilon}^{a}\,V^{\dagger\epsilon\rho}(\bm{z}^{\prime})\,V^{\sigma\beta}(\bm{x})\,t_{\beta\alpha}^{a}\,-\,t_{\varrho\rho}^{a}\,t_{\sigma\beta}^{a}\,V^{\beta\alpha}(\bm{w})}{\vartheta^{2}(1-\vartheta)\bm{R}^{2}\,+\,\xi(\vartheta-\xi)\bm{Z}^{2}}\,-\,\frac{t_{\varrho\rho}^{b}\,U^{ba}(\bm{y})\,V^{\sigma\beta}(\bm{x})\,t_{\beta\alpha}^{a}\,-\,t_{\varrho\rho}^{b}\,t_{\sigma\beta}^{b}\,V^{\beta\alpha}(\bm{w})}{\vartheta^{2}(1-\vartheta)\bm{R}^{2}}\right]\\
&\times\delta^{(2)}\big((1-\vartheta)\bm{x}+\vartheta\bm{y} -\bm{w}
\big)\left|\overline{q}_{\lambda_{3}}^{\rho}((\vartheta-\xi)q^{+},\,\bm{z}^{\prime})\,q_{\lambda_{2}}^{\varrho}(\xi q^{+},\,\bm{z})\,q_{\lambda_{1}}^{\sigma}((1-\vartheta)q^{+},\bm{x})\right\rangle.
\end{split}\end{equation}
Our notations are illustrated in Fig.~\ref{quarinsert}):
$\vartheta$ and $\xi$ are the longitudinal momentum fractions (w.r.t. the 
momentum $q^+$ of the incoming quark) of the intermediate gluon and of the quark
produced by the splitting $g\to q\bar q$, respectively. Also, $\bm{x}$ and 
$\bm{y}$ are the transverse coordinates of the quark and the emitted
gluon, whereas $\bm{z}$ and $\bm{z^{\prime}}$ 
similarly refer to the quark and the antiquark generated by the gluon. 
Notice that $\bm{y}$ is not an independent coordinate, since it must
coincide with the center of energy of the $q\bar q$ pair produced by the gluon decay.
Similarly, the transverse coordinate $\bm{w}$ of the incoming quark must be
the same as the center of energy of the quark-gluon pair in the intermediate state,
or of the three quarks in the final state. Explicitly,
\begin{equation}\label{defyw}
\bm{y}\,\equiv\,\frac{\xi\bm{z}+(\vartheta-\xi)\bm{z}^{\prime}}{\vartheta},\qquad
\bm{w}\,=\,(1-\vartheta)\bm{x}+\vartheta\bm{y}\,=\,
(1-\vartheta)\bm{x}+\xi\bm{z}+(\vartheta-\xi)\bm{z}^{\prime}.
\end{equation}
The $\delta$--function inside the integrand enforces the condition on 
$\bm{w}$ shown in the second equality in \eqn{defyw}, thus introducing a constraint on
 the 5 integrations visible in the r.h.s. of \eqn{qqq_out}.
The transverse coordinates denoted with capital letters, that is,
\begin{equation}\label{defRZ}
\bm{R}\,\equiv\,\bm{x}-\bm{y}\,,\qquad \bm{Z}\,\equiv\,\bm{z}-\bm{z}'\,,
\end{equation}
are the transverse separations between the daughter partons after each of the 
emission vertices. Finally, the spinorial structure of the two branching vertices is encoded in
the following tensors:
  \begin{align}\label{defphi}
\phi_{\lambda_{1}\lambda}^{ij}(\vartheta)&\,\equiv\,\chi_{\lambda_{1}}^{\dagger}\left[(2-\vartheta)\delta^{ij}-i\vartheta\varepsilon^{ij}\sigma^{3}\right]\chi_{\lambda}\,=\,\delta_{\lambda\lambda_{1}}\left[(2-\vartheta)\delta^{ij}-2i\vartheta\varepsilon^{ij}\lambda\right],\nonumber\\*[0.2cm]
\varphi_{\lambda_{1}\lambda}^{ij}(\xi)&\,\equiv\,\chi_{\lambda_{1}}^{\dagger}\left[(2\xi-1)\delta^{ij}+i\varepsilon^{ij}\sigma^{3}\right]\chi_{\lambda}\,=\,\delta_{\lambda\lambda_{1}}\left[(2\xi-1)\delta^{ij}+2i\varepsilon^{ij}\lambda\right].
\end{align}

The two terms within the square brackets in \eqn{qqq_out} represent the contributions
of the two graphs in Fig.~\ref{quarinsert}.a and Fig.~\ref{quarinsert}.c respectively, 
{\it minus} the respective subtraction terms generated by the graph in Fig.~\ref{quarinsert}.b,
as shown in Fig.~\ref{reshuffle}. Thanks to these subtractions, the
numerator in each of these terms vanishes in the limit where the 
transverse separations between the daughter partons (as appearing in the respective
denominator) are shrinking to zero. Consider e.g. the second term: when $\bm{R}=\bm{x}-\bm{y}\to 0$, 
meaning that the three coordinates $\bm{x},\,\bm{y}$ and $\bm{w}$ become coincident with each other, 
one can use the following identity  (cf. \eqn{VUid})
\begin{equation}\label{VUid1}
U^{ba}(\bm{x})\,t_{\beta\alpha}^{a}\,=\,
\left[V^{\dagger}(\bm{x})\,t^{b}\,V(\bm{x})\right]_{\beta\alpha}\,,
\end{equation}
to verify that the respective numerator vanishes, as announced. Similarly, the numerator in
the first line vanishes when both $\bm{R}\to 0$ and $\bm{Z}\to 0$ (meaning that all transverse
coordinates become degenerate: $\bm{w}=\bm{x}=\bm{y}=\bm{z}=\bm{z}'$). 
  Furthermore the two terms within the square brackets in \eqn{qqq_out} cancel each other 
 in the limit $ \bm{Z}\to 0$ (i.e. $\bm{z}=\bm{z}'=\bm{y}$) at fixed $\bm{R}$. To see this,
 one should use  yet another version of the identity \eqn{VUid}, namely,
 \begin{equation}\label{VUid2}
\left[V(\bm{y})\,t^{a}\,V^{\dagger}(\bm{y})\right]_{\alpha\beta}\,=\,t_{\alpha\beta}^{b}\,U^{ba}(\bm{y}),
\end{equation}
which expresses the fact that the scattering cannot distinguish a $q\bar q$ fluctuation of zero size
from its parent gluon (recall the discussion after \eqn{VUid}).

These properties of the LCWF  will be later useful in demonstrating the cancellation of  ``ultraviolet'' (short-distance)  singularities in the calculation of cross-sections for particle
production. Such properties are more difficult to check on the original expression for
the LCWF, i.e. Eq.~(4.9) in  Ref.~\cite{Iancu:2018hwa}.

The previous discussion also implies that the second term in \eqn{qqq_out} can be generated from 
the first term there via the simultaneous replacements $\bm{z}\to \bm{y}$ and $\bm{z}'\to \bm{y}$.
This allows us to introduce a simpler notation, that we shall systematically use throughout
this paper. Specifically, \eqn{qqq_out} can be equivalently written as
\begin{equation}\label{qqq_outgoing}
\begin{split}
&\left|q_{\lambda}^{\alpha}(q^{+},\,\bm{w})\right\rangle _{qq\bar{q}}^{reg}\,=\,\frac{g^{2}\,q^{+}}{2(2\pi)^{4}}\int_{\bm{x},\,\bm{z},\,\bm{z}^{\prime}}\,\int_{0}^{1}d\vartheta\,\int_{0}^{\vartheta}d\xi\:\varphi_{\lambda_{2}\lambda_{3}}^{il}\left(\frac{\xi}{\vartheta}\right)\,\phi_{\lambda_{1}\lambda}^{ij}(\vartheta)\,\frac{\vartheta(1-\vartheta)\,\bm{R}^{j}\,\bm{Z}^{l}}{\bm{Z}^{2}}\\
&\qquad\qquad 
\times\left[\frac{V^{\varrho\delta}(\bm{z})\,t_{\delta\epsilon}^{a}\,V^{\dagger\epsilon\rho}(\bm{z}^{\prime})\,V^{\sigma\beta}(\bm{x})\,t_{\beta\alpha}^{a}\,-\,t_{\varrho\rho}^{a}\,t_{\sigma\beta}^{a}\,V^{\beta\alpha}(\bm{w})}{\vartheta^{2}(1-\vartheta)\bm{R}^{2}\,+\,\xi(\vartheta-\xi)\bm{Z}^{2}}\,-\,\big(
\bm{z}, \,\bm{z}^{\prime}\rightarrow\bm{y}\big)\right]\\
&\qquad\qquad 
\times\delta^{(2)}\big((1-\vartheta)\bm{x}+\vartheta\bm{y} -\bm{w}\big)
\left|\overline{q}_{\lambda_{3}}^{\rho}((\vartheta-\xi)q^{+},\,\bm{z}^{\prime})\,q_{\lambda_{2}}^{\varrho}(\xi q^{+},\,\bm{z})\,q_{\lambda_{1}}^{\sigma}((1-\vartheta)q^{+},\bm{x})\right\rangle .
\end{split}\end{equation}

 \begin{figure}[!t]
\center \includegraphics[scale=0.8]{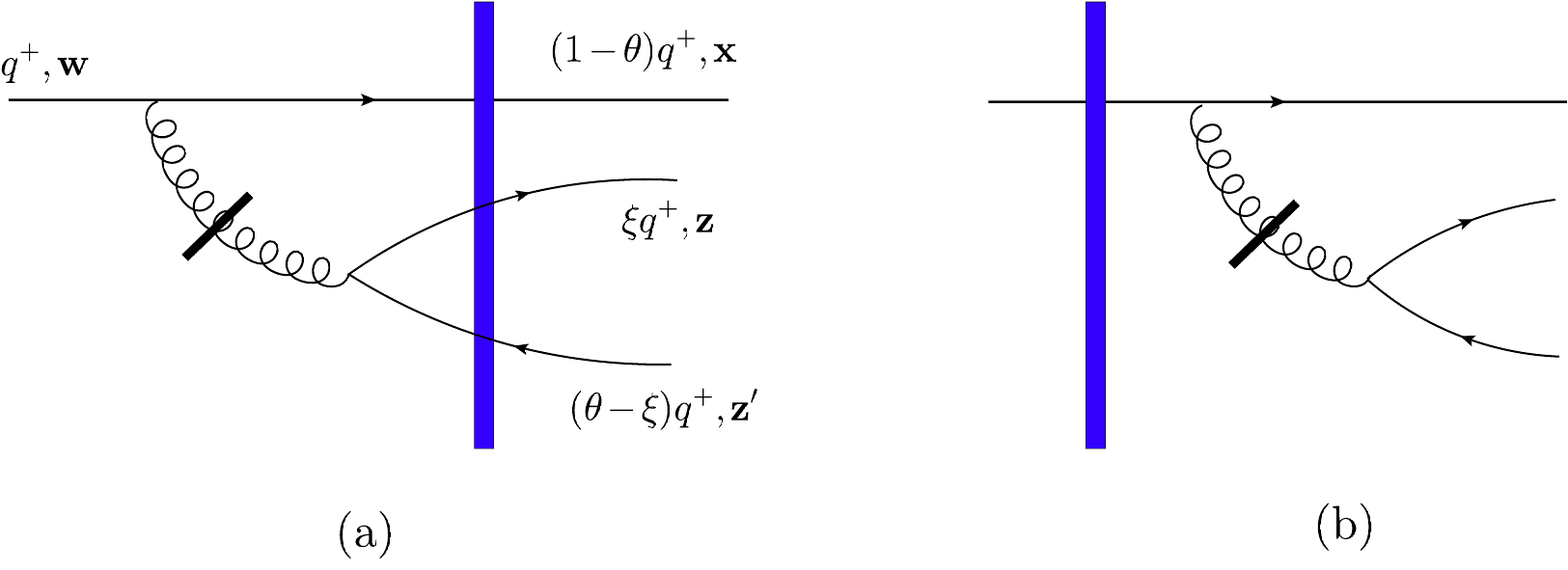}
\caption{The two possible insertions of the shockwave in the case where the final quark antiquark pair has been produced via an instantaneous gluon interaction: (a) initial-state evolution, (b)  final-state evolution.}
\label{quarkinst}
\end{figure}

Besides this ``regular'' contribution, which involves an intermediate state with a propagating
gluon, there is an additional contribution built with the instantaneous piece of the gluon
propagator (see Fig.~\ref{quarkinst}). 
This is given by Eq.~(4.13) in Ref.~\cite{Iancu:2018hwa}, which with the present
conventions reads
\begin{align}\label{fin.qqq2}
\left|q_{\lambda}^{\alpha}(q^{+},\,\bm{w})\right\rangle _{qq\bar{q}}^{inst}&=-\,\frac{g^{2}\,q^{+}}{(2\pi)^{4}}\int_{\bm{x},\,\bm{z},\,\bm{z}^{\prime}}\,\int_{0}^{1}d\vartheta\,\int_{0}^{\vartheta}d\xi\:\frac{(1-\vartheta)\xi(\vartheta-\xi)}{\vartheta\left(\vartheta^{2}(1-\vartheta)\bm{R}^{2}\,+\,\xi(\vartheta-\xi)\bm{Z}^{2}\right)}\nn
&\times\left[V^{\varrho\delta}(\bm{z})\,t_{\delta\epsilon}^{a}\,V^{\dagger\epsilon\rho}(\bm{z}^{\prime})\,V^{\sigma\beta}(\bm{x})\,t_{\beta\alpha}^{a}\,-\,t_{\varrho\rho}^{a}\,t_{\sigma\beta}^{a}\,V^{\beta\alpha}(\bm{w})\right]\nn
&\times\delta^{(2)}\left(\bm{w}-\bm{c}\right)\left|\overline{q}_{\lambda_{3}}^{\rho}((\vartheta-\xi)q^{+},\,\bm{z}^{\prime})\,q_{\lambda_{2}}^{\varrho}(\xi q^{+},\,\bm{z})\,q_{\lambda_{1}}^{\sigma}((1-\vartheta)q^{+},\bm{x})\right\rangle ,
 \end{align}
 where $\bm{c}\equiv (1-\vartheta)\bm{x}+\xi\bm{z}+(\vartheta-\xi)\bm{z}^{\prime}$ is a compact
notation for the center of energy of the final quark triplet. Note that, in this case, 
there is no intermediate gluon state: the cut gluon propagator in Fig.~\ref{quarkinst} together
with the two attached QCD vertices defines a local effective vertex for the decay
$q\to q q\bar q$ of the initial quark into the three final ones. So, in particular, there is no analog
of the second term in \eqn{qqq_out} (the one 
denoted as $(\bm{z}, \,\bm{z}^{\prime}\rightarrow\bm{y})$  in \eqn{qqq_outgoing}). 
That said, it is still convenient
to introduce a compact notation (similar to that in  \eqn{qqq_outgoing})
for the {\it complete} triquark outgoing state, i.e. the sum
of the two contributions in Eqs.~\eqref{qqq_out}  and \eqref{fin.qqq2}. This reads
\begin{equation}\label{qqq_final}
\begin{split}
&\left|q_{\lambda}^{\alpha}(q^{+},\,\bm{w})\right\rangle _{qq\bar{q}}
\,=\,\frac{g^{2}\,q^{+}}{2(2\pi)^{4}}\int_{\bm{x},\,\bm{z},\,\bm{z}^{\prime}}\,\int_{0}^{1}d\vartheta\,\int_{0}^{\vartheta}d\xi\: \,\frac{\bm{R}^{j}\,\bm{Z}^{l}}{\bm{Z}^{2}}\\
&\qquad\qquad 
\times\left[\Phi_{\lambda_{3}\lambda_{2}\lambda_{1}\lambda}^{jl}\,
\frac{V^{\varrho\delta}(\bm{z})\,t_{\delta\epsilon}^{a}\,V^{\dagger\epsilon\rho}(\bm{z}^{\prime})\,V^{\sigma\beta}(\bm{x})\,t_{\beta\alpha}^{a}\,-\,t_{\varrho\rho}^{a}\,t_{\sigma\beta}^{a}\,V^{\beta\alpha}(\bm{w})}{\vartheta^{2}(1-\vartheta)\bm{R}^{2}\,+\,\xi(\vartheta-\xi)\bm{Z}^{2}}\,-\,\big(
\bm{z}, \,\bm{z}^{\prime}\rightarrow\bm{y}\big)\right]\\
&\qquad\qquad 
\times\delta^{(2)}\big(\bm{w}-\bm{c}\big)
\left|\overline{q}_{\lambda_{3}}^{\rho}((\vartheta-\xi)q^{+},\,\bm{z}^{\prime})\,q_{\lambda_{2}}^{\varrho}(\xi q^{+},\,\bm{z})\,q_{\lambda_{1}}^{\sigma}((1-\vartheta)q^{+},\bm{x})\right\rangle .
\end{split}\end{equation}
where $\Phi_{\lambda_{3}\lambda_{2}\lambda_{1}\lambda}^{jl}\equiv \Phi_{\lambda_{3}\lambda_{2}\lambda_{1}\lambda}^{jl}(\bm{x},\,\bm{z},\,\bm{z}',\,\vartheta,\,\xi)$
is an effective vertex for the splitting of the original quark into the three final quarks,
which includes contributions from both propagating and instantaneous intermediate gluons:
  \begin{equation}\label{vertexPhi}
\Phi_{\lambda_{3}\lambda_{2}\lambda_{1}\lambda}^{jl}(\bm{x},\,\bm{z},\,\bm{z}',\,\vartheta,\,\xi)\,\equiv\,\vartheta(1-\vartheta)\varphi_{\lambda_{2}\lambda_{3}}^{il}\left(\frac{\xi}{\vartheta}\right)\,\phi_{\lambda_{1}\lambda}^{ij}(\vartheta)\,-\,\delta_{\lambda_{2}\lambda_{3}}\delta_{\lambda_{1}\lambda}\delta^{jl}
\,\frac{2(1-\vartheta)\xi(\vartheta-\xi)}{\vartheta}\frac{\bm{Z}^{2}}{\bm{R}\cdot\bm{Z}}.
\end{equation}
It is understood that the second piece in \eqn{vertexPhi} vanishes after substituting 
 $\bm{z}\to \bm{y}$ and $\bm{z}'\to \bm{y}$ :
  \begin{equation}\label{vertexGQ-limit}
\Phi_{\lambda_{3}\lambda_{2}\lambda_{1}\lambda}^{jl}(\bm{x},\,\bm{y},\,\bm{y},\,\vartheta,\,\xi)
\,=\,\vartheta(1-\vartheta)\varphi_{\lambda_{2}\lambda_{3}}^{il}\left(\frac{\xi}{\vartheta}\right)\,\phi_{\lambda_{1}\lambda}^{ij}(\vartheta)\,.
\eeq
Hence the subtraction of the contribution from the intermediate gluon state is truly
performed only for the case of a propagating gluon, as it should. Ultimately, one should keep in
mind that the compact expression for the three-quark outgoing state in \eqn{qqq_final}
is merely a convenient {\it notation}, which is rather formal, but will be useful to simplify the 
writing for our final results.

\subsubsection{The final state with one quark and two gluons}
\label{sec:qgg}

In this subsection we shall exhibit the remaining 3-parton Fock space components
of the quark outgoing LCWF: those in which the original quark is accompanied by two gluons.
There are two possible topologies for the gluon emissions, as shown in Figs.~\ref{final-qgg1} and
\ref{final-qgg2}, respectively.  Besides these ``regular'' graphs, where the intermediate parton (quark
or gluon) is propagating, there are similar graphs which involve the instantaneous piece
of the respective propagator. (These are not shown, but can be easily inferred by comparing with 
Fig.~\ref{quarkinst}.) All these graphs have been computed in \cite{Iancu:2018hwa}, with results that
we shall here rewrite by regrouping terms and also correct whenever necessary, 
as explained in the previous subsection on the example of the three-quark final state.

 \begin{figure}[!t]\center
 \includegraphics[scale=0.98]{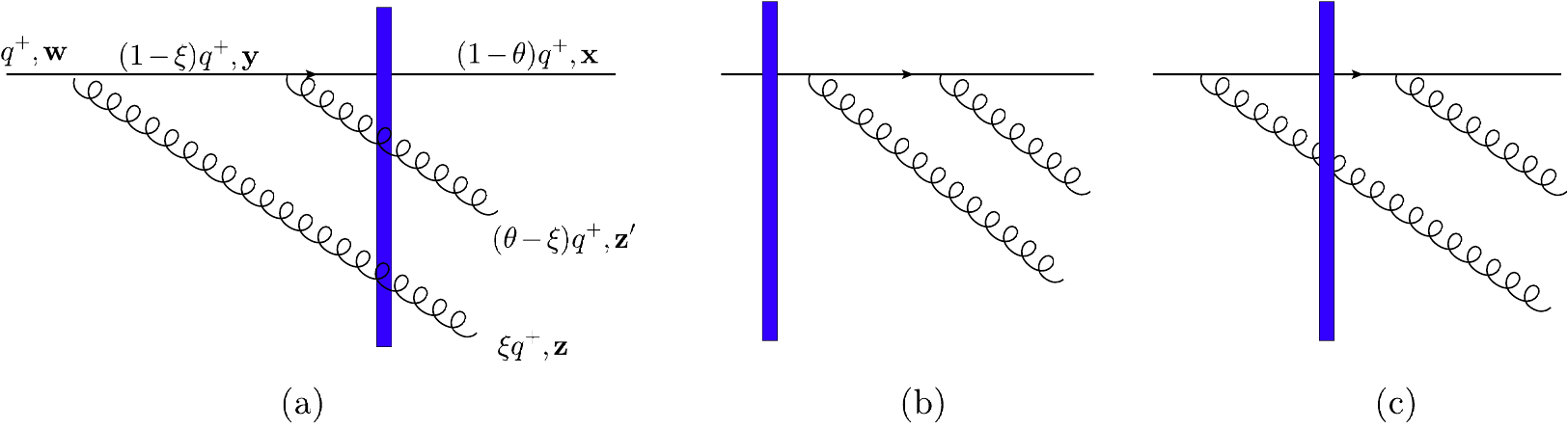}
\caption{The three possible configurations for the interplay between parton branching and scattering, for a final partonic state built with three quarks: (a) initial-state evolution (the first term in the r.h.s. of Eq.~(\ref{out.qqq1})), (b) mixed evolution: the gluon is emitted prior to the scattering, but it splits after the scattering  (the second term in Eq.~(\ref{out.qqq1})), (c) final-state evolution (the third term in Eq.~(\ref{out.qqq1})).}
\label{final-qgg1}
\end{figure}

We start with the graphs where both gluons are emitted by the original quark, cf. Fig.~\ref{final-qgg1}.
Their contribution is shown in Eq.~(4.14) of \cite{Iancu:2018hwa},  that can be rewritten as
  \begin{equation}\label{qgg1_out}
  \begin{split}
&\left|q_{\lambda}^{\alpha}(q^{+},\,\bm{w})\right\rangle _{qgg}^{reg,\,1}\,=-\frac{g^{2}q^{+}}{2(2\pi)^{4}}\int_{\bm{x},\,\bm{z},\,\bm{z}^{\prime}}\int_{0}^{1}d\vartheta\,\int_{0}^{\vartheta}d\xi\,
\frac{\sqrt{\xi}(1-\xi)}{\sqrt{\vartheta-\xi}}\,\tau_{\lambda_{1}\lambda_{2}}^{jm}(\vartheta,\,\xi)\,\phi_{\lambda_{2}\lambda}^{il}(\xi)\,\frac{\bm{Y}^{l}\,(\bm{X}^{\prime})^{m}}{(\bm{X}^{\prime})^{2}}\\
&\qquad\qquad 
\times\left[\frac{V^{\delta\gamma}(\bm{x})\,U^{db}(\bm{z}^{\prime})\,U^{ca}(\bm{z})\,t_{\gamma\beta}^{b}\,t_{\beta\alpha}^{a}-\,t_{\delta\gamma}^{d}\,t_{\gamma\beta}^{c}\,V^{\beta\alpha}(\bm{w})}{(1-\vartheta)(\vartheta-\xi)(\bm{X}^{\prime})^{2}\,+\,\xi(1-\xi)^{2}\bm{Y}^{2}}\,-\,\left(\bm{x},\bm{z}^{\prime}\rightarrow\bm{y}\right)\right]\\
&\qquad\qquad 
\times\delta^{(2)}\left(\bm{w}-\bm{c}\right)\left|q_{\lambda_{1}}^{\delta}((1-\vartheta)q^{+},\,\bm{x})\,g_{i}^{c}(\xi q^{+},\,\bm{z})\,g_{j}^{d}((\vartheta-\xi)q^{+},\,\bm{z}^{\prime})\right\rangle ,
\end{split}\end{equation}
where $\bm{y}$ denotes the transverse position of the intermediate quark and 
$\bm{c}$ is the center of energy of the three final partons (and coincides with the position
$\bm{w}$ of the original quark):
\begin{equation}\label{defyw2}
\bm{y}\,\equiv\,\frac{(1-\vartheta)\bm{x}+(\vartheta-\xi)\bm{z}^{\prime}}{1-\xi}\,,\qquad
\bm{c}\,\equiv\,(1-\xi)\bm{y}+\xi\bm{z}\,=\,
(1-\vartheta)\bm{x}+\xi\bm{z}+(\vartheta-\xi)\bm{z}^{\prime}.
\end{equation}
Furthermore,
\begin{equation}\label{defYX}
\bm{Y}\,\equiv\,\bm{y}-\bm{z}\,,\qquad \bm{X}'\,\equiv\,\bm{x}-\bm{z}'\,,
\end{equation}
are the transverse separations between the daughter partons after each  emission vertex.
The spinorial structure is encoded in $\phi_{\lambda_{1}\lambda}^{il}(\xi)$
(cf. \eqn{defphi}) for the first emission vertex and respectively 
\begin{equation}
\tau_{\lambda_{1}\lambda}^{ij}(\vartheta,\,\xi)\,\equiv\,\chi_{\lambda_{1}}^{\dagger}\left[(2-\xi-\vartheta)\delta^{ij}+i(\xi-\vartheta)\varepsilon^{ij}\sigma^{3}\right]\chi_{\lambda},
\end{equation}
for the second vertex; notice the relation $\tau_{\lambda_{1}\lambda}^{ij}(\vartheta,\,0)\,=\,\phi_{\lambda_{1}\lambda}^{ij}(\vartheta)$.

The structure of \eqn{qgg1_out} is reminiscent of that in \eqn{qqq_outgoing}. The first term within
the squared brackets represents the contribution of the final-state interaction, cf. 
Fig.~\ref{final-qgg1}.a, minus a piece generated by the initial-state interaction, cf. 
Fig.~\ref{final-qgg1}.b. The second term, as obtained from the first term via the replacements
$\bm{x},\bm{z}^{\prime}\rightarrow\bm{y}$, refers to the intermediate-state interaction,  cf. 
Fig.~\ref{final-qgg1}.c, minus the remaining contribution of Fig.~\ref{final-qgg1}.b.

The contribution of the corresponding instantaneous graph, for which there is no analog of
Fig.~\ref{final-qgg1}.c, is found as 
 \begin{equation}\label{qgg1_inst} 
 \begin{split}
&\left|q_{\lambda}^{\alpha}(q^{+},\,\bm{w})\right\rangle _{qgg}^{inst,\,1}\,=\,-\frac{g^{2}q^{+}}{2(2\pi)^{4}}\int_{\bm{x},\,\bm{z},\,\bm{z}^{\prime}}\int_{0}^{1}d\vartheta\,\int_{0}^{\vartheta}d\xi\,(1-\vartheta)\sqrt{\xi(\vartheta-\xi)}\,\chi_{\lambda_{1}}^{\dagger}\,\sigma_{j}\,\sigma_{i}\,\chi_{\lambda}\\
&\qquad\qquad 
\times\frac{V^{\delta\gamma}(\bm{x})\,U^{db}(\bm{z}^{\prime})\,U^{ca}(\bm{z})\,t_{\gamma\beta}^{b}\,t_{\beta\alpha}^{a}-\,t_{\delta\gamma}^{d}\,t_{\gamma\beta}^{c}\,V^{\beta\alpha}(\bm{w})}{(1-\vartheta)(\vartheta-\xi)(\bm{X}^{\prime})^{2}\,+\,\xi(1-\xi)^{2}\bm{Y}^{2}}\\
&\qquad\qquad 
\times\delta^{(2)}\left(\bm{w}-\bm{c}\right)\left|q_{\lambda_{1}}^{\delta}((1-\vartheta)q^{+},\,\bm{x})\,g_{i}^{c}(\xi q^{+},\,\bm{z})\,g_{j}^{d}((\vartheta-\xi)q^{+},\,\bm{z}^{\prime})\right\rangle ,
\end{split}\end{equation}
As in the previous subsection, it is convenient to group together ``regular'' and ``instantaneous'' contributions
in a unique expression involved a non-local effective vertex --- here, for the splitting of the original
quark into a $qgg$ system. Then the sum of  \eqn{qgg1_out}  and \eqn{qgg1_inst}  reads
  \begin{equation}\label{qgg1_tot}
  \begin{split}
&\left|q_{\lambda}^{\alpha}(q^{+},\,\bm{w})\right\rangle _{qgg}^{1}\,=-\frac{g^{2}q^{+}}{2(2\pi)^{4}}\int_{\bm{x},\,\bm{z},\,\bm{z}^{\prime}}\int_{0}^{1}d\vartheta\,\int_{0}^{\vartheta}d\xi\ \frac{\bm{Y}^{m}\,(\bm{X}^{\prime})^{n}}{(\bm{X}^{\prime})^{2}}\\
&\qquad\qquad\times\left[\Xi_{\lambda_{1}\lambda}^{ijmn}\,
\frac{V^{\delta\gamma}(\bm{x})\,U^{db}(\bm{z}^{\prime})\,U^{ca}(\bm{z})\,t_{\gamma\beta}^{b}\,t_{\beta\alpha}^{a}-\,t_{\delta\gamma}^{d}\,t_{\gamma\beta}^{c}\,V^{\beta\alpha}(\bm{w})}{(1-\vartheta)(\vartheta-\xi)(\bm{X}^{\prime})^{2}\,+\,\xi(1-\xi)^{2}\bm{Y}^{2}}\,-\,\left(\bm{x},\bm{z}^{\prime}\rightarrow\bm{y}\right)\right]\\
&\qquad\qquad\times\delta^{(2)}\left(\bm{w}-\bm{c}\right)\left|q_{\lambda_{1}}^{\delta}((1-\vartheta)q^{+},\,\bm{x})\,g_{i}^{c}(\xi q^{+},\,\bm{z})\,g_{j}^{d}((\vartheta-\xi)q^{+},\,\bm{z}^{\prime})\right\rangle ,
\end{split}\end{equation}
with the effective vertex $\Xi_{\lambda_{1}\lambda}^{ijmn}\equiv \Xi_{\lambda_{1}\lambda}^{ijmn}(\bm{x},\,\bm{z},\,\bm{z}',\,\vartheta,\,\xi)$ defined as 
\beq\label{vertexXi}
\Xi_{\lambda_{1}\lambda}^{ijmn}(\bm{x},\,\bm{z},\,\bm{z}',\,\vartheta,\,\xi)\,\equiv\,
\sqrt{\xi(\vartheta-\xi)}\left(
\frac{1-\xi}{\vartheta-\xi}\,
\tau_{\lambda_{1}\lambda_{2}}^{jn}(\vartheta,\,\xi)\,\phi_{\lambda_{2}\lambda}^{im}(\xi)\,
-\,\delta^{mn}
(1-\vartheta)\,\big(\chi_{\lambda_{1}}^{\dagger}\sigma_{j}\sigma_{i}\chi_{\lambda}\big)
\, \frac{(\bm{X}^{\prime})^{2}}{\bm{Y}\cdot \bm{X}'}\right)\,.
\eeq
It is understood that the second piece in \eqn{vertexXi} vanishes when $\bm{X}'=\bm{x}-\bm{z}'\,\to 0$.

The remaining topology for a final state with two gluons and a quark is very similar to that previously
discussed for the tri-quark final state (compare the graphs in Figs.~\ref{final-qgg2} and
\ref{quarinsert}, respectively). Hence, we shall simply present here the final result --- the analog of 
\eqn{qqq_final} --- without further explanations. This reads 
 \begin{equation}\label{qgg2_tot}
 \begin{split}
&\left|q_{\lambda}^{\alpha}(q^{+},\,\bm{w})\right\rangle _{qgg}^{2}\,=\,\frac{ig^{2}\,q^{+}}{2(2\pi)^{4}}\int_{\bm{x},\,\bm{z},\,\bm{z}^{\prime}}\,\int_{0}^{1}d\vartheta\,\int_{0}^{\vartheta}d\xi\,\,\frac{\bm{R}^{m}\,\bm{Z}^{n}}{\bm{Z}^{2}}
\\
&\qquad\qquad\times\left[\Pi_{\lambda_{1}\lambda}^{ijmn}\,
\frac{f^{abe}\,V^{\gamma\beta}(\bm{x})\,U^{de}(\bm{z}^{\prime})\,U^{cb}(\bm{z})\,t_{\beta\alpha}^{a}\,-\,f^{acd}\,t_{\gamma\beta}^{a}\,V^{\beta\alpha}(\bm{w})}{\vartheta^{2}(1-\vartheta)\bm{R}^{2}\,+\,\xi(\vartheta-\xi)\bm{Z}^{2}}\,-\,\left(\bm{z},\bm{z}^{\prime}\rightarrow\bm{y}\right)\right]\\
&\qquad\qquad\times\delta^{(2)}\left(\bm{w}-\bm{c}\right)\left|q_{\lambda_{1}}^{\gamma}((1-\vartheta)q^{+},\,\bm{x})\,g_{i}^{c}(\xi q^{+},\,\bm{z})\,g_{j}^{d}((\vartheta-\xi)q^{+},\,\bm{z}^{\prime})\right\rangle,
\end{split}\end{equation}
where we recall that $\bm{Z}=\bm{z}-\bm{z}'$, $\bm{R}=\bm{x}-\bm{y}$, and  $\bm{c}=(1-\vartheta)\bm{x}+\vartheta\bm{y}$, with $\bm{y}$ defined in Eqs.~\eqref{defyw}.

The effective vertex $\Pi_{\lambda_{1}\lambda}^{ijmn}\equiv \Pi_{\lambda_{1}\lambda}^{ijmn}(\bm{x},\,\bm{z},\,\bm{z}',\,\vartheta,\,\xi)$ is defined as
\beq\label{vertexPi}
\Pi_{\lambda_{1}\lambda}^{ijmn}(\bm{x},\,\bm{z},\,\bm{z}',\,\vartheta,\,\xi)\,\equiv\,
\sqrt{\xi(\vartheta-\xi)}(1-\vartheta)\left(\vartheta\Gamma^{nlij}(\vartheta,\,\xi)\,\phi_{\lambda_{1}\lambda}^{lm}(\vartheta)\,-\,\delta^{ij}\delta^{mn}\delta_{\lambda\lambda_{1}}\,\frac{\vartheta-2\xi}{\vartheta}\,\frac{\bm{Z}^{2}}{\bm{R}\cdot\bm{Z}}\right)
\,,
\eeq
where the first term in the r.h.s. is the three-gluon vertex
\begin{equation}\label{Gamma}
\Gamma^{nlij}(\vartheta,\,\xi)\,\equiv\,\frac{1}{\vartheta}\,\delta_{nl}\delta_{ij}\,-\,\frac{1}{\xi}\,
\delta_{ni}\delta_{lj}\,-\,\frac{1}{\vartheta-\xi}\,\delta_{nj}\delta_{li}\,,
\end{equation}
whereas the second term comes from the graphs with an instantaneous intermediate gluon.

To obtain an explicit expression for the subtracted term denoted as $\left(\bm{z},\bm{z}^{\prime}\rightarrow\bm{y}\right)$ in \eqn{qgg2_tot}, the following identity is useful:
 \begin{equation}
f^{abc}\,U^{dc}(\bm{y})\,U^{eb}(\bm{y})\,=\,f^{bde}\,U^{ba}(\bm{y})\,.
\end{equation}
As before, one can check that, thanks to this subtraction, the two terms within the square brackets
cancel each other in the limit $\bm{Z}\to 0$.

\begin{figure}[!t]\center
 \includegraphics[scale=0.98]{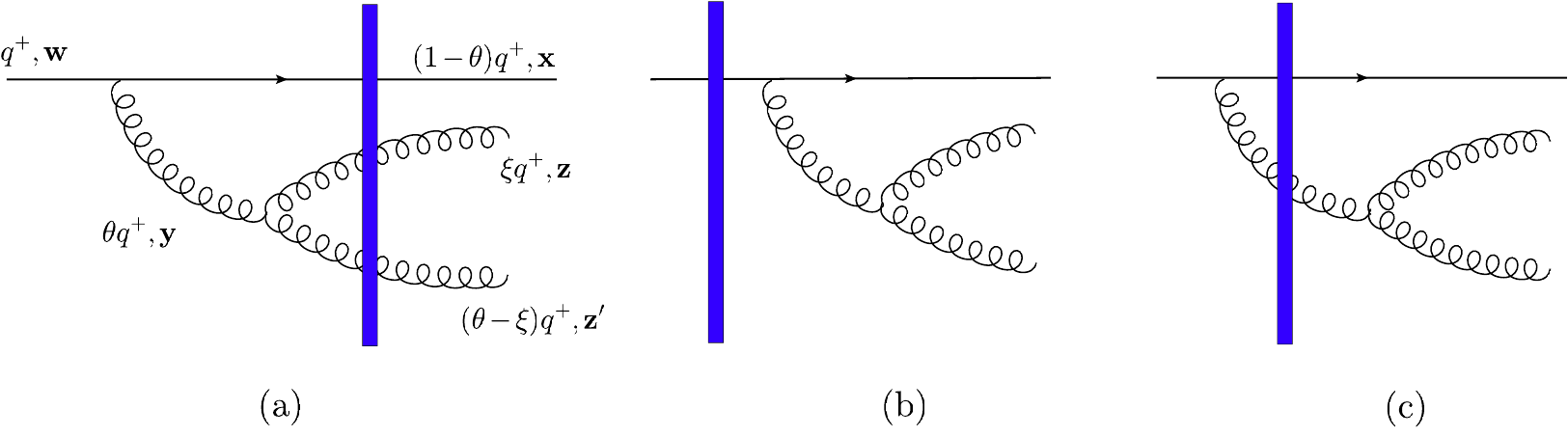}
\caption{The three possible shockwave insertions for a final partonic state built with one quark and two gluons: (a) final-state scattering, (b) initial-state scattering, and
(c) intermediate-state scattering.}
\label{final-qgg2}
\end{figure}

\subsection{The trijet cross-section}
\label{trijet-XSEC}

The leading-order cross-section for the inclusive three partons (``trijet'') production in the quark-nucleus collision
is computed similarly to \eqn{locrosdefin}, that is, as the expectation value of the product of three number-density
operators (themselves built with Fock space operators for bare partons).

\subsubsection{The $qq\bar q$ final state}
\label{trijet-qq}

For the $qq\bar q$ final state, one writes
 (a factor $2\pi\delta(k_{1}^{+}+k_{2}^{+}+k_{3}^{+}-q^{+})$ is implicit in the l.h.s.)
\begin{equation}\begin{split}\label{qqqcross}
\frac{d\sigma^{qA\rightarrow qq\overline{q}+X}}{d^{3}k_{1}\,d^{3}k_{2}\,d^{3}k_{3}}&\,\equiv\,\frac{1}{2N_{c}}\ {}^{out}\!\left\langle q_{\lambda}^{\alpha}(q^{+},\,\bm{q}=\bm{0})\right|\,\hat{\mathcal{N}}_{q}(k_{1})\,\hat{\mathcal{N}}_{q}(k_{2})\,\hat{\mathcal{N}}_{\overline{q}}(k_{3})\,\left|q_{\lambda}^{\alpha}(q^{+},\,\bm{q}=\bm{0})\right\rangle ^{out}\\
&=\,\frac{1}{2N_{c}}\,\int_{\bm{w},\,\overline{\bm{w}}}
\!\!{}_{\ \ \ qq \bar{q}}^{\ \ \ out}\!\left\langle q_{\lambda}^{\alpha}(q^{+},\,\overline{\bm{w}})\right|\,\hat{\mathcal{N}}_{q}(k_{1})\,\hat{\mathcal{N}}_{q}(k_{2})\,\hat{\mathcal{N}}_{\overline{q}}(k_{3})\,\left|q_{\lambda}^{\alpha}(q^{+},\,\bm{w})\right\rangle _{qq\bar{q}}^{out},
 \end{split}\end{equation}
where as shown in the second line, only the $qq\bar q$ Fock-state component of the outgoing quark state,
cf. \eqn{qqq_final}, is involved in this calculation. A priori there are four possible contractions for the product
$\hat{\mathcal{N}}_{q}(k_{1})\,\hat{\mathcal{N}}_{q}(k_{2})$ of quark number density operators. However, 
two of these contractions correspond to non-planar graphs --- these are graphs where the original
parton on one side of the cut is contracted with the quark generated by the gluon decay on the other
side of the cut (see Fig.~13.b in \cite{Iancu:2018hwa}) --- which are suppressed in the multicolour limit
$N_c\to\infty$. In what follows we shall systematically work in this limit, since it allows to
simplify the colour structure of our results and also to render them physically transparent. 
Hence we shall keep only
two of the four possible contractions --- those corresponding to planar graphs as illustrated in Fig.~\ref{S6q}.
We shall explicitly write the result for the case where the
leading quark has momentum $k_1$ and the other quark has momentum $k_2$. The other term
can be simply obtained by permuting these two momenta. 

A straightforward calculation using the outgoing state in \eqn{qqq_final} together with the Fock space
rules summarised in Appendix \ref{fieldef} yields\footnote{In the large-$N_c$ limit under consideration,
the quark Casimir factor  apparent in \eqn{trijetqq}
can be as well approximated as $C_F\simeq N_c/2$; this applies to all the results 
involving $C_F$ to be shown in this paper.}
 \begin{equation}\begin{split}\label{trijetqq}
&\frac{d\sigma^{qA\rightarrow qq\overline{q}+X}}{dk_{1}^{+}\,d^{2}\bm{k}_{1}\,dk_{2}^{+}\,d^{2}\bm{k}_{2}\,dk_{3}^{+}\,d^{2}\bm{k}_{3}}=\frac{\alpha_{s}^{2}\,C_{F}\,N_{f}}{2(2\pi)^{10}(q^{+})^{2}}\,
\delta(q^{+}-k_{1}^{+}-k_{2}^{+}-k_{3}^{+})\\*[0.2cm]
&\qquad\qquad\times
\int_{\overline{\bm{x}},\,\bm{\overline{z}},\,\bm{\overline{z}}^{\prime},\,\bm{x},\,\bm{z},\,\bm{z}^{\prime}}\,e^{-i\bm{k}_{1}\cdot(\bm{x}-\overline{\bm{x}})-i\bm{k}_{2}\cdot(\bm{z}-\overline{\bm{z}})-i\bm{k}_{3}\cdot(\bm{z}^{\prime}-\overline{\bm{z}}^{\prime})}\,\frac{\bm{R}^{i}\,\bm{Z}^{j}\,\bm{\overline{R}}^{\,m}\,\overline{\bm{Z}}^{\,n}}
{\bm{Z}^{2}\,\bm{\overline{Z}}^{2}}
\\*[0.2cm]
&\qquad\qquad\times\left.\Big[\mathcal{K}_{0}^{ijmn}(\bm{x},\,\bm{z},\,\bm{z}^{\prime},\,
\overline{\bm{x}},\,\bm{\overline{z}},\,\bm{\overline{z}}^{\prime},\,\vartheta,\,\xi)
\,\mathcal{W}_{0}\left(\bm{x},\,\bm{z},\,\bm{z}^{\prime},\,
\bm{\overline{x}},\,\bm{\overline{z}},\,\bm{\overline{z}}^{\prime}\right)\right.\\*[0.2cm]
&\qquad\qquad\left.-\,\big(\bm{z}, \,\bm{z}^{\prime}\rightarrow\bm{y}\big)
\,-\,\big(\bm{\overline{z}}, \,\bm{\overline{z}}^{\prime}\rightarrow\bm{\overline{y}}\big)
\,+\,\big(\bm{z}, \,\bm{z}^{\prime}\rightarrow\bm{y}\,
\ \&\ \,\bm{\overline{z}}, \,\bm{\overline{z}}^{\prime}\rightarrow\bm{\overline{y}}\big)
\right.\Big]
\\*[0.2cm] &\qquad\qquad 
\,+\,\left(k_{1}^{+}\leftrightarrow k_{2}^{+},\:\bm{k}_{1}\leftrightarrow\bm{k}_{2}\right).
\end{split}\end{equation}
The notations here are similar to those  in \eqn{qqq_final}. The transverse coordinates
$\bm{x},\,\bm{z},\,\bm{z}^{\prime},\,\bm{y},\,\bm{w}$ refer to the direct amplitude (DA)
and have the same meaning as in Fig.~\ref{quarinsert} and \eqn{defyw}. 
We use a bar to indicate the corresponding coordinates in the 
complex conjugate amplitude (CCA). We use the notations in \eqn{defRZ} for
the transverse separations between the emitted partons, that is,
\begin{equation}\label{defRZ1}
\bm{R}\,\equiv\,\bm{x}-\bm{y}\,,\quad \bm{Z}\,\equiv\,\bm{z}-\bm{z}'\,,\quad
\bm{\overline{R}}\,\equiv\,\bm{\overline{x}}-\bm{\overline{y}}\,,
\quad \bm{\overline{Z}}\,\equiv\,\bm{\overline{z}}-\bm{\overline{z}}'\,,
\end{equation}
The longitudinal momentum fractions  $\vartheta$ and $\xi$ are the same
in the DA and in the CCA, since they are fully fixed by the kinematics of the final state,
as follows:
 \begin{equation}\label{longit1}
 \vartheta=1-\frac{k_{1}^{+}}{q^{+}},\qquad\xi=\frac{k_{2}^{+}}{q^{+}}\,,
 \end{equation}
 The $\delta$-function enforcing longitudinal momentum conservation implies
 $k^+_3=(\vartheta-\xi)q^+$, which in particular requires $\vartheta\ge \xi$ (i.e. $k^+_1+k^+_2\le q^+$).

Let us now explain the new structures occurring in \eqn{trijetqq}.
The tensorial kernel is defined as
 \begin{equation}\label{defK0}
\mathcal{K}_{0}^{ijmn}(\bm{x},\,\bm{z},\,\bm{z}^{\prime},\,
\overline{\bm{x}},\,\bm{\overline{z}},\,\bm{\overline{z}}^{\prime},\,\vartheta,\,\xi)
\equiv\frac{\Phi_{\lambda_{3}\lambda_{2}\lambda_{1}\lambda}^{ij}(\bm{x},\,\bm{z},\,\bm{z}',\,\vartheta,\,\xi)
\,\Phi_{\lambda_{3}\lambda_{2}\lambda_{1}\lambda}^{mn\, *}(\overline{\bm{x}},\,\overline{\bm{z}},\,
\overline{\bm{z}}^\prime,\,\vartheta,\,\xi)}
{\big[\vartheta^{2}(1-\vartheta)\bm{R}^{2}\,+\,\xi(\vartheta-\xi)\bm{Z}^{2}\big]\,
\big[\vartheta^{2}(1-\vartheta)\bm{\overline{R}}^{2}\,+\,\xi(\vartheta-\xi)\bm{\overline{Z}}^{2}\big]}\,,
\end{equation}
with the effective vertex $\Phi_{\lambda_{3}\lambda_{2}\lambda_{1}\lambda}^{ij}$
introduced in \eqn{vertexPhi} (the star in $\Phi^*$ denotes  complex conjugation).
The product of effective vertices in the numerator can be explicitly computed as
\begin{equation}\label{Phi2}
\begin{split}
&\Phi_{\lambda_{3}\lambda_{2}\lambda_{1}\lambda}^{ij}
\Phi_{\lambda_{3}\lambda_{2}\lambda_{1}\lambda}^{mn\,*}\Big|_{\rm non-inst.}\\
&=\,4(1-\vartheta)^{2}\,\left[(\vartheta-2\xi)^{2}\delta^{rj}\delta^{tn}+\vartheta^{2}\left(\delta^{rt}\delta^{jn}-\delta^{rn}\delta^{jt}\right)\right]\,\left[(2-\vartheta)^{2}\delta^{ri}\delta^{tm}+\vartheta^{2}\left(\delta^{rt}\delta^{im}-\delta^{rm}\delta^{it}\right)\right].
\end{split}\end{equation}
As emphasised by our notation, in evaluating this product we have excluded the instantaneous pieces from the two effective vertices. These pieces have a very simple tensorial structure, so they can be easily 
inserted when needed\footnote{These  instantaneous pieces do not contribute to either
the soft, or the collinear, limit that we shall study later; so, a result like \eqn{Phi2}
is indeed sufficient for our purposes in this paper.}.

The effects of the collision are encoded in the function $\mathcal{W}_{0}$,  defined 
as the following linear combination of partonic $S$-matrices: 
\begin{equation}\label{defW0}
\begin{split}
&\mathcal{W}_{0}\left(\bm{x},\,\bm{z},\,\bm{z}^{\prime}, \,\bm{\overline{x}},\,\bm{\overline{z}},\,\bm{\overline{z}}^{\prime}\right)\\
&\equiv\,S_{qq\bar{q}\bar{q}\bar{q}q}\left(\bm{x},\,\bm{z},\,\bm{z}^{\prime},\,
\overline{\bm{x}},\,\overline{\bm{z}},\,\overline{\bm{z}}^{\prime}\right)
\,-\,
S_{qq\bar{q} \bar{q}}\left(\bm{x},\,\bm{z},\,\bm{z}^{\prime}, \overline{\bm{w}}\right)
\,-\,S_{q\bar{q}\bar{q}q}\left(\bm{w},\,\bm{\overline{x}},\,\bm{\overline{z}},\,\bm{\overline{z}}^{\prime}\right)\,+\,\mathcal{S}\left(\bm{w},\,\bm{\overline{w}}\right).
\end{split}\end{equation}
Our notations for the partonic $S$-matrices are intended to describe 
(via the lower scripts) 
the partonic composition of the multi-parton system which scatters off the shockwave 
and to also distinguish (via a bar on the transverse coordinates) between partons in 
the DA and in the CCA, respectively.
The ordering of the transverse coordinates in the argument follows that of the lower subscripts.
The indices/arguments corresponding to partons in the DA appear on the left to those 
representing the CCA. When reading the lower indices,
 one should also keep in mind that a quark in the CCA formally 
counts like an antiquark (from the viewpoint of the direction it its colour charge flow).

According to these conventions, the $S$-matrix 
$S_{qq\bar{q}\bar{q}\bar{q}q}\left(\bm{x},\,\bm{z},\,\bm{z}^{\prime},\,
\overline{\bm{x}},\,\overline{\bm{z}},\,\overline{\bm{z}}^{\prime}\right)$ refers to the
eikonal scattering of  a system made with 6 quarks: 2 quarks ($\bm{x},\,\bm{z}$) and one anti-quark
($\bm{z}^{\prime}$) in the DA, together with  2 ``anti-quarks'' ($\overline{\bm{x}},\,\overline{\bm{z}}$)
and one ``quark'' ($\overline{\bm{z}}^{\prime}$) in the CCA. This describes
the situation where the collision occurs in the final state (i.e. after the second splitting) in both
the DA and in the CCA (see Fig.~\ref{S6q}.left), and reads
\begin{align}\label{wils1}
S_{qq\bar{q}\bar{q}\bar{q}q}\left(\bm{x},\,\bm{z},\,\bm{z}^{\prime},\,
\overline{\bm{x}},\,\overline{\bm{z}},\,\overline{\bm{z}}^{\prime}\right)&\,
\equiv\,\frac{2}{C_{F}\,N_{c}}\,\left\langle 
\mathrm{tr}\left(V^{\dagger}(\bm{\overline{x}})\,V(\bm{x})\,t^{a}\,t^{b}\right)\,\mathrm{tr}\left(V(\bm{\overline{z}}^{\prime})\,t^{b}\,V^{\dagger}(\bm{\overline{z}})\,V(\bm{z})\,t^{a}\,V^{\dagger}(\bm{z}^{\prime})\right)\right\rangle,\nonumber \\*[0.2cm]
& \,\simeq\,\mathcal{Q}(\bm{x},\,\bm{z}^{\prime},\,\overline{\bm{z}}^{\prime},\,\overline{\bm{x}})\,\mathcal{S}(\bm{z},\,\overline{\bm{z}}),
 \end{align}
where the approximate equality holds for large $N_c$: in this limit, the 6-quark $S$-matrix factorises into the product of a dipole, $\mathcal{S}(\bm{z},\,\overline{\bm{z}})$, and a quadrupole, $\mathcal{Q}(\bm{x},\,\bm{z}^{\prime},\,\overline{\bm{z}}^{\prime},\,\overline{\bm{x}})$.

\begin{figure}[!t]\centerline{
 \includegraphics[scale=0.49]{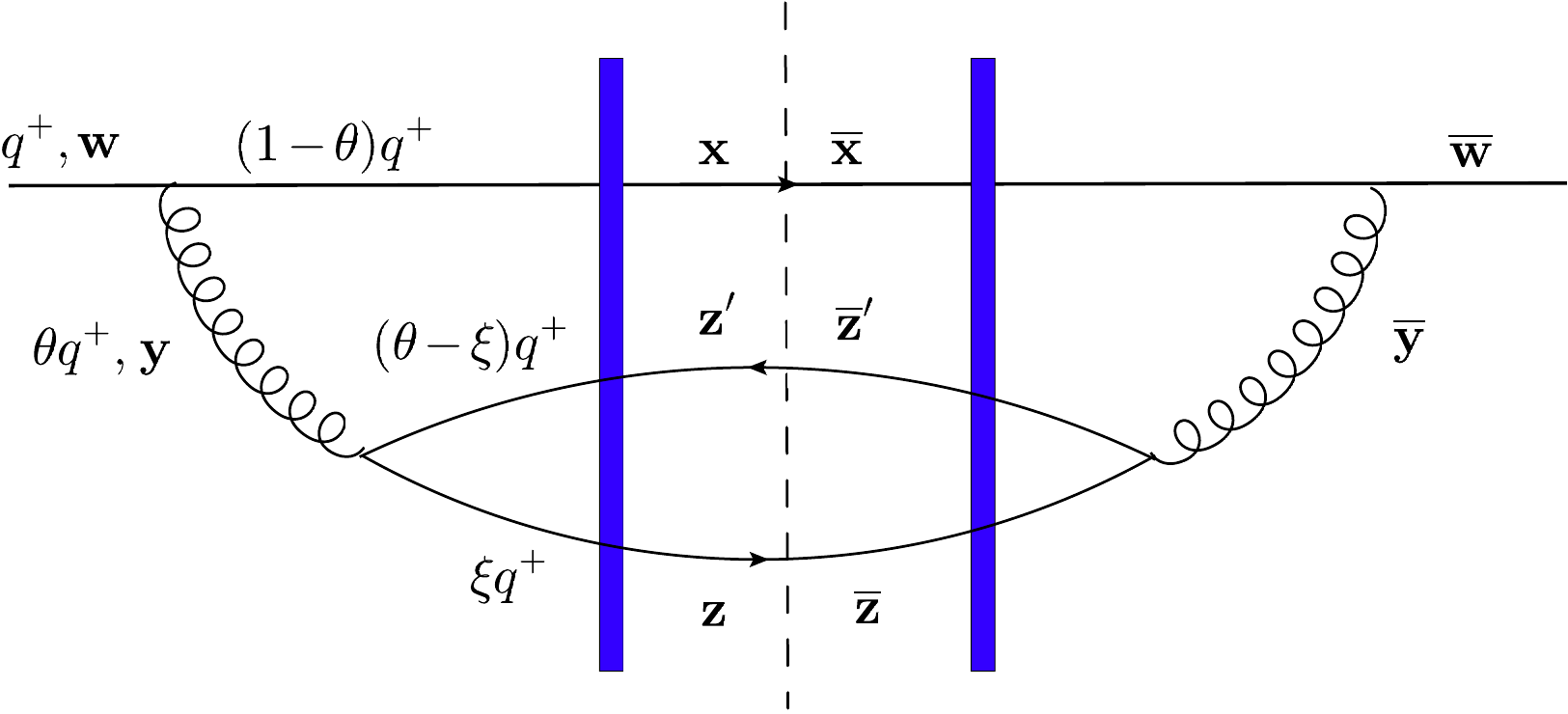}\qquad
 \includegraphics[scale=0.45]{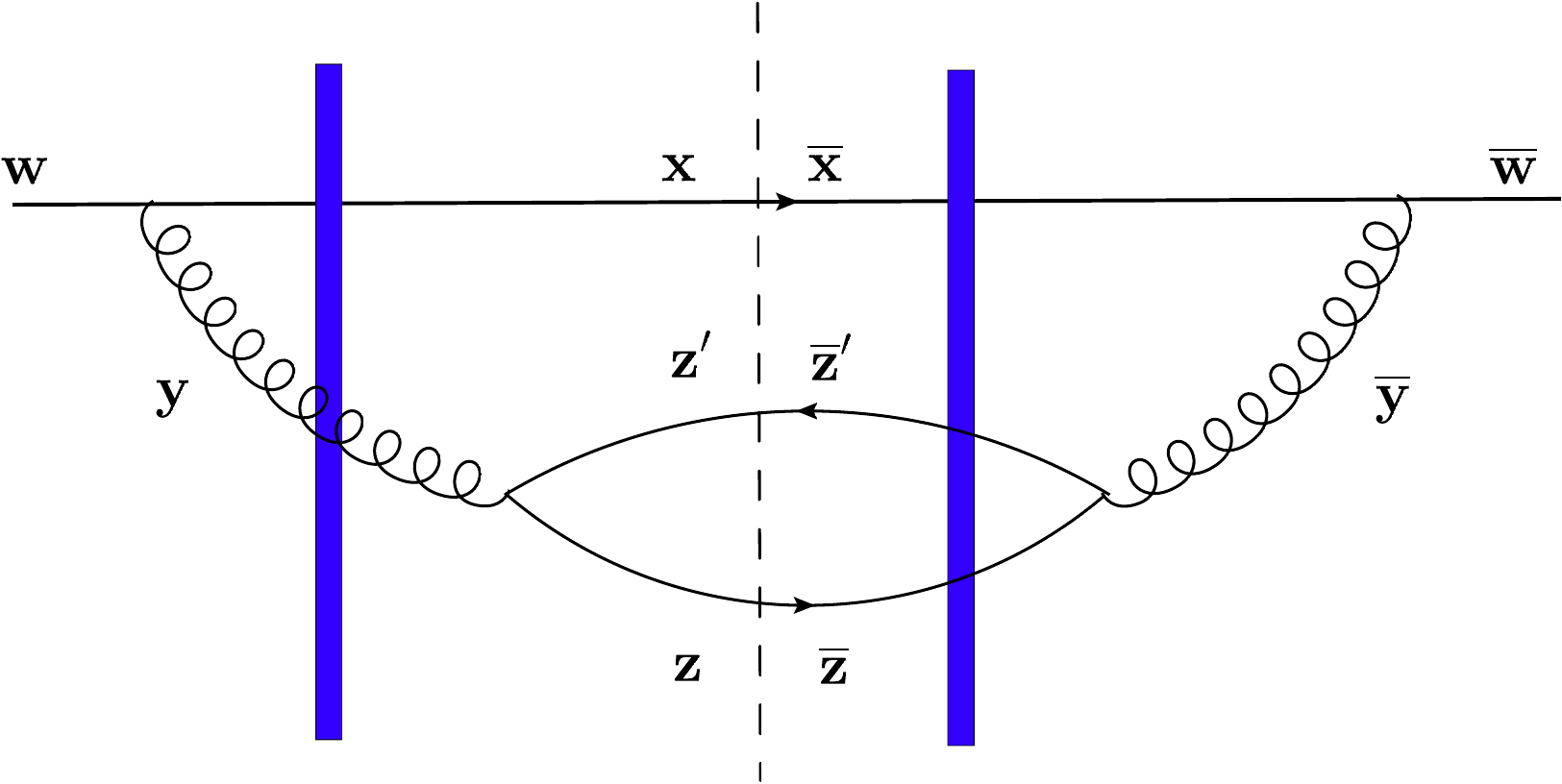}
}
\caption{Left: the graph corresponding to final-state interactions in both the DA and the CCA;
its colour structure is encoded in the 6-quark $S$-matrix shown in \eqn{wils1}. Right: a graph
describing interference between interactions in the intermediate state in the DA and respectively
in the final state in the CCA; the associated $S$-matrix is obtained by replacing 
$\bm{z}\to \bm{y}$ and $\bm{z}^{\prime}\to \bm{y}$ in \eqn{wils1}.
}
\label{S6q}
\end{figure}

Furthermore, 
$S_{qq\bar{q} \bar{q}}\left(\bm{x},\,\bm{z},\,\bm{z}^{\prime}, \overline{\bm{w}}\right)$ and
$S_{q\bar{q}\bar{q}q}\left(\bm{w},\,\bm{\overline{x}},\,\bm{\overline{z}},\,\bm{\overline{z}}^{\prime}\right)$
represent interference terms where the collision with the shockwave occurs in the final
state on one side of the cut, and in the initial state (prior to the first splitting) on the other side; e.g.,
\begin{align}\label{wils2}
S_{qq\bar{q} \bar{q}}\left(\bm{x},\,\bm{z},\,\bm{z}^{\prime}, \overline{\bm{w}}\right)
\,=\,\frac{2}{C_{F}\,N_{c}}\,
\left\langle \mathrm{tr}\left[V(\bm{x})\,t^{a}\,V^{\dagger}(\overline{\bm{w}})\,t^{b}\right]\,\mathrm{tr}\left[V(\bm{z})\,t^{a}\,V^{\dagger}(\bm{z}^{\prime})\,t^{b}\right]\right\rangle
\simeq\,\mathcal{S}(\bm{z},\,\overline{\bm{w}})\,\mathcal{S}(\bm{x},\,\bm{z}^{\prime}).
\end{align}
Finally,  the colour dipole $\mathcal{S}\left(\bm{w},\,\bm{\overline{w}}\right)\equiv S_{q\bar{q}}
(\bm{w},\,\bm{\overline{w}})$ describes initial-state interactions in both the DA and the CCA.

The normalisation factors in the above definitions are such that all the individual $S$-matrices
reduce to unity in the absence of scattering. Accordingly, both the linear combination in \eqn{defW0}
and the cross-section \eqref{trijetqq} vanish in that limit, as expected.
The overall colour structure becomes remarkably simple at large $N_c$:
\beq\label{W1largeNc}
\mathcal{W}_{0}\left(\bm{x},\,\bm{z},\,\bm{z}^{\prime},\,\bm{\overline{x}},\,\bm{\overline{z}},\,\bm{\overline{z}}^{\prime}\right)\,\simeq\,
\mathcal{Q}(\bm{x},\,\bm{z}^{\prime},\,\overline{\bm{z}}^{\prime},\,\overline{\bm{x}})\,\mathcal{S}(\bm{z},\,\overline{\bm{z}}) -\mathcal{S}(\bm{z},\,\overline{\bm{w}})\,\mathcal{S}(\bm{x},\,\bm{z}^{\prime})
 -\mathcal{S}(\bm{w},\,\overline{\bm{z}})\,\mathcal{S}(\overline{\bm{z}}^{\prime}, \overline{\bm{x}})
 +\mathcal{S}\left(\bm{w},\,\bm{\overline{w}}\right).
\eeq

The three other 
terms within the square brackets in \eqn{trijetqq}, which are obtained as various limits of the first term,
refer to situations where the scattering in the final state gets replaced by scattering in the intermediate
state, as explained in relation with \eqn{qqq_outgoing}. E.g. the second term within the
square brackets, denoted as $\big(\bm{z}, \,\bm{z}^{\prime}\rightarrow\bm{y}\big)$, is illustrated in
Fig.~\ref{S6q}.right. We recall that this term can be obtained by letting $\bm{Z}\to 0$
in the kernel \eqref{defK0} (which in particular means using the simplified version of the
effective vertex shown in \eqn{vertexGQ-limit}) and replacing $\bm{z}\to \bm{y}$ and 
$\bm{z}^{\prime}\to \bm{y}$ in the arguments of the partonic $S$-matrices in the r.h.s. of
\eqn{defW0}.


It is quite instructive to display the version of \eqn{defW0} which applies to the last term
 in \eqn{trijetqq}, as obtained  by simultaneously replacing
 $\big(\bm{z}, \,\bm{z}^{\prime}\rightarrow\bm{y}\big)$ and $\big(\bm{\overline{z}}, \,\bm{\overline{z}}^{\prime}\rightarrow\bm{\overline{y}}\big)$. This describes the case where there are no interactions with
 the shockwave in the final $qq \bar q$ state, neither in the DA nor in the CCA. 
 Working at large $N_c$ for definiteness, cf. \eqn{W1largeNc}, one finds
\begin{align}\label{W1qg}
\hspace*{-0.5cm}
\mathcal{W}_{0}\left(\bm{x},\,\bm{y},\,\bm{y},\,\bm{\overline{x}},\,\bm{\overline{y}},\,\bm{\overline{y}}\right)
\,\simeq\,
\mathcal{Q}(\bm{x},\,\bm{y},\,\overline{\bm{y}},\,\overline{\bm{x}})\,\mathcal{S}(\bm{y},\,\overline{\bm{y}}) -\mathcal{S}(\bm{y},\,\overline{\bm{w}})\,\mathcal{S}(\bm{x},\,\bm{y})
 -\mathcal{S}(\bm{w},\,\overline{\bm{y}})\,\mathcal{S}(\overline{\bm{y}}, \overline{\bm{x}})
 +\mathcal{S}\left(\bm{w},\,\bm{\overline{w}}\right).
\end{align}
This $S$-matrix structure is identical to that occurring in the integrand of \eqn{pALOqchannel}
for LO quark-gluon production ($qA\to qg+X$). This is easy
to understand: in both cases, the interactions with the nuclear shockwave refer to the partons
involved in the branching $q\to qg$.
%

So far, we have not specified the $x_g$--argument of the various $S$-matrices in \eqn{defW0},
i.e. the ``minus'' longitudinal momentum fraction of the gluons from the nuclear target which
are involved in the production of the three-parton state. Clearly, this is given by the generalisation
of \eqn{xg} to a final state involving three partons, that is,
\beq\label{xg3}
x_g=\,\frac{\bm{k}_1^2}{x_1s}\,+\,\frac{\bm{k}_2^2}{x_2s}\,+\,\frac{\bm{k}_3^2}{x_3s}\,,
\eeq
where $x_i=k_i^+/Q^+$ are the ``plus'' longitudinal momentum fractions of the produced partons.
Needless to say, this value for $x_g$ applies not only to the $qq\bar q$ final state discussed so far, 
but also to the  $qgg$ final states to which we now turn.

\subsubsection{The $qgg$ final state}
\label{trijet-qgg}

For the $qgg$ final state, the trijet cross-section is computed as
 (once again, a factor $2\pi\delta(k_{1}^{+}+k_{2}^{+}+k_{3}^{+}-q^{+})$ is implicitly understood in the l.h.s.)
\begin{equation}\begin{split}\label{qggcross}
\frac{d\sigma^{qA\rightarrow qgg+X}}{d^{3}k_{1}\,d^{3}k_{2}\,d^{3}k_{3}}&\,\equiv\,\frac{1}{2N_{c}}\ {}^{out}\!\left\langle q_{\lambda}^{\alpha}(q^{+},\,\bm{q}=\bm{0})\right|\,\hat{\mathcal{N}}_{q}(k_{1})\,\hat{\mathcal{N}}_{g}(k_{2})\,\hat{\mathcal{N}}_{g}(k_{3})\,\left|q_{\lambda}^{\alpha}(q^{+},\,\bm{q}=\bm{0})\right\rangle ^{out}\\
&=\,\frac{1}{2N_{c}}\,\int_{\bm{w},\,\overline{\bm{w}}}
\!\!{}_{\ \ \ qgg}^{\ \ \ out}\!\left\langle q_{\lambda}^{\alpha}(q^{+},\,\overline{\bm{w}})\right|\,\hat{\mathcal{N}}_{q}(k_{1})\,\hat{\mathcal{N}}_{g}(k_{2})\,\hat{\mathcal{N}}_{g}(k_{3})\,\left|q_{\lambda}^{\alpha}(q^{+},\,\bm{w})\right\rangle _{qgg}^{out},
 \end{split}\end{equation}
The second line involves  the  $qgg$  Fock component of the quark outgoing state, which, as
explained in Sect.~\ref{sec:qgg}, is the sum of two contributions, corresponding to two different topologies
for the gluon emission vertices, as illustrated in Fig.~\ref{final-qgg1} and Fig.~\ref{final-qgg2},
respectively. It is therefore natural to split the cross-section in \eqn{qggcross} into three contributions. 
In the first one, the topology in Fig.~\ref{final-qgg1} is used for the quark LCWF in both the DA and the CCA.
Similarly, the second contributions involves the topology in Fig.~\ref{final-qgg2} alone. Finally, the third 
contribution represents interferences between the two topologies. Besides, each of these three
contributions is built with two pieces, corresponding to the two possible permutations for the momentum
labels of the produced gluons. As before, we shall work in the limit of a large number of colours, in
which one discards the non-planar graphs. One should however pay attention to the fact that the
symmetry of the triple gluon vertex makes that the graphs generated by
exchanging the two final gluons in Fig.~\ref{final-qgg2} are still planar, and therefore must be kept
in the final result. This will result in appropriate symmetry factors.

 \begin{figure}[!t]\center
 \includegraphics[scale=0.7]{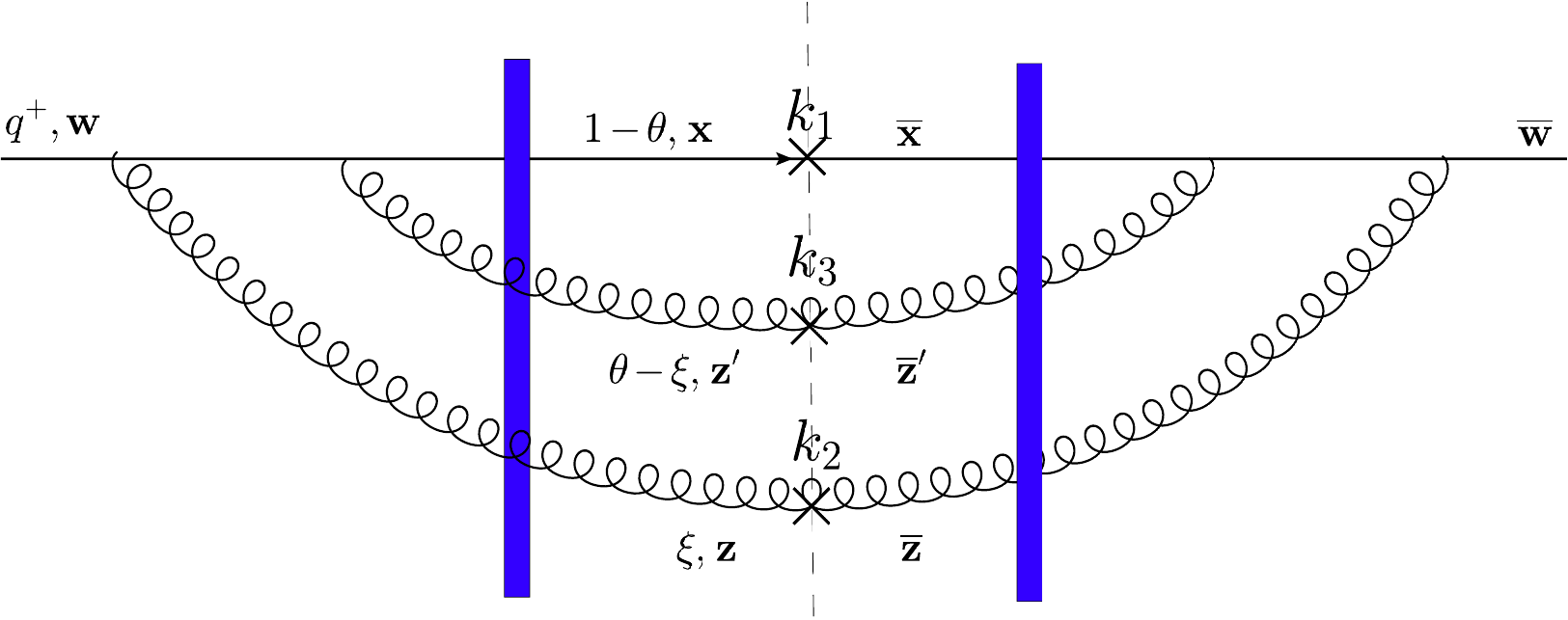}
\caption{A contribution to the $qgg$  cross-section \eqref{trijetqgg1} which features final-state
interactions in both the DA and the CCA; the respective colour structure is shown in \eqn{wils3}.
There is a similar diagram in which the momenta
$k_2$ and $k_3$ of the final gluons are exchanged with each other.}
\label{fig:qgg1}
\end{figure}

A straightforward  calculation using \eqn{qgg1_tot} yields the following expression for the  first piece of the 
cross-section:
 \begin{equation}\begin{split}\label{trijetqgg1}
&\frac{d\sigma_{(1)}^{qA\rightarrow qgg+X}}{dk_{1}^{+}\,d^{2}\bm{k}_{1}\,dk_{2}^{+}\,d^{2}\bm{k}_{2}\,dk_{3}^{+}\,d^{2}\bm{k}_{3}}=\frac{\alpha_{s}^{2}\,C_{F}^2}{2(2\pi)^{10}(q^{+})^{2}}\,
\delta(q^{+}-k_{1}^{+}-k_{2}^{+}-k_{3}^{+})\\*[0.2cm]
&\qquad\qquad\times
\int_{\overline{\bm{x}},\,\bm{\overline{z}},\,\bm{\overline{z}}^{\prime},\,\bm{x},\,\bm{z},\,\bm{z}^{\prime}}\,e^{-i\bm{k}_{1}\cdot(\bm{x}-\overline{\bm{x}})-i\bm{k}_{2}\cdot(\bm{z}-\overline{\bm{z}})-i\bm{k}_{3}\cdot(\bm{z}^{\prime}-\overline{\bm{z}}^{\prime})}\,
\frac{\bm{Y}^{m}\,(\bm{X}^{\prime})^{n}\, \overline{\bm{Y}}^{p}\,(\bm{\overline{X}}^{\prime})^{q}
}{(\bm{X}^{\prime})^{2} (\bm{\overline{X}}^{\prime})^{2}}\\*[0.2cm]
&\qquad\qquad\times\left.\Big[\mathcal{K}_{1}^{mnpq}(\bm{x},\,\bm{z},\,\bm{z}^{\prime},\,
\overline{\bm{x}},\,\bm{\overline{z}},\,\bm{\overline{z}}^{\prime},\,\vartheta,\,\xi)
\,\mathcal{W}_{1}\left(\bm{x},\,\bm{z},\,\bm{z}^{\prime},\,
\bm{\overline{x}},\,\bm{\overline{z}},\,\bm{\overline{z}}^{\prime}\right)\right.\\*[0.2cm]
&\qquad\qquad\left.-\,\big(\bm{x},\bm{z}^{\prime}\rightarrow\bm{y}\big)
\,-\,\big(\bm{\overline{x}},\bm{\overline{z}}^{\prime}\rightarrow\overline{\bm{y}}\big)
\,+\,\big(\bm{x},\bm{z}^{\prime}\rightarrow\bm{y}\,
\ \&\ \,\bm{\overline{x}},\bm{\overline{z}}^{\prime}\rightarrow\overline{\bm{y}}\big)
\right.\Big] 
\\*[0.2cm] &\qquad\qquad 
\,+\,\left(k_{2}^{+}\leftrightarrow k_{3}^{+},\:\bm{k}_{2}\leftrightarrow\bm{k}_{3}\right).
\end{split}\end{equation}
The notations for the transverse coordinates which appear in this equation are easily grasped
by comparing with  \eqn{qgg1_tot}.
The new tensorial kernel $\mathcal{K}_{1}^{mnpq}$  is defined as
 \begin{equation}\label{defK1}
\mathcal{K}_{1}^{mnpq}
\equiv\frac{\Xi_{\lambda_{1}\lambda}^{ijmn}
(\bm{x},\,\bm{z},\,\bm{z}',\,\vartheta,\,\xi)\,
\Xi_{\lambda_{1}\lambda}^{ijpq\,*}
(\overline{\bm{x}},\,\overline{\bm{z}},\,
\overline{\bm{z}}^\prime,\,\vartheta,\,\xi)}
{\big[
(1-\vartheta)(\vartheta-\xi)(\bm{X}^{\prime})^{2}+\xi(1-\xi)^{2}\bm{Y}^{2}\big]\,
\big[(1-\vartheta)(\vartheta-\xi)(\overline{\bm{X}}^{\prime})^{2}+\xi(1-\xi)^{2}\overline{\bm{Y}}^{2}\big]}\,,
\end{equation}
with the effective vertex $\Xi_{\lambda_{1}\lambda}^{ijmn}$ introduced in \eqn{vertexXi} and
the longitudinal momentum fractions  $\vartheta$ and $\xi$  shown in \eqn{longit1}.
The product of effective vertices in the numerator can be explicitly computed:
\begin{align}
\label{XIXI}
\Xi_{\lambda_{1}\lambda}^{ijmn}\,\Xi_{\lambda_{1}\lambda}^{ijpq\,*}\Big|_{\rm non-inst.}&=
\frac{\xi(1-\xi)^2}{\vartheta-\xi}\,
\tau_{\lambda_{1}\lambda_{2}}^{jn}(\vartheta,\,\xi)\,
\tau_{\lambda_{1}\lambda_{3}}^{jq\,*}(\vartheta,\,\xi)\,
\phi_{\lambda_{2}\lambda}^{im}(\xi)\,
\phi_{\lambda_{3}\lambda}^{ip,*}(\xi)\nn
&= 8\delta^{mp}\,\delta^{nq}\,\frac{\xi(1-\xi)^{2}}{\vartheta-\xi}\,
\big[1+(1-\xi)^2 \big]\left[(1-\vartheta)^{2}+(1-\xi)^{2}\right]\,.
\end{align}
Furthermore, $\mathcal{W}_{1}$ denotes the
following linear combination of partonic $S$-matrices: 
\begin{equation}\label{defW1}
\begin{split}
&\mathcal{W}_{1}\left(\bm{x},\,\bm{z},\,\bm{z}^{\prime}, \,\bm{\overline{x}},\,\bm{\overline{z}},\,\bm{\overline{z}}^{\prime}\right)\\
&\equiv\,S_{qgg\bar{q}gg}^{(1)}\left(\bm{x},\,\bm{z},\,\bm{z}^{\prime},\,
\overline{\bm{x}},\,\overline{\bm{z}},\,\overline{\bm{z}}^{\prime}\right)
\,-\,
S_{qgg\bar{q}}^{(1)}\left(\bm{x},\,\bm{z},\,\bm{z}^{\prime}, \overline{\bm{w}}\right)
\,-\,S_{q\bar{q}gg}^{(1)}\left(\bm{w},\,\bm{\overline{x}},\,\bm{\overline{z}},\,\bm{\overline{z}}^{\prime}\right)\,+\,\mathcal{S}\left(\bm{w},\,\bm{\overline{w}}\right).
\end{split}\end{equation}
We use the same notations as explained after \eqn{defW0} in order to synthetically summarize
the partonic content for both the DA and the CCA. 
The first $S$-matrix in the r.h.s., that is,
\begin{align}\label{wils3}
\hspace*{-0.6cm}
S_{qgg\bar{q}gg}^{(1)}\left(\bm{x},\,\bm{z},\,\bm{z}^{\prime},\,
\overline{\bm{x}},\,\overline{\bm{z}},\,\overline{\bm{z}}^{\prime}\right)&
\,\equiv\frac{1}{C_{F}^{2}\,N_{c}}\,\left\langle 
\mathrm{tr}\left[V^{\dagger}(\bm{\overline{x}})\,V(\bm{x})\,t^{b}\,t^{a}\,t^{f}\,t^{e}\right]\,\left[U^{\dagger}(\overline{\bm{z}})\,U(\bm{z})\right]^{fa}\,\left[U^{\dagger}(\bm{\overline{z}}^{\prime})\,U(\bm{z}^{\prime})\right]^{eb}\right\rangle\nonumber\\*[0.2cm]
&\simeq\,\mathcal{Q}\left(\bm{x},\,\bm{z}^{\prime},\,\overline{\bm{z}}^{\prime},\,\bm{\overline{x}}\right)\,
\mathcal{Q}\left(\bm{z}^{\prime},\,\bm{z},\,\overline{\bm{z}},\,\overline{\bm{z}}^{\prime}\right)\,\mathcal{S}
\left(\bm{z}\,,\overline{\bm{z}}\right),
\end{align}
describes final-state interactions in both the DA and the CCA and hence it includes Wilson lines
for  6 partons: one quark ($\bm{x}$) and two gluons ($\bm{z},\,\bm{z}^{\prime}$) in
the DA, and one ``anti-quark'' ($\overline{\bm{x}}$) and two gluons ($\overline{\bm{z}},\,
\overline{\bm{z}}^{\prime}$) in the CCA (see Fig.~\ref{fig:qgg1}). Similarly,
 \begin{align}\label{wils4}
S_{qgg\bar{q}}^{(1)}\left(\bm{x},\,\bm{z},\,\bm{z}^{\prime}, \overline{\bm{w}}\right)
&\,\equiv\,\frac{1}{C_{F}^{2}\,N_{c}}\,\left\langle 
\mathrm{tr}\left[V^{\dagger}(\overline{\bm{w}})\,t^{c}\,t^{d}V(\bm{x})\,t^{b}\,t^{a}\right]\,U^{db}(\bm{z}^{\prime})\,U^{ca}(\bm{z})\right\rangle\nonumber\\*[0.2cm]
&\simeq\,\mathcal{S}(\bm{x},\,\bm{z}^{\prime})\,\mathcal{S}(\bm{z}^{\prime},\,\bm{z})
\,\mathcal{S}(\bm{z},\,\overline{\bm{w}}),
\end{align}
is an interference term between final-state and initial-state interactions.
Once again, the large $N_c$ version of $\mathcal{W}_{1}$ can be fully expressed in terms of colour dipoles and quadrupoles:
\begin{align}\label{W1-largeNc}
\mathcal{W}_{1}\left(\bm{x},\,\bm{z},\,\bm{z}^{\prime}, \,\bm{\overline{x}},\,\bm{\overline{z}},\,\bm{\overline{z}}^{\prime}\right)&\,\simeq\,
\mathcal{Q}\left(\bm{x},\,\bm{z}^{\prime},\,\overline{\bm{z}}^{\prime},\,\bm{\overline{x}}\right)\,
\mathcal{Q}\left(\bm{z}^{\prime},\,\bm{z},\,\overline{\bm{z}},\,\overline{\bm{z}}^{\prime}\right)\,\mathcal{S}
\left(\bm{z}\,,\overline{\bm{z}}\right)+ \mathcal{S}\left(\bm{w},\,\bm{\overline{w}}\right)-\nn
 &\quad - \mathcal{S}(\bm{x},\,\bm{z}^{\prime})\,\mathcal{S}(\bm{z}^{\prime},\,\bm{z})
\,\mathcal{S}(\bm{z},\,\overline{\bm{w}})
- \mathcal{S}(\bm{w},\,\overline{\bm{z}})\,
\mathcal{S}(\overline{\bm{z}},\,\overline{\bm{z}}^{\prime})
\,\mathcal{S}(\overline{\bm{z}}^{\prime},\,\overline{\bm{x}}).
\end{align}
As  a simple check, we notice that after the double replacement $\big(\bm{x},\bm{z}^{\prime}\rightarrow\bm{y}\,
\ \&\ \,\bm{\overline{x}},\bm{\overline{z}}^{\prime}\rightarrow\overline{\bm{y}}\big)$ (which yields the
last term within the square brackets in \eqn{trijetqgg1}), the r.h.s of \eqn{defW1} takes the same form as
for the scattering of a quark-gluon pair\footnote{That is, it reduces to the $S$-matrix structure visible in the integrand of \eqn{LOfinal}, up to some relabelling of variables.}, as expected.

Consider also the second piece in the cross-section \eqref{trijetqgg1}, as obtained by permuting
the momenta $k_2$ and $k_3$ of the final gluons. This can be computed from the first piece
(the one explicitly shown in \eqn{trijetqgg1} and illustrated in Fig.~\ref{fig:qgg1}) 
by exchanging $\bm{k}_{2}\leftrightarrow\bm{k}_{3}$ within the Fourier phases and 
$k_{2}^{+}\leftrightarrow k_{3}^{+}$ within the longitudinal momentum fractions.  Alternatively,
and equivalently, this second term can be written as a Fourier transform which involves exactly 
the same Fourier phases as the first term, but with the following changes of variables in
the remaining part of the integrand:  $\bm{z}\leftrightarrow\bm{z}^\prime$,
 $\overline{\bm{z}}\leftrightarrow \overline{\bm{z}}^\prime$, and $\xi \leftrightarrow\vartheta-\xi$.
This amounts to permuting the transverse coordinates and the longitudinal momentum fractions
assigned to the final gluons in the amplitude in Fig.~\ref{final-qgg1} (and similarly for the CCA).

 \begin{figure}[!t]\center
 \includegraphics[scale=0.7]{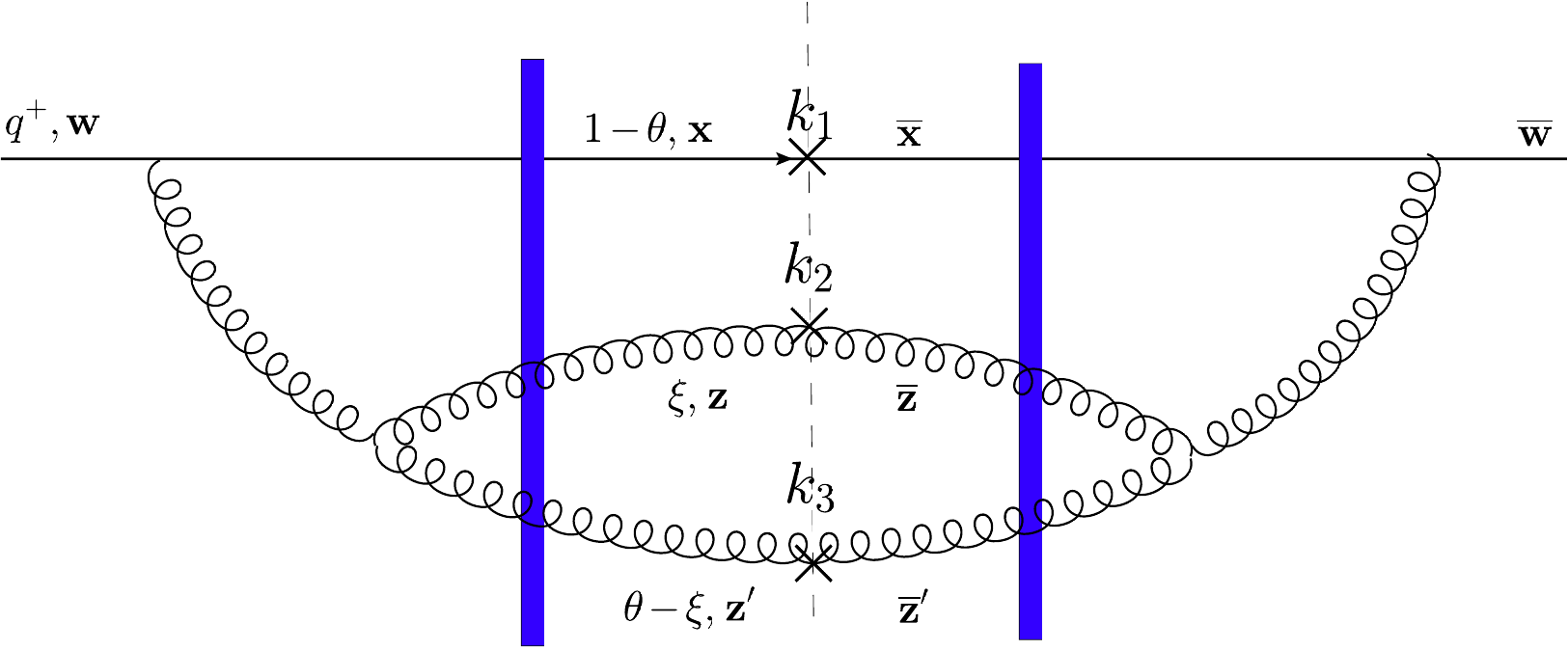}
\caption{A graph contributing to the cross-section in  \eqref{trijetqgg2}; the associated
$S$-matrix is shown in \eqn{wils5}.}
\label{fig:qgg2}
\end{figure}

We now turn to the second contribution to the cross-section for $qgg$ production, that involving the
topology in Fig.~\ref{final-qgg2}. We now use \eqn{qgg2_tot} to deduce
 \begin{equation}\begin{split}\label{trijetqgg2}
&\frac{d\sigma_{(2)}^{qA\rightarrow qgg+X}}{dk_{1}^{+}\,d^{2}\bm{k}_{1}\,dk_{2}^{+}\,d^{2}\bm{k}_{2}\,dk_{3}^{+}\,d^{2}\bm{k}_{3}}=\frac{4\alpha_{s}^{2}\,C_{F}N_c}{2(2\pi)^{10}(q^{+})^{2}}\,
\delta(q^{+}-k_{1}^{+}-k_{2}^{+}-k_{3}^{+})\\*[0.2cm]
&\qquad\qquad\times
\int_{\overline{\bm{x}},\,\bm{\overline{z}},\,\bm{\overline{z}}^{\prime},\,\bm{x},\,\bm{z},\,\bm{z}^{\prime}}\,e^{-i\bm{k}_{1}\cdot(\bm{x}-\overline{\bm{x}})-i\bm{k}_{2}\cdot(\bm{z}-\overline{\bm{z}})-i\bm{k}_{3}\cdot(\bm{z}^{\prime}-\overline{\bm{z}}^{\prime})}\,
\,\frac{\bm{R}^{m}\,\bm{Z}^{n}\,\overline{\bm{R}}^{p}\,\overline{\bm{Z}}^{q}}{\bm{Z}^{2}\,
\overline{\bm{Z}}^{2}}
\\*[0.2cm]
&\qquad\qquad\times\left.\Big[\mathcal{K}_{2}^{mnpq}(\bm{x},\,\bm{z},\,\bm{z}^{\prime},\,
\overline{\bm{x}},\,\bm{\overline{z}},\,\bm{\overline{z}}^{\prime},\,\vartheta,\,\xi)
\,\mathcal{W}_{2}\left(\bm{x},\,\bm{z},\,\bm{z}^{\prime},\,
\bm{\overline{x}},\,\bm{\overline{z}},\,\bm{\overline{z}}^{\prime}\right)\right.\\*[0.2cm]
&\qquad\qquad\left.-\,\big(\bm{z},\bm{z}^{\prime}\rightarrow\bm{y}\big)
\,-\,\big(\bm{\overline{z}},\bm{\overline{z}}^{\prime}\rightarrow\overline{\bm{y}}\big)
\,+\,\big(\bm{z},\bm{z}^{\prime}\rightarrow\bm{y}\,
\ \&\ \,\bm{\overline{z}},\bm{\overline{z}}^{\prime}\rightarrow\overline{\bm{y}}\big)
\right.\Big].
\end{split}\end{equation}
This equation involves the new tensorial kernel (the effective vertex $\Pi_{\lambda_{1}\lambda}^{ijmn}$ has been introduced in \eqn{vertexPi})
 \begin{equation}\label{defK2}
\mathcal{K}_{2}^{mnpq}\,
\equiv\,\frac{\Pi_{\lambda_{1}\lambda}^{ijmn}
(\bm{x},\,\bm{z},\,\bm{z}',\,\vartheta,\,\xi)\,
\Pi_{\lambda_{1}\lambda}^{ijpq\,*}
(\overline{\bm{x}},\,\overline{\bm{z}},\,
\overline{\bm{z}}^\prime,\,\vartheta,\,\xi)}
{\big[\vartheta^{2}(1-\vartheta)\bm{R}^{2}\,+\,\xi(\vartheta-\xi)\bm{Z}^{2}\big]\,
\big[\vartheta^{2}(1-\vartheta)\overline{\bm{R}}^{2}\,+\,\xi(\vartheta-\xi)\overline{\bm{Z}}^{2}\big]
}\,,
\end{equation}
whose numerator can be explicitly computed as (as before, we exclude the contribution of the instantaneous 
pieces of the vertices, to simplify writing)
\begin{align}\label{Pi2}
\hspace*{-1.cm}
\Pi_{\lambda_{1}\lambda}^{ijmn}\Pi_{\lambda_{1}\lambda}^{ijpq\,*}\Big|_{\rm non-inst.}
&=\,2\xi(\vartheta-\xi)\vartheta^{2}(1-\vartheta)^{2}\nonumber\\
&\times \left[(2-\vartheta)^{2}\delta^{lm}\delta^{rp}+\vartheta^{2}(\delta^{lr}\delta^{mp}-\delta^{lp}\delta^{mr})\right]
\left(\frac{2\delta^{nl}\,\delta^{qr}}{\vartheta^{2}}+\frac{\delta^{nq}\,\delta^{lr}}{\xi^{2}}+\frac{\delta^{nq}\,\delta^{lr}}{(\vartheta-\xi)^{2}}\right)\nonumber\\*[0.2cm]
&=\,4\xi(\vartheta-\xi)\vartheta^{2}(1-\vartheta)^{2}\bigg[\frac{(2-\vartheta)^{2}\delta^{nm}\,\delta^{qp}+\vartheta^{2}\delta^{mp}\delta^{nq}-\vartheta^{2}\delta^{np}\,\delta^{qm}}{\vartheta^{2}}\,+
\nonumber\\*[0.2cm]
&\qquad\,+\delta^{nq}\,\delta^{mp}\big[1+(1-\vartheta)^{2}\big]\left(\frac{1}{\xi^{2}}+\frac{1}{(\vartheta-\xi)^{2}}\right)\bigg].
\end{align}

Furthermore, $\mathcal{W}_{2}$ is the following linear combination of partonic $S$-matrices: 
\begin{equation}\label{defW2}
\begin{split}
&\mathcal{W}_{2}\left(\bm{x},\,\bm{z},\,\bm{z}^{\prime}, \,\bm{\overline{x}},\,\bm{\overline{z}},\,\bm{\overline{z}}^{\prime}\right)\\
&\equiv\,S_{qgg\bar{q}gg}^{(2)}\left(\bm{x},\,\bm{z},\,\bm{z}^{\prime},\,
\overline{\bm{x}},\,\overline{\bm{z}},\,\overline{\bm{z}}^{\prime}\right)
\,-\,
S_{qgg\bar{q}}^{(2)}\left(\bm{x},\,\bm{z},\,\bm{z}^{\prime}, \overline{\bm{w}}\right)
\,-\,S_{q\bar{q}gg}^{(2)}\left(\bm{w},\,\bm{\overline{x}},\,\bm{\overline{z}},\,\bm{\overline{z}}^{\prime}\right)\,+\,\mathcal{S}\left(\bm{w},\,\bm{\overline{w}}\right),
\end{split}\end{equation}
where the 6-parton $S$-matrix describes final-state interactions in both the DA and the CCA,
 \begin{equation}\begin{split}\label{wils5}
&S_{qgg\bar{q}gg}^{(2)}\left(\bm{x},\,\bm{z},\,\bm{z}^{\prime},\,
\overline{\bm{x}},\,\overline{\bm{z}},\,\overline{\bm{z}}^{\prime}\right)\equiv \frac{1}{C_{F} N_{c}^{2}}\,f^{rmn}f^{abc}\left\langle 
\left[U^{\dagger}(\overline{\bm{z}})U(\bm{z})\right]^{nc}\left[U^{\dagger}(\overline{\bm{z}}^{\prime})U(\bm{z}^{\prime})\right]^{mb}\mathrm{tr}\left[V^{\dagger}(\overline{\bm{x}})V(\bm{x})t^{a}t^{r}\right]
\right\rangle\\
&\simeq\,\frac{1}{2}\left[\mathcal{Q}\left(\bm{z},\,\bm{z}^{\prime},\,\overline{\bm{z}}^{\prime},\,\overline{\bm{z}}\right)\,\mathcal{Q}\left(\bm{x},\,\bm{z},\,\overline{\bm{z}},\,\overline{\bm{x}}\right)\,\mathcal{S}\left(\bm{z}^{\prime},\,\overline{\bm{z}}^{\prime}\right)+\mathcal{Q}\left(\bm{z}^{\prime},\,\bm{z},\,\overline{\bm{z}},\,\overline{\bm{z}}^{\prime}\right)\,\mathcal{Q}\left(\bm{x},\,\bm{z}^{\prime},\,\overline{\bm{z}}^{\prime},\,\overline{\bm{x}}\right)\,\mathcal{S}\left(\bm{z},\,\overline{\bm{z}}\right)\right],
  \end{split}\end{equation}
  (the two terms correspond to permutations of the final gluons),
whereas the 4-parton $S$-matrix describes the interference between scattering in the final 
state and the initial state, respectively:
\begin{align}\label{wils6}
S_{qgg\bar{q}}^{(2)}\left(\bm{x},\,\bm{z},\,\bm{z}^{\prime}, \overline{\bm{w}}\right)
&\,\equiv\,\frac{1}{C_{F}N_{c}^{2}}\,f^{abc}f^{rde} \left\langle 
U^{ec}(\bm{z}^{\prime})U^{db}(\bm{z})\,\mathrm{tr}\left[V^{\dagger}(\overline{\bm{w}})t^{r}V(\bm{x})t^{a}\right]\right\rangle\nonumber\\*[0.2cm]
&\simeq\,\frac{1}{2}\left[\mathcal{S}(\bm{x},\,\bm{z}^{\prime})\,\mathcal{S}(\bm{z}^{\prime},\,\bm{z})
\,\mathcal{S}(\bm{z},\,\overline{\bm{w}})\,+\,\mathcal{S}(\bm{x},\,\bm{z})\,\mathcal{S}(\bm{z},\,\bm{z}^{\prime})
\,\mathcal{S}(\bm{z}^{\prime},\,\overline{\bm{w}})\right].
  \end{align}
As a check, one can see that after performing the double replacement  $\big(\bm{z},\bm{z}^{\prime}\rightarrow\bm{y}\,
\ \&\ \,\bm{\overline{z}},\bm{\overline{z}}^{\prime}\rightarrow\overline{\bm{y}}\big)$ in \eqn{defW2},
one recovers the colour structure describing the scattering of a quark-gluon pair
(the structure shown in \eqn{W1qg} at large $N_c$).

Note that, as compared to  \eqn{trijetqgg1}, \eqn{trijetqgg2} contains an additional factor of 4 (besides
the modified colour factor, which reflects the different structure of the partonic $S$-matrices). This factor
of 4 is related to the symmetry of the amplitude in Fig.~\ref{final-qgg2} under the exchange of the two
final gluons. This in turn implies that the integrand of \eqn{trijetqgg2} is symmetric under the simultaneous exchanges $\bm{z}\leftrightarrow\bm{z}^\prime$,
 $\overline{\bm{z}}\leftrightarrow \overline{\bm{z}}^\prime$, and $\xi \leftrightarrow\vartheta-\xi$,
 as it can be easily verified. We used this symmetry property twice: \texttt{(i)} for a given
 assignment of the momenta of the two gluons (e.g., $k_{2}^{+}=\xi q^+$ and $k_{3}^{+}=
 (\vartheta-\xi)q^+$, as in Fig.~\ref{fig:qgg2}), there are two possible Wick contractions in the calculation
 of the expectation value $\hat{\mathcal{N}}_{g}(k_{2})\,\hat{\mathcal{N}}_{g}(k_{3})$, cf. \eqn{qggcross},
 which give identical results;  \texttt{(ii)} the graphs obtained by
 permuting the gluon momenta, $k_2 \leftrightarrow k_{3}$, give identical results
 as well, because we can undo the effect of this permutation via a change of
 variables which leaves the integrand unchanged.

 \begin{figure}[!t]\center
 \includegraphics[scale=0.7]{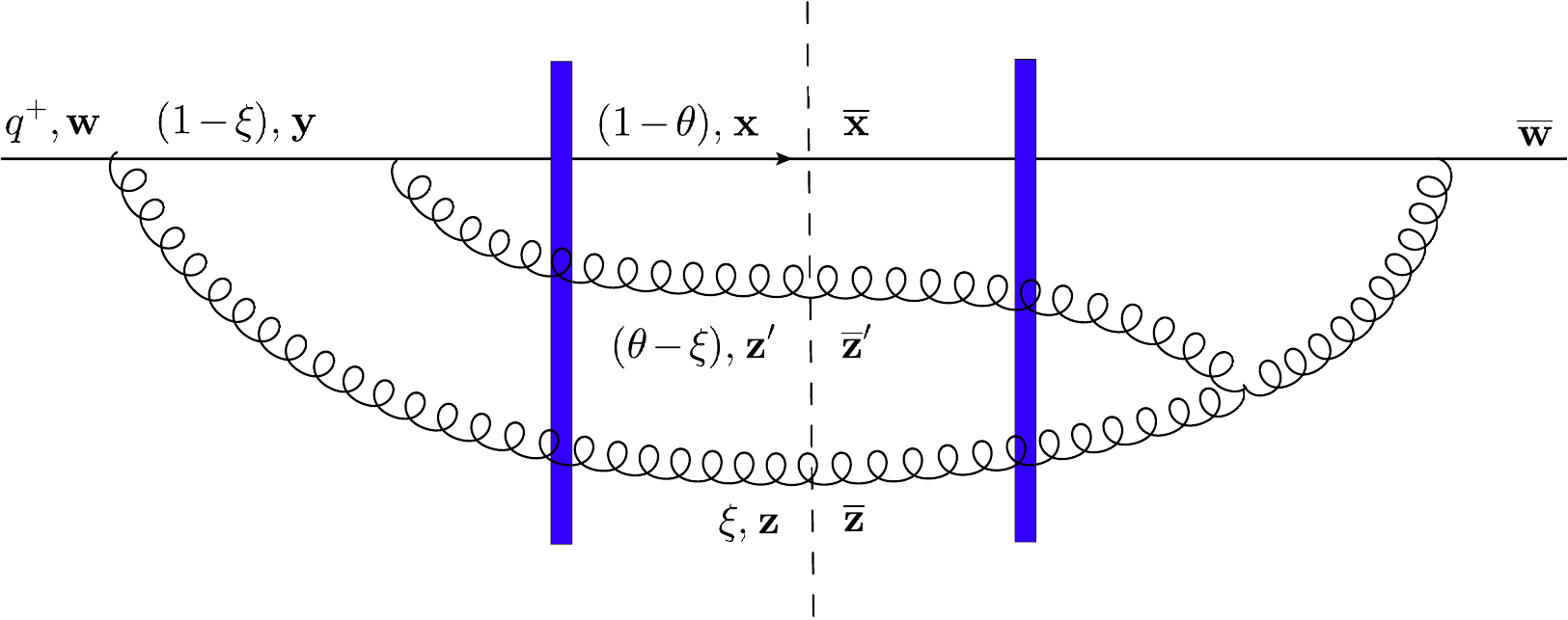}
\caption{A particular graph contributing to the ``interference'' cross-section \eqref{trijetqgg3};
its colour structure is exhibited in \eqn{wilsinterf}.}
\label{qgg-interf}
\end{figure}

The remaining terms refer to interferences between the two topologies shown in Fig.~\ref{final-qgg1} and Fig.~\ref{final-qgg2}, respectively. We shall explicitly show the piece where the topology in Fig.~\ref{final-qgg1}
counts for the DA and that in Fig.~\ref{final-qgg2}, for the CCA (see Fig.~\ref{qgg-interf}
for an illustration). The total result can be then obtained by taking
the double of the real part of the shown contribution. One finds
\begin{equation}\begin{split}\label{trijetqgg3}
&\frac{d\sigma_{(3)}^{qA\rightarrow qgg+X}}{dk_{1}^{+}\,d^{2}\bm{k}_{1}\,dk_{2}^{+}\,d^{2}\bm{k}_{2}\,dk_{3}^{+}\,d^{2}\bm{k}_{3}}=\frac{2\alpha_{s}^{2}\,C_{F}N_c}{4(2\pi)^{10}(q^{+})^{2}}\,
\delta(q^{+}-k_{1}^{+}-k_{2}^{+}-k_{3}^{+})\\*[0.2cm]
&\qquad\qquad\times 2\,{\mathcal Re}
\int_{\overline{\bm{x}},\,\bm{\overline{z}},\,\bm{\overline{z}}^{\prime},\,\bm{x},\,\bm{z},\,\bm{z}^{\prime}}\,e^{-i\bm{k}_{1}\cdot(\bm{x}-\overline{\bm{x}})-i\bm{k}_{2}\cdot(\bm{z}-\overline{\bm{z}})-i\bm{k}_{3}\cdot(\bm{z}^{\prime}-\overline{\bm{z}}^{\prime})}\,
\,\frac{\bm{Y}^{m}\,(\bm{X}^{\prime})^{n}\,\overline{\bm{R}}^{p}\,\overline{\bm{Z}}^{q}}{(\bm{X}^{\prime})^{2} \,
\overline{\bm{Z}}^{2}}
\\*[0.2cm]
&\qquad\qquad\times\left.\Big[\mathcal{K}_{3}^{mnpq}(\bm{x},\,\bm{z},\,\bm{z}^{\prime},\,
\overline{\bm{x}},\,\bm{\overline{z}},\,\bm{\overline{z}}^{\prime},\,\vartheta,\,\xi)
\,\mathcal{W}_{3}\left(\bm{x},\,\bm{z},\,\bm{z}^{\prime},\,
\bm{\overline{x}},\,\bm{\overline{z}},\,\bm{\overline{z}}^{\prime}\right)\right.\\*[0.2cm]
&\qquad\qquad\left.-\,\big(\bm{x},\bm{z}^{\prime}\rightarrow\bm{y}\big)
\,-\,\big(\bm{\overline{z}},\bm{\overline{z}}^{\prime}\rightarrow\overline{\bm{y}}\big)
\,+\,\big(\bm{x},\bm{z}^{\prime}\rightarrow\bm{y}\,
\ \&\ \,\bm{\overline{z}},\bm{\overline{z}}^{\prime}\rightarrow\overline{\bm{y}}\big)
\right.\Big]
\\*[0.2cm] &\qquad\qquad 
\,+\,\left(k_{2}^{+}\leftrightarrow k_{3}^{+},\:\bm{k}_{2}\leftrightarrow\bm{k}_{3}\right).
\end{split}\end{equation}
It is important to notice that, in the subtracted terms, $\bm{y}$ is the function of 
$\bm{x}$ and $\bm{z}^{\prime}$ defined in \eqn{defyw2}, whereas 
$\overline{\bm{y}}$ is rather a function of 
$\bm{\overline{z}}$ and $\bm{\overline{z}}^{\prime}$ defined by the ``bared'' version of \eqn{defyw}.
Furthermore,
 \begin{equation}\label{defK3}
\mathcal{K}_{3}^{mnpq}
\equiv\frac{\Xi_{\lambda_{1}\lambda}^{ijmn}
(\bm{x},\,\bm{z},\,\bm{z}',\,\vartheta,\,\xi)\,
\Pi_{\lambda_{1}\lambda}^{ijpq\,*}
(\overline{\bm{x}},\,\overline{\bm{z}},\,
\overline{\bm{z}}^\prime,\,\vartheta,\,\xi)}
{\big[
(1-\vartheta)(\vartheta-\xi)(\bm{X}^{\prime})^{2}+\xi(1-\xi)^{2}\bm{Y}^{2}\big]\,
\big[\vartheta^{2}(1-\vartheta)\overline{\bm{R}}^{2}\,+\,\xi(\vartheta-\xi)\overline{\bm{Z}}^{2}\big]
}\,.
\end{equation}
The combination of Wilson lines describing the scattering has the same general structure, that is,
\begin{equation}\label{defW3}
\begin{split}
&\mathcal{W}_{3}\left(\bm{x},\,\bm{z},\,\bm{z}^{\prime}, \,\bm{\overline{x}},\,\bm{\overline{z}},\,\bm{\overline{z}}^{\prime}\right)\\
&\equiv\,S_{qgg\bar{q}gg}^{(3)}\left(\bm{x},\,\bm{z},\,\bm{z}^{\prime},\,
\overline{\bm{x}},\,\overline{\bm{z}},\,\overline{\bm{z}}^{\prime}\right)
\,-\,
S_{qgg\bar{q}}^{(3)}\left(\bm{x},\,\bm{z},\,\bm{z}^{\prime}, \overline{\bm{w}}\right)
\,-\,S_{q\bar{q}gg}^{(3)}\left(\bm{w},\,\bm{\overline{x}},\,\bm{\overline{z}},\,\bm{\overline{z}}^{\prime}\right)\,+\,\mathcal{S}\left(\bm{w},\,\bm{\overline{w}}\right),
\end{split}\end{equation}
with however some new ingredients, namely, 
 \begin{align}\label{wilsinterf}\hspace*{-0.5cm}
  S_{qgg\bar{q}gg}^{(3)}\left(\bm{x},\,\bm{z},\,\bm{z}^{\prime},\,
\overline{\bm{x}},\,\overline{\bm{z}},\,\overline{\bm{z}}^{\prime}\right)&
\,\equiv\,\frac{2i}{C_{F}\,N_{c}^{2}}f^{cde}\,\left\langle 
\left[U^{\dagger}(\overline{\bm{z}})\,U(\bm{z})
\right]^{da}\,\left[U^{\dagger}(\overline{\bm{z}}^{\prime})\,U(\bm{z}^{\prime})\right]^{eb}\,\mathrm{tr}\left[V^{\dagger}(\overline{\bm{x}})\,V(\bm{x})\,t^{b}\,t^{a}\,t^{c}\right]\right\rangle\nonumber\\*[.2cm]
&\simeq\,\mathcal{Q}\left(\bm{x},\,\bm{z}^{\prime},\,\overline{\bm{z}}^{\prime},\,\overline{\bm{x}}\right)\,
\mathcal{Q}\left(\bm{z}^{\prime},\,\bm{z},\,\overline{\bm{z}},\,\overline{\bm{z}}^{\prime}\right)\,\mathcal{S}\left(\bm{z},\,\overline{\bm{z}}\right),
  \end{align}
 \begin{equation}\begin{split}\label{wils7}
S_{qgg\bar q}^{(3)}\left(\bm{x},\,\bm{z},\,\bm{z}^{\prime},\,\overline{\bm{w}},\,\right)&\,\equiv\,
\frac{2i}{C_{F}\,N_{c}^{2}}\,f^{ecd}\,\left\langle 
U^{ca}(\bm{z})\,U^{db}(\bm{z}^{\prime})\,\mathrm{tr}\left[V^{\dagger}(\overline{\bm{w}})\,t^{e}\,V(\bm{x})\,t^{b}\,t^{a}\right]\right\rangle\\*[.2cm]
&\simeq\,\mathcal{S}(\bm{x},\,\bm{z}^{\prime})\,\mathcal{S}(\bm{z}^{\prime},\,\bm{z})\,\mathcal{S}(\bm{z},\,\overline{\bm{w}}),
  \end{split}\end{equation}
 \begin{equation}\begin{split}\label{wils8}
S_{q\bar{q}gg}^{(3)}\left(\bm{w},\,\bm{\overline{x}},\,\bm{\overline{z}},\,\bm{\overline{z}}^{\prime}\right)
&\,\equiv\,
\frac{2i}{C_{F}\,N_{c}^{2}}\,f^{ecd}\,\left\langle 
U^{ca}(\overline{\bm{z}})\,U^{bd}(\overline{\bm{z}}^{\prime})\,
\mathrm{tr}\left[V^{\dagger}(\overline{\bm{x}})\,t^{b}\,t^{a}\,V(\bm{w})\,t^{e}\right]\right\rangle\\*[.2cm]
&\simeq\,\mathcal{S}(\bm{w},\,\overline{\bm{z}})\,
\mathcal{S}(\overline{\bm{z}},\,\overline{\bm{z}}^{\prime})
\,\mathcal{S}(\overline{\bm{z}}^{\prime},\,\overline{\bm{x}})\,.
  \end{split}\end{equation}
It might be interesting to notice that the last term in \eqn{trijetqgg3}, as obtained via the double replacement
$\big(\bm{x},\bm{z}^{\prime}\rightarrow\bm{y}\,
\ \&\ \,\bm{\overline{z}},\bm{\overline{z}}^{\prime}\rightarrow\overline{\bm{y}}\big)$, does not has
the same colour structure as expected for a final quark-gluon state (unlike the respective
limits for  all the other colour functions $\mathcal{W}_{0}$,
$\mathcal{W}_{1}$, and  $\mathcal{W}_{2}$). One finds indeed
  \begin{align}\label{W3qg}
  \hspace*{-0.5cm}
\mathcal{W}_{3}\left(\bm{y},\,\bm{z},\,\bm{y},\,\bm{\overline{x}},\,\bm{\overline{y}},\,\bm{\overline{y}}\right)
\,\simeq\,\mathcal{S}(\bm{y},\,\bm{z})
\mathcal{S}(\overline{\bm{y}},\,\overline{\bm{x}})\,\mathcal{S}(\bm{z},\,\overline{\bm{y}}) -\mathcal{S}(\bm{y},\,\bm{z})\,\mathcal{S}(\bm{z},\,\bm{\overline{w}})
 -\mathcal{S}(\bm{w},\,\overline{\bm{y}})\,\mathcal{S}(\overline{\bm{y}}, \overline{\bm{x}})
 +\mathcal{S}\left(\bm{w},\,\bm{\overline{w}}\right),
\end{align}
which should be compared to \eqn{W1qg}.  This difference is due to the fact that
the pattern of the colour flow in the interference graphs is different as compared to the direct graphs.

Note finally that, in the large-$N_c$ limit, the three functions encoding the colour structure
in the various  contributions to the cross-section, namely   $\mathcal{W}_{1}$, $\mathcal{W}_{2}$, 
and $\mathcal{W}_{3}$,  become very similar to each other:   $\mathcal{W}_{1}$ and $\mathcal{W}_{3}$
take exactly the same form, shown in  \eqn{W1-largeNc}, whereas  $\mathcal{W}_{2}$ differs only through
the additional symmetrisation with respect to the exchange of the two final gluons. 
 
  \section{Next-to-leading order corrections: the real terms}
\label{sec:NLOreal}

As mentioned in the Introduction, the next-to-leading order (NLO) corrections to the
cross-section for forward dijet production in $pA$ collisions can be divided into two classes: 
{\it real} and {\it virtual}.  In the remaining part of this paper, we shall compute the
{\it real} corrections --- those associated with a final state which involves three partons (``jets''),
out of which only two are measured. By integrating out the kinematics of the unmeasured
parton, one generates a loop correction to the cross-section for the two measured jets.
This loop opens in the direct amplitude (DA) and closes back in the complex conjugate
amplitude (CCA). The virtual NLO corrections, on the other hand, refer to loop corrections
to the amplitude itself. They will be addressed in a subsequent paper.

\subsection{The di-quark jet production}

We first consider final states which include two measured fermions: two quarks, or a quark-antiquark
pair. The cross-section for di-quark jet production\footnote{This particular channel, i.e. $qA\rightarrow qq+X$,
does not exist at leading-order, unless one considers double quark scattering, that is, the simultaneous
scattering and production of two collinear quarks from the wavefunction of the incoming proton.
The respective contribution counts at zeroth order in $\alpha_s$, but it is proportional to the double-quark
distribution in the proton (that can be roughly estimated as the product of two standard, single-quark,
distributions).} is obtained by ``integrating out'' the final antiquark
in our general formula for the three-quark final state, cf. \eqn{trijetqq}:
\begin{align}
\label{2q-v0}
\frac{d\sigma^{qA\rightarrow qq+X}_\rnlo}{dk_{1}^{+}\,d^{2}\bm{k}_{1}\,dk_{2}^{+}\,d^{2}\bm{k}_{2}}=\int dk^+_3 d^2\bm{k}_{3}\,
\frac{d\sigma^{qA\rightarrow qq\overline{q}+X}}{dk_{1}^{+}\,d^{2}\bm{k}_{1}\,dk_{2}^{+}\,d^{2}\bm{k}_{2}\,dk_{3}^{+}\,d^{2}\bm{k}_{3}}\,.
\end{align}
(The subscript ``rNLO'' stays for real next-to-leading order corrections.)
With reference to \eqn{trijetqq}, it is quite clear that the integral over $k^+_3 $ can be trivially performed 
by using the $\delta$-function for longitudinal momentum conservation, whereas the integral over 
$\bm{k}_{3}$ yields a factor $(2\pi)^2\delta^{(2)}(\bm{z}^{\prime}-\overline{\bm{z}}^{\prime})$, 
which allows one to identify the coordinates $\bm{z}^{\prime}$ and $\overline{\bm{z}}^{\prime}$
of the unmeasured antiquark in the DA and the CCA, respectively. The result of \eqn{2q-v0}
can be succinctly written as
\begin{align}
\label{2quarks}
\frac{d\sigma_\rnlo^{qA\rightarrow qq+X}}{dk_{1}^{+}\,d^{2}\bm{k}_{1}\,dk_{2}^{+}\,d^{2}\bm{k}_{2}}=\,
(2\pi)^2\,\frac{d\sigma^{qA\rightarrow qq\overline{q}+X}}{dk_{1}^{+}\,d^{2}\bm{k}_{1}\,dk_{2}^{+}\,d^{2}\bm{k}_{2}\,dk_{3}^{+}\,d^{2}\bm{k}_{3}}\,\Bigg |_{k^+_3=q^+-k^+_1-k^+_2\,,\ 
\bm{z}^{\prime}=\overline{\bm{z}}^{\prime}},
\end{align}
where it is understood that the trijet cross-section  in the r.h.s. is given by \eqn{trijetqq}, but without
the $\delta$-function expressing the conservation of longitudinal momentum. 

Some care
must be taken, concerning the order of limits: the identification $\bm{z}^{\prime}=\overline{\bm{z}}^{\prime}$
must be made only {\it after} performing the subtractions which occur 
in the integrand of  \eqn{trijetqq}. For instance,
in the subtracted term denoted as $\big(\bm{\overline{z}}, \,\bm{\overline{z}}^{\prime}
\rightarrow\bm{\overline{y}}\big)$, one must first perform the replacements $\bm{\overline{z}}
\rightarrow\bm{\overline{y}}$ and $\bm{\overline{z}}^{\prime}\rightarrow\bm{\overline{y}}$ at {\it fixed}
$\bm{z}^{\prime}$ and only {\it then} identify $\bm{\overline{z}}^{\prime}$ with $\bm{z}^{\prime}$.
(The two limits do not commute with each other, as one can easily check.)

The fact that the antiquark is not measured brings some simplifications in the structure of the
6-quark $S$-matrix in \eqn{wils1} (which describes final-state interactions, cf. Fig.~\ref{S6q}-left):
when $\bm{z}^{\prime}=\overline{\bm{z}}^{\prime}$, the Wilson lines describing
the scattering of the antiquark compensate
each other by unitarity, $V^{\dagger}(\bm{z}^{\prime}) V(\bm{z}^{\prime}) =1$, and then
the quadrupole appearing in the second line of  \eqn{wils1} reduces to a dipole 
(we consider large $N_c$, for simplicity):
\begin{equation}\label{6q1}
S_{qq\bar{q}\bar{q}\bar{q}q}\left(\bm{x},\,\bm{z},\,\bm{z}^{\prime},\,
\overline{\bm{x}},\,\overline{\bm{z}},\,\overline{\bm{z}}^{\prime}=\bm{z}^{\prime}\right)
\,\simeq\,\mathcal{S}(\bm{x},\,\overline{\bm{x}})\,\mathcal{S}(\bm{z},\,\overline{\bm{z}}),
 \end{equation}
However, this simplification refers only to the first term within the square brackets of \eqn{trijetqq}.  
Consider e.g. the second term, denoted as $\big(\bm{z}, \,\bm{z}^{\prime}\rightarrow\bm{y}\big)$.
If one first replaces both $\bm{z}$ and $\bm{z}^{\prime}$ by $\bm{y}$, and only {\it then} one
identifies $\bm{z}^{\prime}=\overline{\bm{z}}^{\prime}$, then the quadrupole structure survives
in  \eqn{wils1}:
 \begin{equation}\label{6q2}
S_{qq\bar{q}\bar{q}\bar{q}q}\left(\bm{x},\,\bm{y},\,\bm{y},\,
\overline{\bm{x}},\,\overline{\bm{z}},\,\overline{\bm{z}}^{\prime}=\bm{z}^{\prime}\right)
\,\simeq\,\mathcal{O}(\bm{x},\,\bm{y},\,{\bm{z}}^{\prime},\,\overline{\bm{x}})\,\mathcal{S}(\bm{y},\,\overline{\bm{z}}).
 \end{equation}
 In this equation, $\bm{y}$ is understood as the function of $\bm{z}$ and  ${\bm{z}}^{\prime}$
shown in \eqn{defyw}.

The longitudinal momentum fractions $\vartheta$ and $\xi$ which implicitly appear in
the r.h.s. of \eqn{2quarks} are fixed by \eqn{longit1},  which for the present purposes
should be rewritten in terms of the longitudinal momentum fractions 
$x_1\equiv k_{1}^{+}/Q^+$ and  $x_2\equiv k_{2}^{+}/Q^+$ of the measured partons 
(``jets'') and the respective fraction $x_p\equiv q^+/Q^+$ of the original quark:
 \begin{equation}\label{longit0}
 \vartheta=1-\frac{k_{1}^{+}}{q^{+}}\,=1-\frac{x_1}{x_p}\,,
 \qquad\xi=\frac{k_{2}^{+}}{q^{+}}\,=\frac{x_2}{x_p}\,.
 \end{equation}
As discussed in relation with  \eqn{pALO}, the physical  dijet cross-section (to the
order of interest) is obtained by averaging over $x_p$ with the quark distribution inside the proton:
\begin{equation}\label{pANLO-qq}
\frac{d\sigma_{\rnlo}^{pA\rightarrow 2jet+X}}{d^{3}k\,d^{3}p} \bigg |_{q\to qq}
\,=\,\int dx_{p}\,q_f(x_{p},\mu^{2})\,\Theta(x_p-x_1-x_2)\,
\frac{d\sigma_\rnlo^{qA\rightarrow qq+X}}{dk_{1}^{+}\,d^{2}\bm{k}_{1}\,dk_{2}^{+}\,d^{2}\bm{k}_{2}}\,.
 \end{equation}
It is furthermore important to keep in mind that the cross-section \eqref{pANLO-qq}
 depends upon the ``plus'' longitudinal fractions $x_1$, $x_2$, and $x_p$ not only via its dependence upon $\vartheta$ 
and $\xi$, as explicit in \eqn{defK0} for the kernel $\mathcal{K}_{0}$, but also via the dependence
of the various $S$-matrices upon the ``rapidity'' variable\footnote{More precisely,
the evolution ``rapidity'' related to $x_g$ is $Y_g\equiv\ln(1/x_g)$.} $x_g$, as introduced
by their high-energy evolution. We recall that $x_g$ is the  ``minus'' longitudinal momentum fraction 
carried by the gluons from the nucleus which participate in the scattering. This is indeed a function of
$x_1$, $x_2$ and $x_3=x_p-x_1-x_2$, as shown in \eqn{xg3}.

\subsection{The quark-antiquark di-jets}

The case where the measured dijet is made with a quark-antiquark pair\footnote{This case
can be viewed as a NLO correction to the $\order{\alpha_s}$--process in which a gluon collinear 
with the incoming parton splits into a $q\bar q$ pair while scattering off the nuclear target
\cite{Iancu:2013dta,Iancu:2018hwa}. The respective cross-section involves
the gluon distribution, which is however suppressed at large $x$, i.e. for
the forward kinematics of interest for us here.} is similarly obtained 
by integrating \eqn{trijetqq} over the kinematics  of one of the two final quarks, e.g. over $k_1$. This gives
\begin{align}
\label{qbarq0}
\frac{d\sigma^{qA\rightarrow q\bar q+X}_\rnlo}{dk_{2}^{+}\,d^{2}\bm{k}_{2}\,dk_{3}^{+}\,d^{2}\bm{k}_{3}}=\int dk^+_1 d^2\bm{k}_{1}\,
\frac{d\sigma^{qA\rightarrow qq\overline{q}+X}}{dk_{1}^{+}\,d^{2}\bm{k}_{1}\,dk_{2}^{+}\,d^{2}\bm{k}_{2}\,dk_{3}^{+}\,d^{2}\bm{k}_{3}}\,.
\end{align}
The above r.h.s. involves two contributions: the unmeasured quark $k_1$ can be either the 
incoming quark, or the quark produced by the decay of the intermediate gluon. The 
contribution of the first case to the $ qq\bar q$ cross-section is explicitly shown in \eqn{trijetqq}, 
while that of the second case can be obtained by exchanging $k_1$ and $k_2$ within the first contribution.

Specifically, in the first case ($k_1$ corresponds to the leading quark), the integral over $\bm{k}_{1}$
enforces $\bm{x}=\overline{\bm{x}}$ and the longitudinal momentum fractions should be evaluated as
\begin{equation}\label{longit3}
 \vartheta 
 \,=1-\frac{x_2+x_3}{x_p}\,,
 \qquad\xi
 \,=\,\frac{x_2}{x_p}\,.
 \end{equation}
The fact that  $\bm{x}=\overline{\bm{x}}$ simplifies the colour structure of the cross-sections in all
 the situations where the leading quark interacts in both the DA and the CCA: the respective Wilson
 lines compensate each other  by unitarity.
In what follows we shall exhibit these simplifications directly in the large $N_c$ limit.
 
Specifically, the 6-quark $S$-matrix $S_{qq\bar{q}\bar{q}\bar{q}q}$
 \eqref{wils1} reduces to a product of two dipoles (one for each of the two measured partons):
\begin{equation}\label{6q4}
S_{qq\bar{q}\bar{q}\bar{q}q}\left(\bm{x},\,\bm{z},\,\bm{z}^{\prime},\,
\overline{\bm{x}}=\bm{x},\,\overline{\bm{z}},\,\overline{\bm{z}}^{\prime}\right)
\,\simeq\,\mathcal{S}(\overline{\bm{z}}^{\prime},\,\bm{z}^{\prime})\,\mathcal{S}(\bm{z},\,\overline{\bm{z}}).
 \end{equation}
A similar simplification occurs in all the four terms within the squared brackets in  \eqn{trijetqq}; e.g.,
for the last term, as obtained via the double replacement $\big(\bm{z},\bm{z}^{\prime}\rightarrow\bm{y}\,
\ \&\ \,\bm{\overline{z}},\bm{\overline{z}}^{\prime}\rightarrow\overline{\bm{y}}\big)$, one finds
 \begin{equation} 
S_{qq\bar{q}\bar{q}\bar{q}q}\left(\bm{x},\,\bm{y},\,\bm{y},\,
\overline{\bm{x}}=\bm{x},\,\overline{\bm{y}},\,\overline{\bm{y}}\right)
\,\simeq\,\mathcal{S}(\overline{\bm{y}},\,\bm{y})\,\mathcal{S}(\bm{y},\,\overline{\bm{y}}).
 \end{equation}
This is recognised as the large-$N_c$ version of the $S$-matrix for a colour dipole made with
two gluons.

In the second case, where $k_1$ refers to the quark originating from the gluon decay, 
the integral over $\bm{k}_{1}$
enforces $\bm{z}=\overline{\bm{z}}$ and the longitudinal momentum fractions are evaluated as
\begin{equation}\label{longit2}
 \vartheta 
 \,=1-\frac{x_2}{x_p}\,,
 \qquad\xi 
 \,=1-\frac{x_2+x_3}{x_p}\,.
 \end{equation}
 As in the case of \eqn{2quarks}, the limit  $\bm{z}\to \overline{\bm{z}}$ does not commute
 with the various replacements in \eqn{trijetqq}, like  $\big(\bm{z}, \bm{z}^{\prime}\rightarrow\bm{y}\big)$,
 and must be performed {\it after} them. Accordingly, there is only one simplification
 in the colour structure, corresponding to the unique topology (in terms of
 shockwave insertions) in which the unmeasured gluon $k_1$ scatters both in the DA and in the CCA.
 One then has
 \begin{equation}\label{6q3}
S_{qq\bar{q}\bar{q}\bar{q}q}\left(\bm{x},\,\bm{z},\,\bm{z}^{\prime},\,
\overline{\bm{x}},\,\overline{\bm{z}}=\bm{z},\,\overline{\bm{z}}^{\prime}\right)
 \,\simeq\,\mathcal{Q}(\bm{x},\,\bm{z}^{\prime},\,\overline{\bm{z}}^{\prime},\,\overline{\bm{x}})\,.
 \end{equation}

\subsection{The two-gluon di-jets}


The cross-section for the  final state involving two measured gluons 
is obtained by ``integrating out'' the final quark in the trijet cross-section corresponding to the
$qgg$ final state. As discussed in Sect.~\ref{trijet-qgg}, there are three contributions to the
$qgg$  cross-section,  illustrated in Figs.~\ref{fig:qgg1}, \ref{fig:qgg2}, and \ref{qgg-interf}. 
 By integrating out these three contributions
over  the kinematics $(k_{1}^{+},\,\bm{k}_{1}$) of the unmeasured quark, one finds
\begin{align}
\label{real-gg}
\frac{d\sigma^{qA\rightarrow gg+X}_\rnlo}{dk_{2}^{+}\,d^{2}\bm{k}_{2}\,dk_{3}^{+}\,d^{2}\bm{k}_{3}}=\,
(2\pi)^2\sum_{i=1,2,3}\,\frac{d\sigma_{(i)}^{qA\rightarrow qgg+X}}{dk_{1}^{+}\,d^{2}\bm{k}_{1}\,dk_{2}^{+}\,d^{2}\bm{k}_{2}\,dk_{3}^{+}\,d^{2}\bm{k}_{3}}\,\Bigg |_{k^+_1=q^+-k^+_2-k^+_3\,,\ 
\bm{x}=\overline{\bm{x}}}.
\end{align}
We again used the compact but somewhat formal notation introduced in \eqn{2quarks}, where it
is understood that the $\delta$-function expressing longitudinal momentum conservation
is excluded from the trijet cross-sections.
In turn, each of the three terms in the r.h.s. is made with two pieces, corresponding to the permutations
of the measured gluons $k_2$ and $k_3$. 

It is further instructive to notice the simplifications in the colour structure of the trijet cross-sections
which appear due to the fact that the quark is not measured  ($\bm{x}=\overline{\bm{x}}$). 
These simplifications refer to the $S$-matrices denoted as $S_{qgg\bar{q}gg}^{(i)}$ with
$i=1,\,2,\,3$, which describe situations where the final quark scatters both in the DA and in the CCA. 
As usual, we show the ensuing simplifications only at large $N_c$. In the three cases, 
one of the two quadrupole factors reduces to a dipole. 
One finds\footnote{As implied by its notation, the colour dipole
$\mathcal{S}\left(\overline{\bm{z}}^{\prime},\,\bm{z}^{\prime}\right)$ is built with a quark located at
$\overline{\bm{z}}^{\prime}$ in the CCA and an anti-quark located at $\bm{z}^{\prime}$ in the DA; recall \eqn{lowils2}.},
\begin{align}\label{wils3-gg}
S_{qgg\bar{q}gg}^{(1)}\left(\bm{x},\,\bm{z},\,\bm{z}^{\prime},\,
\overline{\bm{x}}=\bm{x},\,\overline{\bm{z}},\,\overline{\bm{z}}^{\prime}\right)
&\,\simeq\,
S_{qgg\bar{q}gg}^{(3)}\left(\bm{x},\,\bm{z},\,\bm{z}^{\prime},\,
\overline{\bm{x}}=\bm{x},\,\overline{\bm{z}},\,\overline{\bm{z}}^{\prime}\right)\nn
&\, \simeq\,
\mathcal{S}\left(\overline{\bm{z}}^{\prime},\,\bm{z}^{\prime}\right)\,
\mathcal{Q}\left(\bm{z}^{\prime},\,\bm{z},\,\overline{\bm{z}},\,\overline{\bm{z}}^{\prime}\right)\,\mathcal{S}
\left(\bm{z}\,,\overline{\bm{z}}\right).
\end{align}
and respectively 
 \begin{align}\label{wils5-gg}
S_{qgg\bar{q}gg}^{(2)}\left(\bm{x},\,\bm{z},\,\bm{z}^{\prime},\,
\overline{\bm{x}}=\bm{x},\,\overline{\bm{z}},\,\overline{\bm{z}}^{\prime}\right)
&\,\simeq\,\frac{1}{2}\Big[\mathcal{S}\left(\overline{\bm{z}},\,\bm{z}\right)\,
\mathcal{Q}\left(\bm{z},\,\bm{z}^{\prime},\,\overline{\bm{z}}^{\prime},\,\overline{\bm{z}}\right)\,\mathcal{S}\left(\bm{z}^{\prime},\,\overline{\bm{z}}^{\prime}\right)\nn
&\qquad +
\mathcal{S}\left(\overline{\bm{z}}^{\prime},\,\bm{z}^{\prime}\right)\,\mathcal{Q}\left(\bm{z}^{\prime},\,\bm{z},\,\overline{\bm{z}},\,\overline{\bm{z}}^{\prime}\right)\,\mathcal{S}\left(\bm{z},\,\overline{\bm{z}}\right)\Big].
 \end{align}

\subsection{The quark-gluon dijets}

The quark-gluon dijet final state already exists at leading order, as discussed in Sect.~\ref{sec:LO}.
The corresponding real NLO corrections are obtained by integrating out one of the two final gluons
from the $qgg$ trijet cross-section computed in Sect.~\ref{trijet-qgg}, say the one with
momentum $k_3$.  The complete result is
therefore  the sum of three terms,
\begin{align}
\label{qgrNLO}
\frac{d\sigma_{\rnlo}^{qA\rightarrow qg+X}}{dk_{1}^{+}\,d^{2}\bm{k}_{1}\,dk_{2}^{+}\,d^{2}\bm{k}_{2}}
&\,=\int dk^+_3 d^2\bm{k}_{3}\,\sum_{i=1,2,3}\,
\frac{d\sigma_{(i)}^{qA\rightarrow qgg+X}}{dk_{1}^{+}\,d^{2}\bm{k}_{1}\,dk_{2}^{+}\,d^{2}\bm{k}_{2}\,dk_{3}^{+}\,d^{2}\bm{k}_{3}}\,.
\end{align}
In turn, each of these three terms is made with two pieces, since the unmeasured gluon with momentum 
$k_3$ can be any of the two final gluons in Figs.~\ref{fig:qgg1}, \ref{fig:qgg2}, and \ref{qgg-interf}.
Consider for definiteness the case where this is the gluon with longitudinal momentum $\vartheta -\xi$
and transverse coordinates $\bm{z}^\prime$ and $\overline{\bm{z}}^\prime$ in the DA and the CCA,
respectively. Then the integral over $\bm{k}_{3}$ enforces $\bm{z}^\prime=\overline{\bm{z}}^\prime$
and the longitudinal  fractions should be computed as in \eqn{longit2}, that is,
 \begin{equation}\label{longit21}
 \vartheta=1-\frac{k_{1}^{+}}{q^{+}} \,=1-\frac{x_1}{x_p}\,,
\qquad\xi=\frac{k_{2}^{+}}{q^{+}}\,=\frac{x_2}{x_p} \,.
 \end{equation}

It is again instructive to display the main simplifications in the colour structure due to the fact
that one of the final gluons is not measured ($\bm{z}^\prime=\overline{\bm{z}}^\prime$).  One finds
\begin{align}\label{wils3-qg2}
S_{qgg\bar{q}gg}^{(1)}\left(\bm{x},\,\bm{z},\,\bm{z}^{\prime},\,
\overline{\bm{x}},\,\overline{\bm{z}},\,\overline{\bm{z}}^{\prime}=\bm{z}^{\prime}\right)
\simeq S_{qgg\bar{q}gg}^{(3)}\left(\bm{x},\,\bm{z},\,\bm{z}^{\prime},\,
\overline{\bm{x}},\,\overline{\bm{z}},\,\overline{\bm{z}}^{\prime}=\bm{z}^{\prime}\right)
\simeq\mathcal{S}\left(\bm{x},\,\bm{\overline{x}}\right)\,
\mathcal{S}\left(\overline{\bm{z}},\,\bm{z}\right)
\mathcal{S}\left(\bm{z},\,\overline{\bm{z}}\right)\,,
\end{align}
\begin{align}\label{wils5-qg2}
S_{qgg\bar{q}gg}^{(2)}\left(\bm{x},\,\bm{z},\,\bm{z}^{\prime},\,
\overline{\bm{x}},\,\overline{\bm{z}},\,\overline{\bm{z}}^{\prime}=\bm{z}^{\prime}\right)
\simeq\,\frac{1}{2}\Big[\mathcal{S}\left(\bm{x},\,\bm{\overline{x}}\right)\,
\mathcal{S}\left(\overline{\bm{z}},\,\bm{z}\right)
\mathcal{S}\left(\bm{z},\,\overline{\bm{z}}\right)
 +\mathcal{S}\left(\bm{z},\,\overline{\bm{z}}\right)\,\mathcal{Q}\left(\bm{x},\,\bm{z},\,\overline{\bm{z}},\,\overline{\bm{x}}\right)\Big].
  \end{align}

\section{Soft gluon emissions: recovering the B-JIMWLK evolution}
\label{sec:JIMWLK}

In this section we shall study the behaviour of the previous results in the limit where one of the two final 
gluons is very soft, that is, either $\xi\to 0$  or $\vartheta-\xi\to 0$. This is interesting since, in this limit,
our general results should reduce to one-step in the B-JIMWLK (or BK) evolution of the leading-order dijet cross-section. As we shall see, this expectation is indeed verified and it provides a rather strong test of the correctness of our previous calculation.
For more generality, we shall first address the soft gluon limit for the trijet ($qgg$) cross-section, before
eventually specialising to the (real) NLO corrections to the dijet ($qg$) cross-section.


\subsection{Direct emissions by the quark}
\label{sec:eikonal1}

We start with the contribution to the cross-section shown in \eqn{trijetqgg1},
where the two gluons are directly emitted by the quark, both in the DA and in the CCA
 (recall Fig.~\ref{fig:qgg1}).  When (at least) one of these gluons is soft, there are two
 types of simplifications. 
 
 First one can neglect the recoil of the emitter (here, the quark) at the respective emission vertex,
 meaning that its transverse coordinate  is not modified by the soft emission. 
 With reference to Fig.~\ref{final-qgg1}, when $\xi\to 0$
 one can approximate  $\bm{y}\simeq \bm{w}\simeq (1-\vartheta)\bm{x}+\vartheta\bm{z}'$,
 whereas for $\vartheta-\xi\to 0$, one has $\bm{y}\simeq \bm{x}$ and 
  $\bm{w}\simeq (1-\vartheta)\bm{x}+\vartheta\bm{z}$. 
  
  Second, one can simplify the
  dependence of the kernel \eqref{defK1} --- which, we recall, encodes information
  about both the emission vertices and the energy denominators ---
  upon the longitudinal momentum fractions $\vartheta$ and $\xi$. It is easy
  to check that the instantaneous piece of the effective vertex $\Xi$ (cf.  \eqn{vertexXi})
   does not contribute to
  the would-be singularity in the soft limit. Hence, one can use the expression 
  \eqref{XIXI} for the square of the effective vertex. This gives the following soft limits:
 \begin{align}\label{K1-num}
\Xi_{\lambda_{1}\lambda}^{ijmn}
(\vartheta,\,\xi)
\Xi_{\lambda_{1}\lambda}^{ijpq\,*}
(\vartheta,\,\xi)
&\ \longrightarrow \ \delta^{mp}\delta^{nq}\,\frac{16\vartheta (1-\vartheta)^4\,[1+(1-\vartheta)^2]}
{\vartheta-\xi}\,,\quad\mbox{when}\quad \vartheta-\xi\to 0,\,\nonumber\\*[0.2cm]
&\ \longrightarrow \ \delta^{mp}\delta^{nq}\, \frac{16\xi[1+(1-\vartheta)^2]}{\vartheta}\,
,\quad\mbox{when}\quad \xi\to 0
\,.
\end{align}
Concerning the denominator of $\mathcal{K}_{1}$, the limit $\vartheta-\xi\to 0$ poses no
special difficulty and yields
\begin{align}\label{K1-eik1}
\mathcal{K}_{1}^{mnpq} 
&\ \longrightarrow \ \delta^{mp}\delta^{nq}\, \frac{16[1+(1-\vartheta)^2]}
{\vartheta(\vartheta-\xi)\bm{Y}^{2}\overline{\bm{Y}}^{2}}\,\quad\mbox{when}\quad \vartheta-\xi\to 0\,.
\end{align}
The other eikonal limit, namely $\xi\to 0$, is a bit more subtle, as it does not
commute with the equal transverse points limits which define three of the four terms 
within the square brackets in \eqn{trijetqgg1}. Keeping the dominant terms as $\xi\to 0$ in each of
the respective kernels, one finds
\begin{align}\label{K1-eik2}
\mathcal{K}_{1}^{mnpq}\Big |_{\xi\to 0} 
&\ \longrightarrow \ 0 ,\quad\mbox{when both $\bm{X}^{\prime}\ne 0$
and $\overline{\bm{X}}^{\prime}\ne 0$}  \,, 
\nonumber\\*[0.2cm]
&\ \longrightarrow \ \delta^{mp}\delta^{nq}\, \frac{16[1+(1-\vartheta)^2]}
{\vartheta^2(1-\vartheta)\bm{Y}^{2}(\overline{\bm{X}}^{\prime})^{2}}\,,\quad\mbox{when
$\bm{X}^{\prime}=0$, but $\overline{\bm{X}}^{\prime}\ne 0$}
\,,
\nonumber\\*[0.2cm]
&\ \longrightarrow \ \delta^{mp}\delta^{nq}\, \frac{16[1+(1-\vartheta)^2]}
{\vartheta^2(1-\vartheta)(\bm{X}^{\prime})^{2}\overline{\bm{Y}}^{2}}\,,\quad\mbox{when
$\bm{X}^{\prime}\ne 0$, but $\overline{\bm{X}}^{\prime}= 0$}
\,,
\nonumber\\*[0.2cm]
&\ \longrightarrow \ \delta^{mp}\delta^{nq}\,  \frac{16[1+(1-\vartheta)^2]}
{\xi\vartheta\bm{Y}^{2}\overline{\bm{Y}}^{2}}\,,\quad\mbox{when
$\bm{X}^{\prime}=\overline{\bm{X}}^{\prime}=0$}\,,
\end{align}
These expressions exhibit single poles at  $\vartheta=\xi$ and respectively $\xi= 0$, 
which are the expected infrared singularities associated with the bremsstrahlung of very soft 
gluons. What is however quite remarkable and also important for what follows, is the special way how
these singularities appear within the structure of the cross-section \eqref{trijetqgg1}:
\texttt{(i)} the pole at  $\vartheta=\xi$ is common to all the four terms in \eqn{trijetqgg1} and,
moreover, the respective kernels become degenerate in this limit;
\texttt{(ii)} the pole at $\xi= 0$ refers only
to the last term in \eqn{trijetqgg1},  as obtained via the double replacement 
$\big(\bm{x},\bm{z}^{\prime}\rightarrow\bm{y}\,
\ \&\ \,\bm{\overline{x}},\bm{\overline{z}}^{\prime}\rightarrow\overline{\bm{y}}\big)$.
These properties are essential in order to recover the JIMWLK evolution,
as we now explain.

 \begin{figure}[!t]\center
 \includegraphics[scale=0.95]{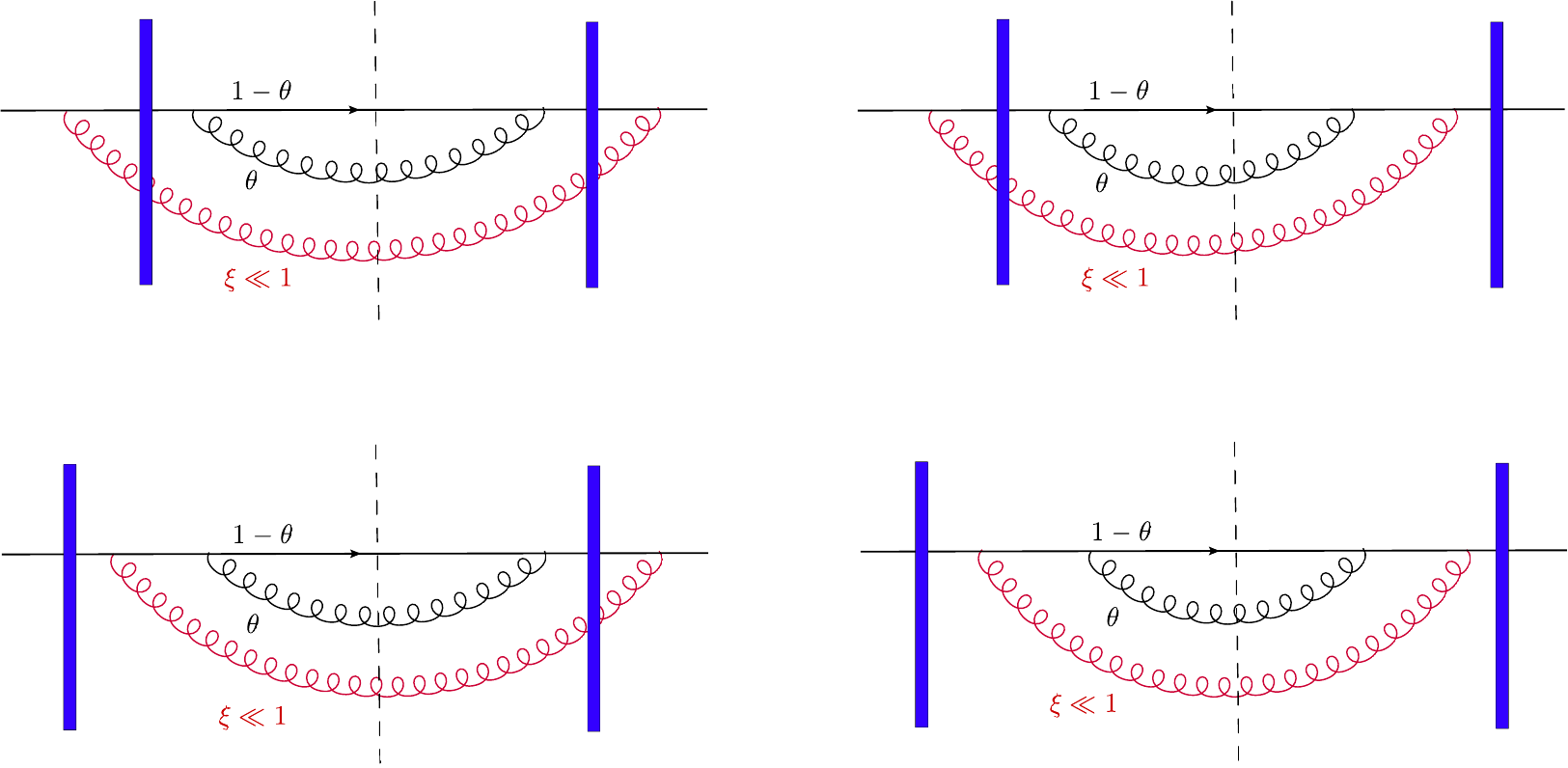}
\caption{The four diagrams which dominate the cross-section \eqref{trijetqgg1} in
the eikonal limit $\xi\to 0$ in which the first emitted gluon (shown in red) is soft.}
\label{qgg1-first-soft}
\end{figure}

The limit $\xi\to 0$ corresponds to the case where the soft gluon is the first one to be
emitted (see Fig.~\ref{final-qgg1}). Property \texttt{(ii)} above tells us that the dominant
graphs in this limit are the four graphs shown in Fig.~\ref{qgg1-first-soft}, where the interaction with
the shockwave occurs either in the initial, or in the intermediate, state 
(the soft gluon is drawn in red). Notice that this is only a small subset of all
the possible topologies which contribute to the cross-section \eqref{trijetqgg1} in general:
4 out of 16. (Two of the 12 excluded diagrams are illustrated in Fig.~\ref{qgg1-BKfake1}.)
What is special about the four graphs in Fig.~\ref{qgg1-first-soft} is the fact that there is
no intermediate emission vertex between the  shockwave and the emission of the soft gluon,
neither in the DA, nor in the CCA (compare in this respect  Fig.~\ref{qgg1-first-soft}
to Fig.~\ref{qgg1-BKfake1}). In other terms, the parent quark radiates the soft gluon right
before, or right after, its collision with the nucleus. This is in agreement with the fact that a soft
gluon has a short formation time, hence it is first one to be emitted after a collision,
or the last one to be emitted prior to it.

 \begin{figure}[!t]\center
 \includegraphics[scale=0.9]{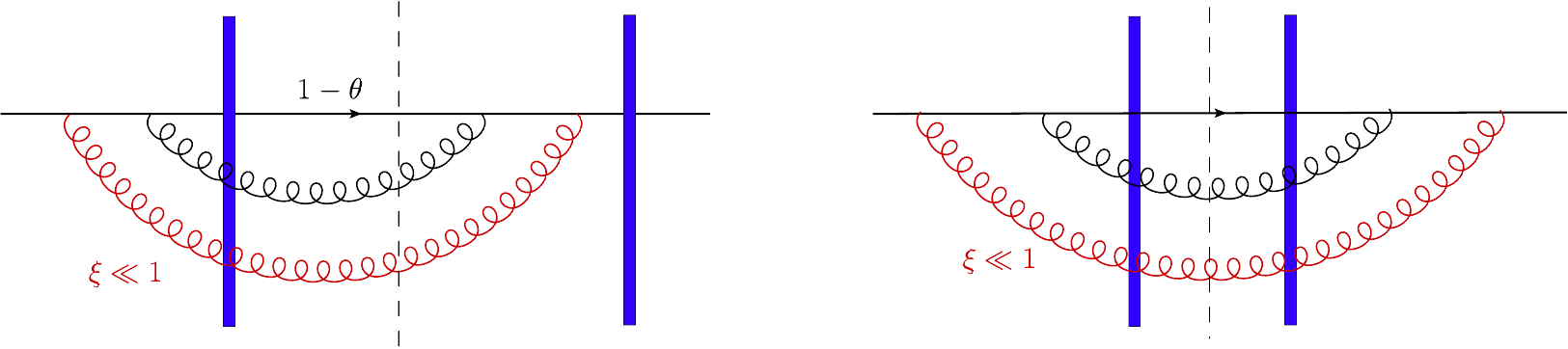}
\caption{Diagrams which contribute to the $qgg$ cross-section \eqref{trijetqgg1}
in general, but which do not survive in the eikonal (soft) limit for the gluon
with  longitudinal momentum fraction $\xi$ (shown in red).
These graphs do not contribute to the JIMWLK evolution of any of the $S$-matrices
which enter  the LO cross-section \eqref{LOfinal}.}
\label{qgg1-BKfake1}
\end{figure}

The graphs shown  in Fig.~\ref{qgg1-first-soft} have the right structure to describe
one step in the JIMWLK evolution of the LO dijet cross-section: they precisely encode the BK 
evolution of the last piece in \eqn{LOfinal} --- the dipole $S$-matrix $\mathcal{S}\left(\bm{w},\,\overline{\bm{w}}\right)$ ---, as we now demonstrate. Let us use the momentum assignments shown
 in Fig.~\ref{fig:qgg1}, i.e. $k_2^+=\xi q^+$ and
$k_3^+=(\vartheta-\xi)q^+$, with $\xi\to0$.
In this limit, \eqn{trijetqgg1} simplifies to
 \begin{align}\label{qgg1-eik1}
 \hspace*{-0.6cm}
&\frac{d\sigma_{(1)}^{qA\rightarrow qgg+X}}{dk_{1}^{+}\,d^{2}\bm{k}_{1}\,dk_{2}^{+}\,d^{2}\bm{k}_{2}\,dk_{3}^{+}\,d^{2}\bm{k}_{3}}\,\bigg |_{\xi\to 0} =\frac{8\alpha_{s}^{2}\,C_{F}^2}{(2\pi)^{10}}\,
\frac{1+(1-\vartheta)^2}{\xi\vartheta(q^{+})^{2}}\,
\delta(q^{+}-k_{1}^{+}-k_{3}^{+})\nonumber \\*[0.2cm]
&\qquad\qquad\times
\int_{\overline{\bm{x}},\,\bm{\overline{z}},\,\bm{\overline{z}}^{\prime},\,\bm{x},\,\bm{z},\,\bm{z}^{\prime}}\,e^{-i\bm{k}_{1}\cdot(\bm{x}-\overline{\bm{x}})-i\bm{k}_{2}\cdot(\bm{z}-\overline{\bm{z}})-i\bm{k}_{3}\cdot(\bm{z}^{\prime}-\overline{\bm{z}}^{\prime})}\,
\frac{\left(\bm{Y}\cdot\overline{\bm{Y}}\right)\left(\bm{X}^{\prime}\cdot \bm{\overline{X}}^{\prime}\right)
}{\bm{Y}^{2}\overline{\bm{Y}}^{2}(\bm{X}^{\prime})^{2} (\bm{\overline{X}}^{\prime})^{2}}\nonumber\\*[0.2cm]
&\qquad\qquad\times\Big[
\mathcal{Q}(\bm{w},\,\bm{z},\,\overline{\bm{z}},\,\overline{\bm{w}})\,\mathcal{S}(\bm{z},\,\overline{\bm{z}}) -\mathcal{S}(\bm{w},\,\bm{z})\,\mathcal{S}(\bm{z},\,\overline{\bm{w}})
 -\mathcal{S}(\bm{w},\,\overline{\bm{z}})\,\mathcal{S}(\overline{\bm{z}}, \overline{\bm{w}})
 +\mathcal{S}\left(\bm{w},\,\bm{\overline{w}}\right)\Big],
 \end{align}
where we have used  $\bm{w}\simeq\bm{y}\simeq (1-\vartheta)\bm{x}+\vartheta\bm{z}'$ and we recall
that $\bm{Y}=\bm{y}-\bm{z}\simeq \bm{w}-\bm{z}$  and  $\bm{X}^{\prime}= \bm{x}-\bm{z}^{\prime}$ (and similarly for the 
respective coordinates with a bar). Also, we have ignored the soft momentum $k_2^+$ in the
$\delta$--function for longitudinal momentum conservation. This is in the spirit of the soft gluon
approximation to the high-energy evolution, which effectively violates energy conservation, albeit only
marginally (by ignoring the soft gluons in the energy balance).

As expected, \eqn{qgg1-eik1} is the same as the result of acting with the ``production'' version of
the JIMWLK Hamiltonian \cite{Kovner:2006ge,Kovner:2006wr,Iancu:2013uva}
on the last term (the dipole $\mathcal{S}\left(\bm{w},\,\bm{\overline{w}}\right)$)
in the LO $qg$ cross-section \eqref{LOfinal}. The ``production'' Hamiltonian
is a generalised version of the JIMWLK Hamiltonian which, when acting on the cross-section for particle production in dilute-dense ($pA$ or $eA$) collisions, leads to the emission of an additional, soft,
gluon, which is measured in the final state. 
If on the other hand this soft gluon is {\it not} measured,
i.e. if one integrates out its kinematics, then one generates the ``real'' piece of the standard
JIMWLK evolution. By integrating  \eqn{qgg1-eik1}  over $k_2^+=\xi q^+$ and $\bm{k}_{2}$,
one finds (with $2\alpha_s C_F
\simeq \alpha_s N_c \equiv\pi\abar$ at large $N_c$)
\begin{align}
\label{qgevolu1}
\frac{d\sigma_{\nlo, 1}^{qA\rightarrow qg+X}}{dk_{1}^{+}\,d^{2}\bm{k}_{1}\,dk_{3}^{+}\,d^{2}\bm{k}_{3}}
&\,\simeq \,\frac{\abar}{(2\pi)^5}\,\frac{1+(1-\vartheta)^2}{2\vartheta q^{+}}\,
\delta(q^{+}-k_{1}^{+}-k_{3}^{+}) \nonumber \\*[0.4cm] 
&\ \times
\int_{\overline{\bm{x}},\,\bm{\overline{z}}^{\prime},\,\bm{x},\,\bm{z}^{\prime}}\,e^{-i\bm{k}_{1}\cdot(\bm{x}-\overline{\bm{x}})-i\bm{k}_{3}\cdot(\bm{z}^{\prime}-\overline{\bm{z}}^{\prime})}\,
\frac{(\bm{x}-\bm{z}')\cdot (\overline{\bm{x}}-\overline{\bm{z}}')}
{(\bm{x}-\bm{z}')^{2} (\overline{\bm{x}}-\overline{\bm{z}}')^{2}}
\nonumber
\\*[0.4cm] 
&\ \times \frac{\abar}{2\pi} \int_0^1\frac{d\xi}{\xi}\,\int_{\bm{z}}
\frac{2(\bm{w}-\bm{z})\cdot (\bm{\overline{w}}-\bm{z})}{(\bm{w}-\bm{z})^2 (\bm{\overline{w}}-\bm{z})^{2}}\,
\Big[\mathcal{S}\left(\bm{w},\,\bm{\overline{w}}\right) -
\mathcal{S}(\bm{w},\,\bm{z})\,\mathcal{S}(\bm{z},\,\overline{\bm{w}})
 \Big].
\end{align}
Notice the simplifications in the colour structure due to the fact that the integral over $\bm{k}_{2}$ has  
identified the coordinates $\bm{z}$ and $\bm{\overline{z}}$ of the unmeasured gluon: 
$\bm{z}=\bm{\overline{z}}$. 

The expression in the third line of \eqn{qgevolu1}
 is recognised as the ``real'' part of the BK equation for the
evolution of the dipole $\mathcal{S}\left(\bm{w},\,\bm{\overline{w}}\right)$. 
More precisely, it exhibits  the
{\it integral} version of the BK equation, that is, its formal solution as obtained by
integrating the r.h.s. of that equation over $\xi$. This integral  
exhibits a logarithmic divergence at its lower limit ($\xi=0$). 
In reality, this divergence is cut off by the condition of energy conservation. 
Indeed, this integral can be equivalently rewritten as $\int({d\xi}/{\xi})=
\int(dx_2/x_2)$, with $x_2=k_2^+/Q^+$; then \eqn{xg3} together with the constraint
$x_g \le 1$ clearly implies a lower-limit on $x_2$, hence on $\xi$.

This discussion also shows that the soft ($\xi\ll 1$) part of the integral over the unmeasured gluon
contributes to the high-energy (B-JIMWLK) evolution of the LO dijet cross-section, so it is only
the remaining contribution at large $\xi\sim 1$ which should be viewed as a genuine NLO correction
to the ``hard impact factor'', i.e. to the partonic cross-section for the $qA\to qg+X$ process.
That said, it turns out that the separation of the full integral over $\xi$ between a ``soft part''
describing LO evolution and a ``hard part'' describing NLO corrections to the impact factor is
generally subtle, due to the fact that the various $S$-matrices in \eqn{qgevolu1}
do implicitly depend upon $\xi$. (Indeed, their rapidity evolution scale is $x_g$,
which is a function of $x_2$, hence of $\xi$, as shown in \eqn{xg3}.) This dependence
is essential, as argued in Refs.~\cite{Iancu:2016vyg,Ducloue:2017dit}: 
attempts to separate the ``soft'' evolution
from the ``hard'' NLO impact factor which ignore this evolution may lead into troubles, like
negative values for the NLO cross-section, as observed in the context of single inclusive
particle production in both $pA$ \cite{Chirilli:2012jd,Stasto:2013cha,Ducloue:2017mpb}
 and $eA$ collisions \cite{Ducloue:2017ftk}. A detailed discussion of such issues 
in the context of dijet production goes well beyond the purposes of this paper. But one should keep
them in mind when trying to actually estimate the NLO corrections (say, for the purposes
of the phenomenology).

  \begin{figure}[!t]\center
 \includegraphics[scale=0.95]{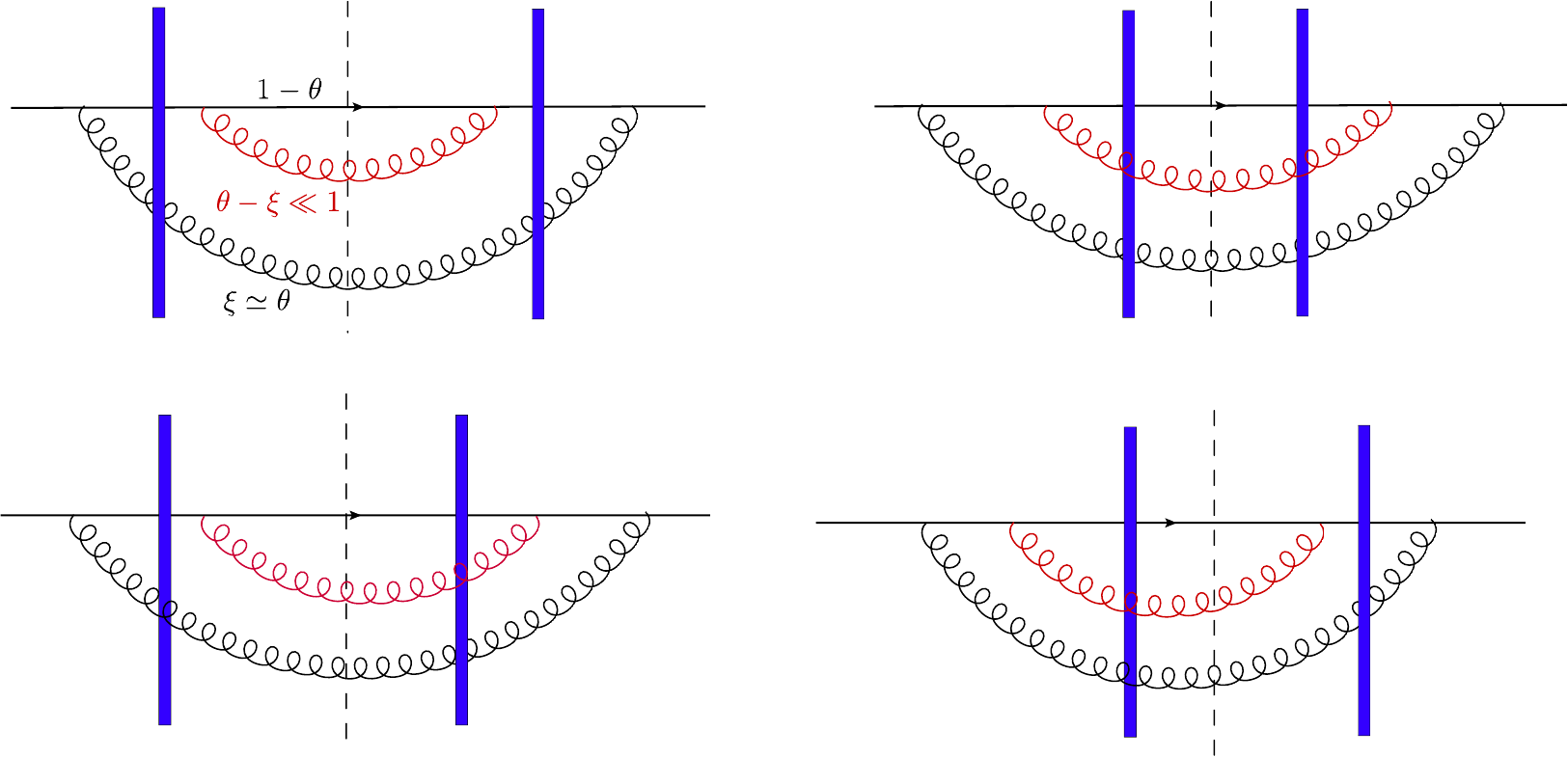}
\caption{The four diagrams contributing to the cross-section \eqref{trijetqgg1}
which survive in  the eikonal limit $\vartheta-\xi\to 0$ in which the second 
emitted gluon (shown in red) is soft.}
\label{qgg1-second-soft}
\end{figure}

 \begin{figure}[!t]\center
 \includegraphics[scale=0.9]{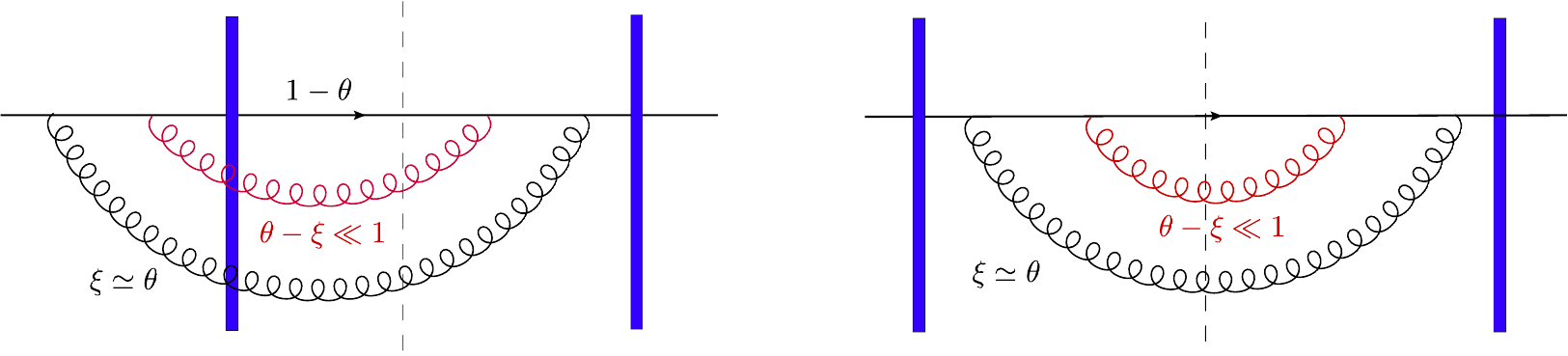}
\caption{Diagrams which contribute to the $qgg$ cross-section \eqref{trijetqgg1}
in general, but which do not survive in the eikonal (soft) limit for the gluon with 
longitudinal momentum fraction $\vartheta-\xi$ (shown in red).}
\label{qgg1-BKfake2}
\end{figure}

Consider similarly the eikonal limit for the second emitted gluon, that with longitudinal momentum 
$k_3^+=(\vartheta-\xi)q^+$. \eqn{K1-eik1} implies that, when this gluon is soft  ($\vartheta-\xi\ll 1$),
all the four terms within the squared brackets in \eqn{trijetqgg1} are multiplied by the same kernel, hence the
respective colour structures can be simply added to each other. Remarkably, 12 of the 16 partonic
$S$-matrices which are {\it a priori} contained in these structures mutually cancel in their sum,
and we are left with 
 \begin{align}\label{qgg1-eik2}
 \hspace*{-0.6cm}
&\frac{d\sigma_{(1)}^{qA\rightarrow qgg+X}}{dk_{1}^{+}\,d^{2}\bm{k}_{1}\,dk_{2}^{+}\,d^{2}\bm{k}_{2}\,dk_{3}^{+}\,d^{2}\bm{k}_{3}}\,\bigg |_{\vartheta-\xi\to 0}
 =\frac{8\alpha_{s}^{2}\,C_{F}^2}{(2\pi)^{10}}\,
\frac{1+(1-\vartheta)^2}{(\vartheta-\xi)\vartheta(q^{+})^{2}}\,
\delta(q^{+}-k_{1}^{+}-k_{2}^{+})\nonumber\\*[0.2cm]
&\qquad\qquad\times
\int_{\overline{\bm{x}},\,\bm{\overline{z}},\,\bm{\overline{z}}^{\prime},\,\bm{x},\,\bm{z},\,\bm{z}^{\prime}}\,e^{-i\bm{k}_{1}\cdot(\bm{x}-\overline{\bm{x}})-i\bm{k}_{2}\cdot(\bm{z}-\overline{\bm{z}})-i\bm{k}_{3}\cdot(\bm{z}^{\prime}-\overline{\bm{z}}^{\prime})}\,
\frac{\left(\bm{X}\cdot\overline{\bm{X}}\right)\left(\bm{X}^{\prime}\cdot \bm{\overline{X}}^{\prime}\right)
}{\bm{X}^{2}\overline{\bm{X}}^{2}(\bm{X}^{\prime})^{2} (\bm{\overline{X}}^{\prime})^{2}}\nonumber\\*[0.2cm]
&\qquad\qquad\times\Big[\mathcal{Q}\left(\bm{x},\,\bm{z}^{\prime},\,\overline{\bm{z}}^{\prime},\,\bm{\overline{x}}\right)
\mathcal{Q}\left(\bm{z}^{\prime},\,\bm{z},\,\overline{\bm{z}},\,\overline{\bm{z}}^{\prime}\right)\,-\,
\mathcal{S}\left(\overline{\bm{z}}^{\prime},\,\bm{\overline{x}}\right)
\mathcal{Q}\left(\bm{x},\,\bm{z},\,\overline{\bm{z}},\,\overline{\bm{z}}^{\prime}\right)
\nonumber\\*[0.2cm]
&\qquad\qquad\quad - \,\mathcal{S}\left(\bm{x},\,\bm{z}^{\prime}\right)
\mathcal{Q}\left(\bm{z}^{\prime},\,\bm{z},\,\overline{\bm{z}},\,\overline{\bm{x}}\right)\, +\,
\mathcal{Q}\left(\bm{x},\,\bm{z},\,\overline{\bm{z}},\,\overline{\bm{x}}\right)\Big]\mathcal{S}
\left(\bm{z}\,,\overline{\bm{z}}\right)\,,
 \end{align}
where $\bm{X}= \bm{x}-\bm{z}$ and $\bm{X}^{\prime}= \bm{x}-\bm{z}^{\prime}$
 (we have used $\bm{y}\simeq \bm{x}$). The four surviving $S$-matrices correspond to the situations
 where the shockwave is inserted either in the final $qgg$ state, or in the intermediate $qg$ state (see Fig.~\ref{qgg1-second-soft}),
 so that at least one gluon participates in the scattering
 (together with the quark, of course). For more clarity,
 we also show in  Fig.~\ref{qgg1-BKfake2} two graphs whose contributions have cancelled
 in the overall sum. 
 
%

 The four surviving graphs in Fig.~\ref{qgg1-second-soft} follow the same pattern
 as that illustrated in Fig.~\ref{qgg1-first-soft}: the soft gluon is the last (first) one
 to be emitted prior to (after) the shockwave. In this case too, they describe the JIMWLK evolution
 of one of the $S$-matrices  from \eqn{LOfinal} --- here, the colour quadrupole 
 which enters in the structure of the first term there, cf. \eqn{lowils1} and
  \eqref{pALOqchannel}. Specifically, these are the
 four diagrams describing the evolution of the quadrupole $\mathcal{Q}\left(\bm{x},\,\bm{z},\,\overline{\bm{z}},\,\bm{\overline{x}}\right)$ via the emission of a soft gluon which is emitted at $\bm{x}$ in the DA
 and reabsorbed at $\bm{\overline{x}}$ in the CCA.  One can check (e.g. by comparing to Eq.~(A.5) in 
 Ref.~\cite{Iancu:2013uva}) that the various factors in \eqn{qgg1-eik2} are precisely as needed 
 in order to describe this particular evolution\footnote{In the case where the soft gluon $k_3$
 is not measured (which amounts to identifying $\bm{z}^{\prime}=\bm{z}$), we precisely
  recover the relevant ``real'' terms from the JIMWLK equation for the high-energy evolution
  of the quadrupole $S$-matrix  \cite{Dominguez:2011gc,Iancu:2011ns,Iancu:2011nj}.}.
Of course, the JIMWLK evolution of the quadrupole also involves
 other diagrams, in which the soft gluon is emitted/absorbed by some other pair of external legs
   \cite{Dominguez:2011gc,Iancu:2011ns,Iancu:2011nj,Iancu:2013uva}.
 These additional diagrams will be generated by the remaining contributions to the $qgg$ cross-section,
 whose soft limit will be explored in the next two subsections.

\subsection{Direct contributions involving the 3-gluon vertex}
\label{sec:eikonal2}

The discussion of the soft gluon limit for the second contribution to the $qgg$ cross-section,
 \eqn{trijetqgg2}, is somewhat simpler, due to the manifest symmetry of this contribution under
the exchange of the two final gluons. It is therefore enough to consider only one of the two 
eikonal limits aforementioned, say $\xi\to 0$. In this limit, the triple-gluon vertex in \eqn{Gamma} can
be approximated as
\beq
\Gamma^{nlij}(\vartheta,\,\xi\to 0)\,\simeq\,-\frac{1}{\xi}\,\delta_{ni}\delta_{lj}\,.
\eeq
It is then easily to see, by using the expressions \eqref{vertexPi} for the relevant effective vertex and
\eqref{defK2} for the kernel, that the latter admits the same limit for all the four terms
within the square brackets in  \eqn{trijetqgg2}, that is,
\begin{equation}\label{K2eik}
\mathcal{K}_{2}^{mnpq}\Big |_{\xi\to 0}\,
\,=\,\frac{4\left[1+(1-\vartheta)^{2}\right]\,\delta^{mp}\,\delta^{nq}}{\vartheta\xi\bm{R}^{2}\,\overline{\bm{R}}^{2}}\,.
\end{equation}
This expression 
shows the expected soft singularity: a single pole at $\xi=0$. As before, this degeneracy of the
four kernels implies that only four topologies survive in the final result (among the 16 
 which were originally present in the r.h.s. of \eqn{trijetqgg2}): those in which the shockwave is
inserted in either the final $qgg$ state, or in the intermediate $qg$ state.  The surviving
configurations are shown in Fig.~\ref{qgg2-soft}. 
At large $N_c$, each of the $S$-matrices associated with these four graphs 
is in turn built with two colour structures, corresponding to permutations of the two gluons, 
as shown e.g. in \eqn{wils5}.
Still for $\xi\to 0$, one can also identify
$\bm{z}^{\prime}=\bm{y}$ and $ \overline{\bm{z}}^\prime=\overline{\bm{y}}$
in the arguments of the surviving $S$-matrices. 

We are thus led to the following result
(we use the conventions in Fig.~\ref{fig:qgg2}, i.e. the $k_2^+=\xi q^+$ and
$k_3^+=(\vartheta-\xi)q^+$): 
 \begin{align}\label{qgg2-eik}
 \hspace*{-0.6cm}
&\frac{d\sigma_{(2)}^{qA\rightarrow qgg+X}}{dk_{1}^{+}\,d^{2}\bm{k}_{1}\,dk_{2}^{+}\,d^{2}\bm{k}_{2}\,dk_{3}^{+}\,d^{2}\bm{k}_{3}}\,\bigg |_{\xi\to 0}
 =\frac{4\alpha_{s}^{2}\,C_{F}N_c}{(2\pi)^{10}}\,
\frac{1+(1-\vartheta)^2}{\xi\vartheta(q^{+})^{2}}\,
\delta(q^{+}-k_{1}^{+}-k_{3}^{+})\nonumber \\*[0.2cm]
&\qquad\qquad\times
\int_{\overline{\bm{x}},\,\bm{\overline{z}},\,\bm{\overline{y}},\,\bm{x},\,\bm{z},\,\bm{y}}\,e^{-i\bm{k}_{1}\cdot(\bm{x}-\overline{\bm{x}})-i\bm{k}_{2}\cdot(\bm{z}-\overline{\bm{z}})-i\bm{k}_{3}\cdot(\bm{y}-\overline{\bm{y}})}\,
\frac{\left(\bm{R}\cdot\overline{\bm{R}}\right)\left(\bm{Z}\cdot \bm{\overline{Z}}\right)
}{\bm{R}^{2}\,\overline{\bm{R}}^{2}\bm{Z}^{2}\,\overline{\bm{Z}}^{2}}
\nonumber\\*[0.2cm]
&\qquad\qquad\times
\bigg\{\Big[\mathcal{Q}\left(\bm{x},\,\bm{z},\,\overline{\bm{z}},\,\bm{\overline{x}}\right)
\mathcal{Q}\left(\bm{z},\,\bm{y},\,\overline{\bm{y}},\,\overline{\bm{z}}\right)\,-\,
\mathcal{Q}\left(\bm{x},\,\bm{y},\,\overline{\bm{z}},\,\overline{\bm{x}}\right)
\mathcal{S}\left(\overline{\bm{y}},\,\bm{\overline{z}}\right)
\nonumber\\*[0.2cm]
&\qquad\qquad\quad - 
\mathcal{Q}\left(\bm{x},\,\bm{z},\,\overline{\bm{y}},\,\overline{\bm{x}}\right)
\mathcal{S}\left(\bm{z},\,\bm{y}\right)\, +\,
\mathcal{Q}\left(\bm{x},\,\bm{y},\,\overline{\bm{y}},\,\overline{\bm{x}}\right)\Big]\mathcal{S}
\left(\bm{y}\,,\overline{\bm{y}}\right)\,,
\nonumber\\*[0.2cm]
&\qquad\qquad\quad + 
\Big[\mathcal{Q}\left(\bm{y},\,\bm{z},\,\overline{\bm{z}},\,\bm{\overline{y}}\right)
\mathcal{S}\left({\bm{z}},\,\bm{\overline{z}}\right)
- \mathcal{S}\left(\overline{\bm{z}},\,\bm{\overline{y}}\right)
\mathcal{S}\left({\bm{y}},\,\bm{\overline{z}}\right)
\nonumber\\*[0.2cm]
&\qquad\qquad\quad - 
\mathcal{S}\left(\bm{y},\,\bm{z}\right)
\mathcal{S}\left({\bm{z}},\,\bm{\overline{y}}\right)
+\mathcal{S}\left(\bm{y}\,,\overline{\bm{y}}\right)
\Big]\mathcal{Q}\left(\bm{x},\,\bm{y},\,\overline{\bm{y}},\,\overline{\bm{x}}\right)
\bigg\}
 \end{align}
where $\bm{Z}=\bm{z}-\bm{y}$, $\bm{R}=\bm{x}-\bm{y}$ and similarly for the coordinates with a bar.
Both the first term within the first square bracket and the first term within the second square bracket
are generated by the first graph in Fig.~\ref{qgg2-soft}, via its decomposition at large
$N_c$, which is graphically illustrated in Fig.~\ref{qgg2-largeNc}. This figure also makes clear
that the four terms within the first square bracket describe the emission of a soft gluon from the
quadrupole $\mathcal{Q}\left(\bm{x},\,\bm{y},\,\overline{\bm{y}},\,\overline{\bm{x}}\right)$,
such that the soft gluon is emitted at $\bm{y}$ in the DA and at $\overline{\bm{y}}$ in the CCA.
Similarly, the four terms within the second square bracket describe the emission of a soft gluon 
from the dipole $\mathcal{S}\left(\bm{y}\,,\overline{\bm{y}}\right)$.

 \begin{figure}[!t]\center
 \includegraphics[scale=0.95]{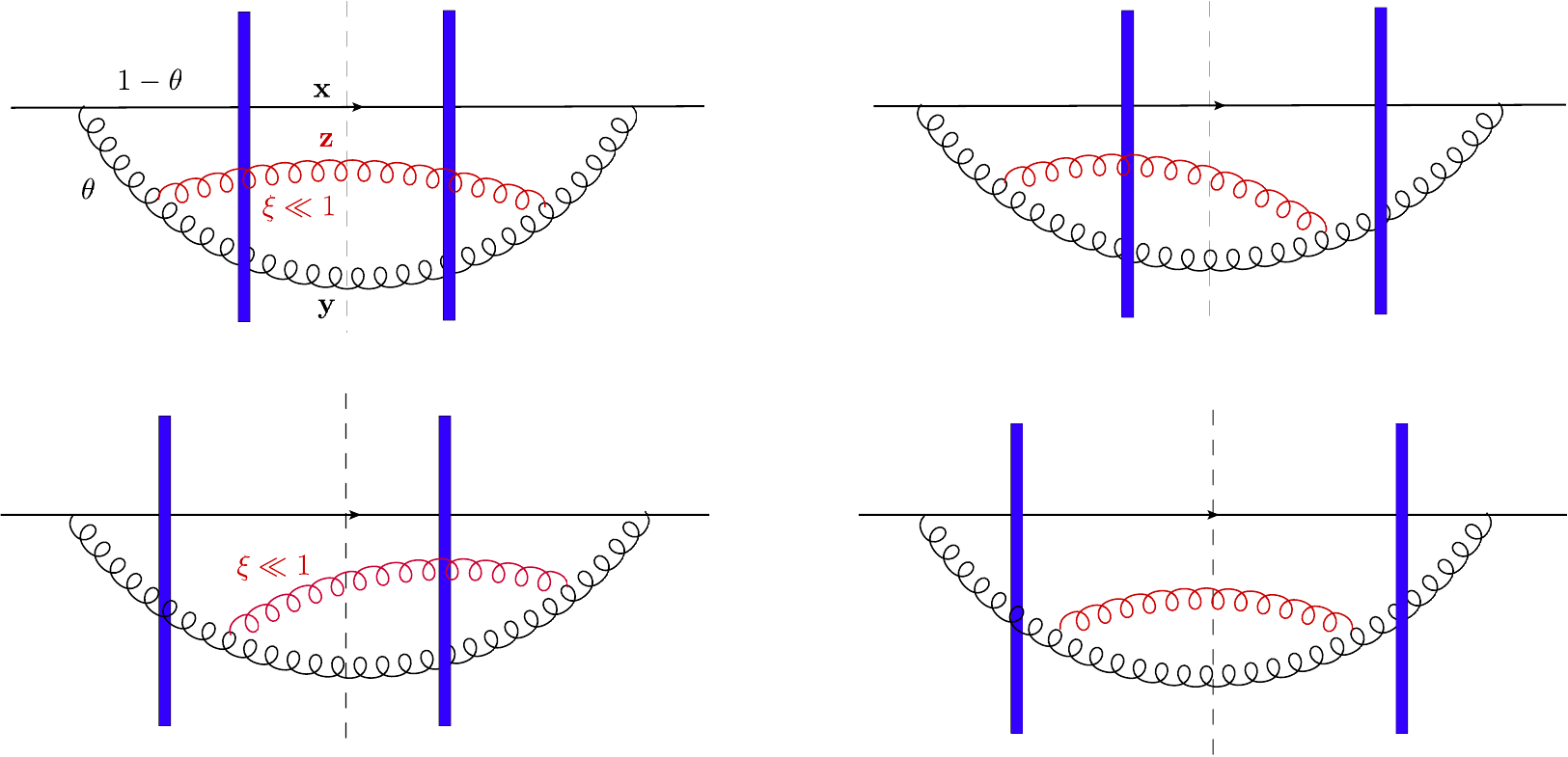}
\caption{The four diagrams contributing to the cross-section \eqref{trijetqgg2} in
the eikonal limit $\xi\to 0$ for one of the two emitted gluons (the one shown in red).}
\label{qgg2-soft}
\end{figure}

%
%
%
%

The result \eqref{qgg2-eik} takes a more familiar form in the case when the soft gluon $k_2$
is not measured, meaning that one can integrate over its kinematics (which in particular allows
us to identify $\bm{z}=\bm{\overline{z}}$).
\begin{align}
\label{qgevolu2}
\frac{d\sigma_{\nlo, 2}^{qA\rightarrow qg+X}}{dk_{1}^{+}\,d^{2}\bm{k}_{1}\,dk_{3}^{+}\,d^{2}\bm{k}_{3}}
&\,\simeq \,\frac{\abar}{(2\pi)^5}\,\frac{1+(1-\vartheta)^2}{2\vartheta q^{+}}\,
\delta(q^{+}-k_{1}^{+}-k_{3}^{+}) \nonumber \\*[0.4cm] 
&\ \times
\int_{\overline{\bm{x}},\,\bm{\overline{y}},\,\bm{x},\,\bm{y}}\,e^{-i\bm{k}_{1}\cdot(\bm{x}-\overline{\bm{x}})-i\bm{k}_{3}\cdot(\bm{y}-\overline{\bm{y}})}\,
\frac{(\bm{x}-\bm{y})\cdot (\overline{\bm{x}}-\overline{\bm{y}})}
{(\bm{x}-\bm{y})^{2} (\overline{\bm{x}}-\overline{\bm{y}})^{2}}
\nonumber \\*[0.4cm] 
&\ \times \frac{\abar}{2\pi} \int_0^1\frac{d\xi}{\xi}\,\int_{\bm{z}}
\frac{(\bm{y}-\bm{z})\cdot (\bm{\overline{y}}-\bm{z})}{(\bm{y}-\bm{z})^2 (\bm{\overline{y}}-\bm{z})^{2}}\,
 \nonumber\\*[0.2cm]
&\qquad\qquad\times
\bigg\{\Big[\mathcal{S}\left(\bm{x},\,\bm{\overline{x}}\right)
\mathcal{S}\left(\bm{y},\,\overline{\bm{y}}\right)\,-\,
\mathcal{Q}\left(\bm{x},\,\bm{y},\,{\bm{z}},\,\overline{\bm{x}}\right)
\mathcal{S}\left(\overline{\bm{y}},\,\bm{{z}}\right)
\nonumber\\*[0.2cm]
&\qquad\qquad\quad - 
\mathcal{Q}\left(\bm{x},\,\bm{z},\,\overline{\bm{y}},\,\overline{\bm{x}}\right)
\mathcal{S}\left(\bm{z},\,\bm{y}\right)\, +\,
\mathcal{Q}\left(\bm{x},\,\bm{y},\,\overline{\bm{y}},\,\overline{\bm{x}}\right)\Big]\mathcal{S}
\left(\bm{y}\,,\overline{\bm{y}}\right)\,,
\nonumber\\*[0.2cm]
&\qquad\qquad\quad + 
2\Big[\mathcal{S}\left(\bm{y},\,\bm{\overline{y}}\right)
-\mathcal{S}\left({\bm{y}},\,\bm{{z}}\right)
 \mathcal{S}\left({\bm{z}},\,\bm{\overline{y}}\right)
\Big]\mathcal{Q}\left(\bm{x},\,\bm{y},\,\overline{\bm{y}},\,\overline{\bm{x}}\right)
\bigg\}.
 \end{align}
 This result is recognised as a piece of the JIMWLK evolution of the first term
 $ S_{qg\bar{q}g}\left(\bm{x},\,\bm{y},\,\overline{\bm{x}},\,\overline{\bm{y}}\right)\simeq
\mathcal{Q}(\bm{x},\,\bm{y},\,\overline{\bm{y}},\,\overline{\bm{x}})\,\mathcal{S}(\bm{y},\,\overline{\bm{y}})$
in the expression  \eqref{pALOqchannel} or the LO cross-section. More precisely, we have the real part
of the BK equation for the dipole $\mathcal{S}(\bm{y},\,\overline{\bm{y}})$ together with the
relevant part of the real part of the JIMWLK equation for the quadrupole $\mathcal{Q}(\bm{x},\,\bm{y},\,\overline{\bm{y}},\,\overline{\bm{x}})$. To complete the (real terms in the) evolution of the latter,
we also need the interference contribution in  \eqn{trijetqgg3}. This will be discussed in the next section.

 \begin{figure}[!t]\center
 \includegraphics[scale=0.95]{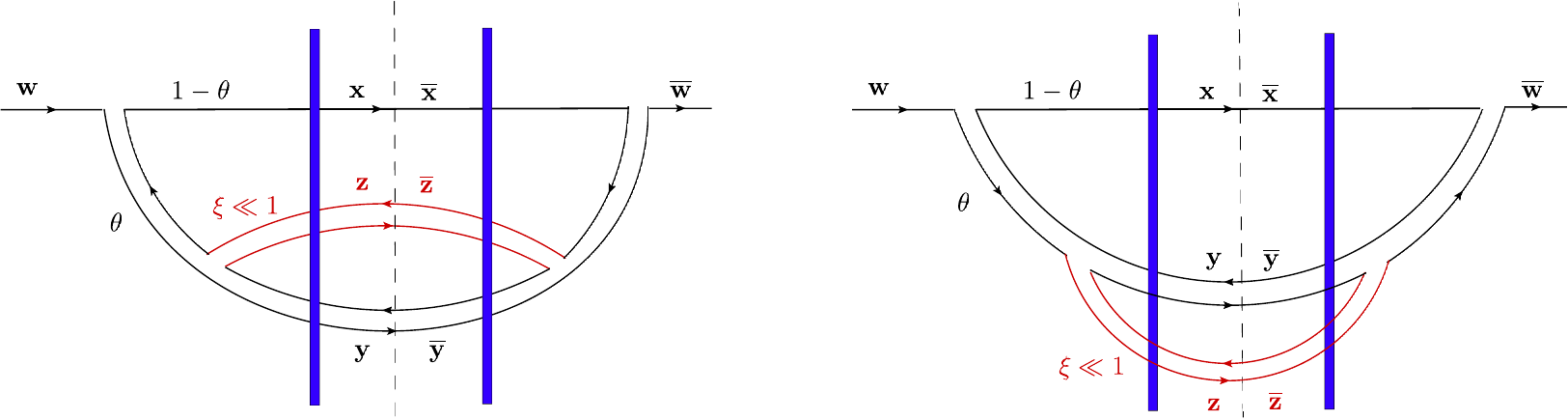}
\caption{A redrawing of the first graph in Fig.~\ref{qgg2-soft} 
valid at large $N_c$ which exhibits the 2 colour structures hidden 
in this graph: each gluon is replaced by a double line denoting a quark-antiquark
pair of zero transverse size. These two large-$N_c$
graphs correspond to the first terms in the first and, respectively, second
squared bracket in \eqn{qgg2-eik}. }
\label{qgg2-largeNc}
\end{figure}

\subsection{Interference graphs}
\label{sec:eikonal3}

We finally consider the eikonal limits for the contributions to the $qgg$ cross-section associated
with interference graphs, cf.  \eqn{trijetqgg3} and  Fig.~\ref{qgg-interf}. The relevant limits for
the product of effective vertices in the numerator of \eqn{defK3} read as follows:
\begin{align}\label{K3-num}
\Xi_{\lambda_{1}\lambda}^{ijmn}
(\vartheta,\,\xi)\,
\Pi_{\lambda_{1}\lambda}^{ijpq\,*}
(\vartheta,\,\xi)
&\ \longrightarrow \ - \delta^{mp}\delta^{nq}\,\frac{8\vartheta^2 (1-\vartheta)^3[1+(1-\vartheta)^2]}
{\vartheta-\xi}\,,\quad\mbox{when}\quad \vartheta-\xi\to 0,\,\nonumber\\*[0.2cm]
&\ \longrightarrow \  - 8
 \delta^{mq}\delta^{np}\,\vartheta (1-\vartheta)[1+(1-\vartheta)^2]
\,,\quad\mbox{when}\quad \xi\to 0
\,.
\end{align}
When $\vartheta-\xi\to 0$, all the four kernels in  \eqn{trijetqgg3} take the same form, which reads
 \begin{align}\label{K3-eik1}
\mathcal{K}_{3}^{mnpq} 
&\ \longrightarrow \ - \delta^{mp}\delta^{nq}\,\frac{8[1+(1-\vartheta)^2]}
{\vartheta(\vartheta-\xi)\bm{Y}^{2}\overline{\bm{R}}^{2}}\,\quad\mbox{when}\quad \vartheta-\xi\to 0\,.
\end{align}
Once again, this means that only four topologies survive in the cross-section in that limit --- those
in which at least one of the final gluons interacts with the shockwave and which are
illustrated in Fig.~\ref{qgg-interf-eik2}.
Still for  $\vartheta-\xi\to 0$, we can also approximate the transverse coordinates as
$\bm{y}\simeq \bm{x}$ and   $\overline{\bm{y}}\simeq \overline{\bm{z}}$.

Concerning the limit $\xi\to 0$, we need to distinguish between two cases, depending upon the fact that
$\bm{X}^{\prime}=\bm{x}-\bm{z}^{\prime}$ vanishes, or not:
\begin{align}\label{K3-eik2}
\mathcal{K}_{3}^{mnpq}\Big |_{\xi\to 0} 
&\ \longrightarrow \ - \delta^{mq}\delta^{np}\,
\frac{8[1+(1-\vartheta)^2]}{\vartheta^2 (1-\vartheta)
(\bm{X}^{\prime})^{2}\overline{\bm{R}}^{2}}\,
 ,\quad\mbox{when $\bm{X}^{\prime}\ne 0$}  \,, 
\nonumber\\*[0.2cm]
&\ \longrightarrow \ - \delta^{mq}\delta^{np}\,\frac{8[1+(1-\vartheta)^2]}
{\xi\vartheta\bm{Y}^{2}\overline{\bm{R}}^{2}}\,,\quad\mbox{when
$\bm{X}^{\prime}=0$}
\,.
\end{align}
Clearly, the soft singularity at $\xi=0$ is present only in the case where $\bm{X}^{\prime}=0$. i.e.
for the two terms within the square brackets in   \eqn{trijetqgg3} where the coordinates
$\bm{x}$ and $\bm{z}^{\prime}$ get replaced by $\bm{y}$. The eight $S$-matrices within the
structure of these two terms can now be simply added to each other (since they are all multiplied by the
same kernel) and it is easy to see that only four of them survive in the final result: those
illustrated in Fig.~\ref{qgg-interf-eik1}. Still for $\xi\to 0$,  one can
 use $\bm{w}\simeq \bm{y}\simeq (1-\vartheta)\bm{x}+\vartheta\bm{z}'$ and
$\overline{\bm{y}}\simeq \overline{\bm{z}}^\prime$.

 \begin{figure}[!t]\center
 \includegraphics[scale=0.95]{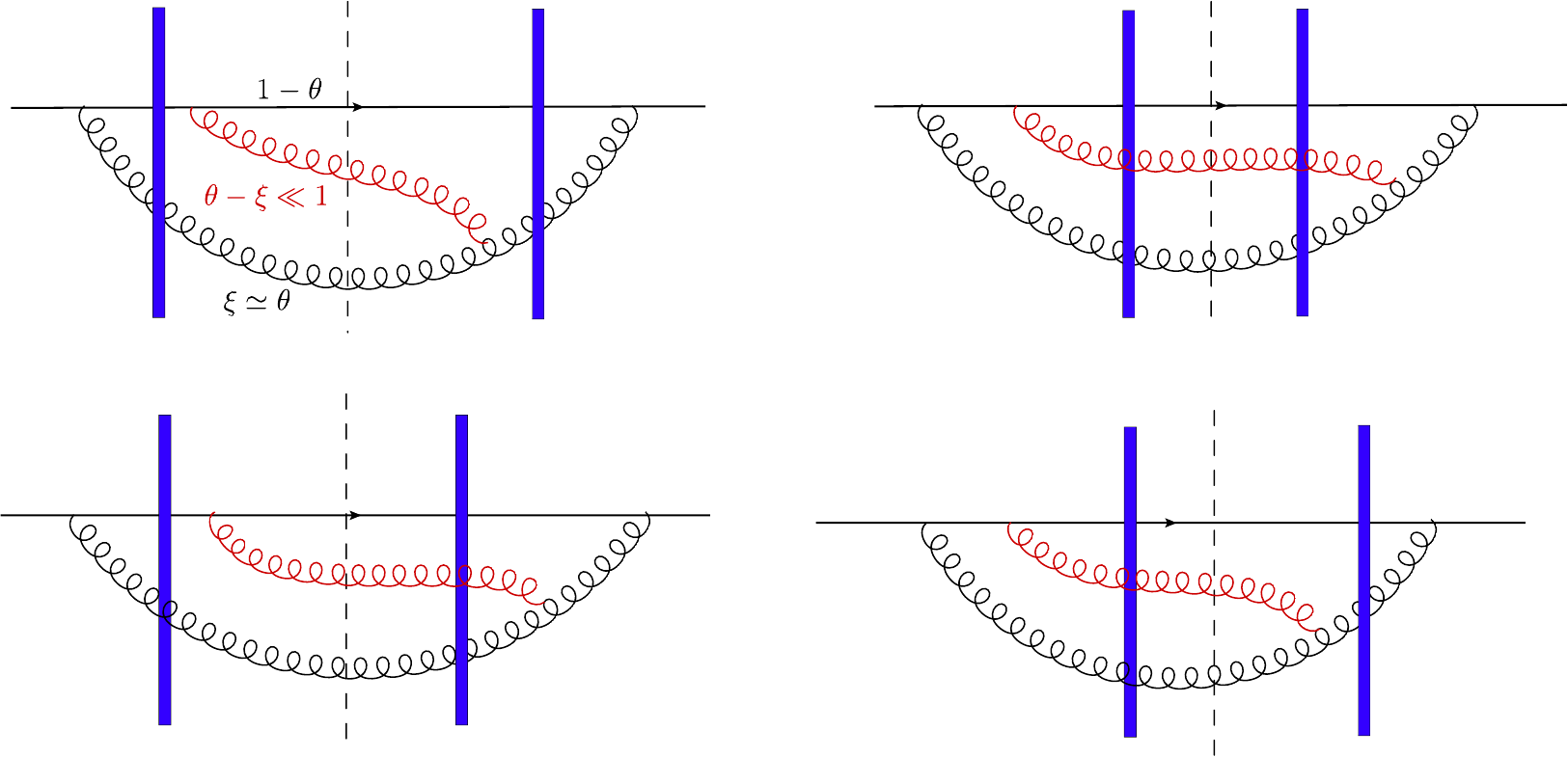}
\caption{The four diagrams contributing to the cross-section \eqref{trijetqgg3} in
the eikonal limit $\vartheta-\xi\to 0$ (the soft gluon is shown in red).}
\label{qgg-interf-eik2}
\end{figure}

 \begin{figure}[!t]\center
 \includegraphics[scale=0.95]{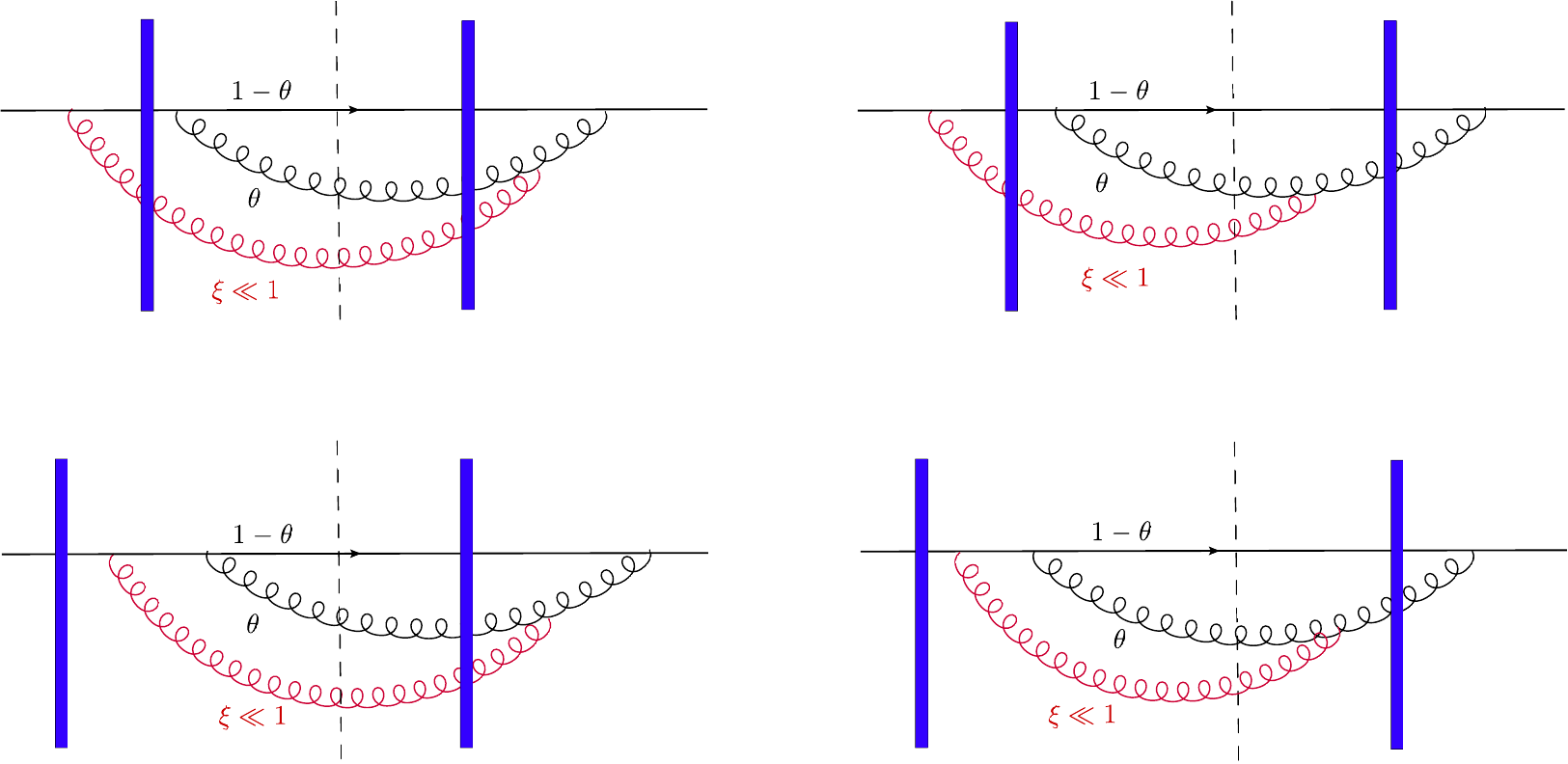}
\caption{The four diagrams contributing to the cross-section \eqref{trijetqgg3} in
the eikonal limit $\xi\to 0$ (the soft gluon is shown in red).}
\label{qgg-interf-eik1}
\end{figure}

To show explicit results, we shall use the momenta assignments in Fig.~\ref{qgg-interf}, which in
particular imply $k_2^+=\xi q^+$ and $k_3^+=(\vartheta-\xi)q^+$.
For the case $\vartheta-\xi\to 0$, one easily finds
 \begin{align}\label{qgginterf-eik2}
 \hspace*{-0.6cm}
&\frac{d\sigma_{(3)}^{qA\rightarrow qgg+X}}{dk_{1}^{+}\,d^{2}\bm{k}_{1}\,dk_{2}^{+}\,d^{2}\bm{k}_{2}\,dk_{3}^{+}\,d^{2}\bm{k}_{3}}\,\bigg |_{\vartheta-\xi\to 0}
 = - \frac{4\alpha_{s}^{2}\,C_{F} N_c}{(2\pi)^{10}}\,
\frac{1+(1-\vartheta)^2}{(\vartheta-\xi)\vartheta(q^{+})^{2}}\,
\delta(q^{+}-k_{1}^{+}-k_{2}^{+})\nonumber \\*[0.2cm]
&\qquad\qquad\times
\int_{\overline{\bm{x}},\,\bm{\overline{z}},\,\bm{\overline{z}}^{\prime},\,\bm{x},\,\bm{z},\,\bm{z}^{\prime}}\,e^{-i\bm{k}_{1}\cdot(\bm{x}-\overline{\bm{x}})-i\bm{k}_{2}\cdot(\bm{z}-\overline{\bm{z}})-i\bm{k}_{3}\cdot(\bm{z}^{\prime}-\overline{\bm{z}}^{\prime})}\,
\frac{\left(\bm{X}\cdot\overline{\bm{X}}\right)\left(\bm{X}^{\prime}\cdot \bm{\overline{Z}}\right)
}{\bm{X}^{2}\overline{\bm{X}}^{2}(\bm{X}^{\prime})^{2} \bm{\overline{Z}}^{2}}\nonumber\\*[0.2cm]
&\qquad\qquad\times\Big[\mathcal{Q}\left(\bm{x},\,\bm{z}^{\prime},\,\overline{\bm{z}}^{\prime},\,\bm{\overline{x}}\right)
\mathcal{Q}\left(\bm{z}^{\prime},\,\bm{z},\,\overline{\bm{z}},\,\overline{\bm{z}}^{\prime}\right)\,-\,
\mathcal{Q}\left(\bm{x},\,\bm{z},\,\overline{\bm{z}},\,\overline{\bm{z}}^{\prime}\right)
\mathcal{S}\left(\overline{\bm{z}}^{\prime},\,\bm{\overline{x}}\right)
\nonumber\\*[0.2cm]
&\qquad\qquad\quad - \,\mathcal{Q}\left(\bm{x},\,\bm{z}^{\prime},\,\overline{\bm{z}},\,\overline{\bm{x}}\right)
\mathcal{S}\left(\bm{z}^{\prime},\,\bm{z}\right)\, +\,
\mathcal{S}\left(\bm{x},\,\bm{z}\right)
\mathcal{S}\left(\overline{\bm{z}},\,\overline{\bm{x}}\right)\Big]\mathcal{S}
\left(\bm{z}\,,\overline{\bm{z}}\right)\,,
 \end{align}
where $\bm{X}^{\prime}=\bm{x}-\bm{z}^{\prime}$, 
$\overline{\bm{Z}}=\overline{\bm{z}}-\overline{\bm{z}}^\prime$,
and we have also used $\bm{y}= \bm{x}$ and   $\overline{\bm{y}}=\overline{\bm{z}}$ to
replace $\bm{Y}\to \bm{X}= \bm{x}-\bm{z}$ and $\overline{\bm{R}}\to 
\overline{\bm{X}}= \overline{\bm{x}}-\overline{\bm{z}}$. These alternative notations facilitate
the comparison with \eqn{qgg1-eik2}, which refers to
the direct contribution of the amplitude in Fig.~\ref{final-qgg1}. Both \eqn{qgg1-eik2} and 
\eqn{qgginterf-eik2} express the effect of producing one soft gluon when acting with the ``production'' version
of the JIMWLK Hamiltonian on the quadrupole 
$\mathcal{Q}(\bm{x},\,\bm{z},\,\overline{\bm{z}},\,\overline{\bm{x}})$ in the first
term in the r.h.s. of  \eqn{pALOqchannel} for the LO dijet cross-section\footnote{Notice the change of notation 
$\bm{y}\to \bm{z}$ when going from  \eqn{LOfinal} to \eqn{qgg1-eik2} or \eqref{qgginterf-eik2}.}.
In \eqn{qgg1-eik2}, the soft gluon is emitted and reabsorbed by the same quark, with transverse
coordinates $\bm{x}$ and $\overline{\bm{x}}$ in the DA and the CCA, respectively.  \eqn{qgginterf-eik2} is
the interference term where the soft gluon is emitted by the quark $\bm{x}$ in the DA and absorbed
by the ``other quark'', at  $\overline{\bm{z}}$, in the CCA. (This ``other'' quark is truly a component in the
large-$N_c$ decomposition of the harder gluon with energy fraction $\vartheta$.)  
Notice the difference in sign between 
 Eqs.~\eqref{qgg1-eik2} and \eqref{qgginterf-eik2}. Except for the sign, the respective numerical factors are identical at large $N_c$, as they should.
 
 Besides Eqs.~\eqref{qgg1-eik2} and \eqref{qgginterf-eik2}, there are two more contributions to
 the JIMWLK evolution of the quadrupole (here, in the sense of soft gluon production) 
  \cite{Iancu:2013uva}. One of them was already shown  \eqn{qgg2-eik},
  although with somewhat different notations for the transverse
coordinates\footnote{The difference 
in notations between \eqn{qgg2-eik} on one hand and  Eqs.~\eqref{qgg1-eik2} and \eqref{qgginterf-eik2}
on the other hand comes from the fact that, in Sect.~\ref{sec:eikonal2}, we preferred to work in
 the other soft limit, that is, $\xi\to 0$.}.  The other one is the second interference term, 
  in which one exchanges the topologies 
 between the DA and the CCA; this can be easily inferred from \eqref{qgginterf-eik2} via
hermitian conjugation.
 When the soft gluon is not measured, the sum of these four contributions reproduces the ``real'' terms 
 in the B-JIMWLK evolution of the quadrupole \cite{Dominguez:2011gc,Iancu:2011ns,Iancu:2011nj}.

Consider now  the other eikonal limit, namely $\xi\to 0$. Using the second line in \eqn{K3-eik2}
together with the relevant terms from  \eqn{trijetqgg3}, one finds
 \begin{align}\label{qgginterf-eik1}
 \hspace*{-0.6cm}
&\frac{d\sigma_{(3)}^{qA\rightarrow qgg+X}}{dk_{1}^{+}\,d^{2}\bm{k}_{1}\,dk_{2}^{+}\,d^{2}\bm{k}_{2}\,dk_{3}^{+}\,d^{2}\bm{k}_{3}}\,\bigg |_{\xi\to 0}
 =\frac{4\alpha_{s}^{2}\,C_{F} N_c}{(2\pi)^{10}}\,
\frac{1+(1-\vartheta)^2}{\xi\vartheta(q^{+})^{2}}\,
\delta(q^{+}-k_{1}^{+}-k_{3}^{+})\nonumber\\*[0.2cm]
&\qquad\qquad\times
\int_{\overline{\bm{x}},\,\bm{\overline{z}},\,\overline{\bm{y}}\,\bm{x},\,\bm{z},\,\bm{z}^{\prime}}\,e^{-i\bm{k}_{1}\cdot(\bm{x}-\overline{\bm{x}})-i\bm{k}_{2}\cdot(\bm{z}-\overline{\bm{z}})-i\bm{k}_{3}\cdot(\bm{z}^{\prime}-\overline{\bm{y}})}\,
\frac{\left(\bm{X}^{\prime}\cdot \bm{\overline{R}}\right)\left(\bm{Y}\cdot \bm{\overline{Z}}\right)
}{(\bm{X}^{\prime})^{2} \bm{\overline{R}}^{2}
\bm{Y}^{2} \bm{\overline{Z}}^{2}}\nonumber\\*[0.2cm]
&\qquad\qquad\times\Big[\mathcal{S}\left(\bm{z},\,\overline{\bm{z}}\right)
\mathcal{Q}\left(\bm{y},\,\bm{z},\,\overline{\bm{z}},\,\overline{\bm{y}}\right)\,-\,
\mathcal{S}\left(\bm{y},\,\bm{z}\right)
\mathcal{S}\left({\bm{z}},\,\overline{\bm{y}}\right)
\nonumber\\*[0.2cm]
&\qquad\qquad\quad - \,\mathcal{S}\left(\bm{y},\,\bm{\overline{z}}\right)
\mathcal{S}\left(\bm{\overline{z}},\,\overline{\bm{y}}\right)
\, +\,
\mathcal{S}\left(\bm{y},\,\overline{\bm{y}}\right)\Big]\mathcal{S}
\left(\overline{\bm{y}}\,,\overline{\bm{x}}\right)\,,
 \end{align}
where  $\bm{Y}= \bm{y}-\bm{z}$, $\bm{X}^{\prime}=\bm{x}-\bm{z}^{\prime}$, 
$\overline{\bm{Z}}=\overline{\bm{z}}-\overline{\bm{z}}^\prime$, and
$\overline{\bm{R}}=\overline{\bm{x}}-\overline{\bm{y}}$ and
 we have replaced $\bm{w}\to  \bm{y}$ and $\overline{\bm{z}}^{\prime}\to \overline{\bm{y}}$,
as appropriate in this limit. The $S$-matrices within the square brackets have the right structure
to describe the production of a soft gluon from the dipole $\mathcal{S}\left(\bm{y},\,\overline{\bm{y}}\right)$
(see Eq.~(A.2) in Ref.~\cite{Iancu:2013uva}). 

If the soft gluon is not measured in the final state, one must  integrate  \eqn{qgginterf-eik1}  over $k_2^+=\xi q^+$ and $\bm{k}_{2}$; this gives (with $2\alpha_s C_F \simeq \pi\abar$)
\begin{align}
\label{qgevolu3}
\frac{d\sigma_{\nlo, 3}^{qA\rightarrow qg+X}}{dk_{1}^{+}\,d^{2}\bm{k}_{1}\,dk_{3}^{+}\,d^{2}\bm{k}_{3}}
&\,\simeq \,-\frac{\abar}{(2\pi)^5}\,\frac{1+(1-\vartheta)^2}{2\vartheta q^{+}}\,
\delta(q^{+}-k_{1}^{+}-k_{3}^{+}) \nonumber \\*[0.4cm] 
&\ \times
\int_{\overline{\bm{x}},\,\overline{\bm{y}},\,\bm{x},\,\bm{z}^{\prime}}\,e^{-i\bm{k}_{1}\cdot(\bm{x}-\overline{\bm{x}})-i\bm{k}_{3}\cdot(\bm{z}^{\prime}-\overline{\bm{y}})}\,
\frac{(\bm{x}-\bm{z}')\cdot (\overline{\bm{x}}-\overline{\bm{y}})}
{(\bm{x}-\bm{z}')^{2} (\overline{\bm{x}}-\overline{\bm{y}})^{2}}\,
\mathcal{S}\left(\overline{\bm{y}}\,,\overline{\bm{x}}\right)
\nonumber
\\*[0.4cm] 
&\ \times \frac{\abar}{2\pi} \int_0^1\frac{d\xi}{\xi}\,\int_{\bm{z}}
\frac{2(\bm{y}-\bm{z})\cdot (\bm{\overline{y}}-\bm{z})}{(\bm{y}-\bm{z})^2 (\bm{\overline{y}}-\bm{z})^{2}}\,
\Big[\mathcal{S}\left(\bm{y},\,\bm{\overline{y}}\right) -
\mathcal{S}(\bm{y},\,\bm{z})\,\mathcal{S}(\bm{z},\,\overline{\bm{y}})
 \Big],
 \end{align}
 where the change of global sign occurred because $\overline{\bm{Z}}=-(\bm{\overline{y}}-\bm{z})$.
 This is indeed the right sign to describe the BK evolution of the dipole $\mathcal{S}(\bm{w},\,\overline{\bm{y}})
$ from the product $\mathcal{S}(\bm{w},\,\overline{\bm{y}})\,
\mathcal{S}(\overline{\bm{y}},\,\overline{\bm{x}})$ which appears with a negative sign
in the third term in the r.h.s. of
\eqn{pALOqchannel}. (Recall that $\bm{w}\simeq  \bm{y}$ in the soft limit at hand.)

To summarise, the discussion of the eikonal limits of our results for the $qgg$ trijet cross-section has 
 allowed us to reconstruct the ``real'' terms in the B-JIMWLK evolution of the
LO cross-section \eqref{pALOqchannel} for the $qg$ dijet. Specifically, 
we have identified the ``real'' part of the evolution equations for the first term 
(the product $\mathcal{Q}(\bm{x},\,\bm{y},\,\overline{\bm{y}},\,\overline{\bm{x}})\,\mathcal{S}(\bm{y},\,\overline{\bm{y}})$ between a quadrupole and a dipole) and for the 
last term (the dipole $\mathcal{S}\left(\bm{w},\,\overline{\bm{w}}\right)$) in \eqn{pALOqchannel},
and also for the dipoles $\mathcal{S}(\bm{y},\,\overline{\bm{w}})$ and
$\mathcal{S}(\bm{w},\,\overline{\bm{y}})$ which appears in the two intermediate terms there.
On the other hand, we have not yet encountered the evolution of the two other dipoles which
enter these intermediate terms, namely $\mathcal{S}(\bm{x},\,\bm{y})$ and
$\mathcal{S}(\overline{\bm{y}},\,\overline{\bm{x}})$. There is a good reason for that:
these two other dipoles are built with quark legs which ``live'' on the same side of the 
cut, that is, they both ``live'' in the DA for  $\mathcal{S}(\bm{x},\,\bm{y})$, 
and respectively in the CCA for 
$\mathcal{S}(\overline{\bm{y}},\,\overline{\bm{x}})$. Accordingly, the BK evolution of these
dipoles involves graphs like those shown in Fig.~\ref{qgg-virtual}, which represent
{\it virtual} corrections to the $qg$ dijet cross-section. The contributions of such graphs
will be considered in a companion paper devoted to virtual corrections.

\begin{figure}[t]\center
  \includegraphics[scale=0.9]{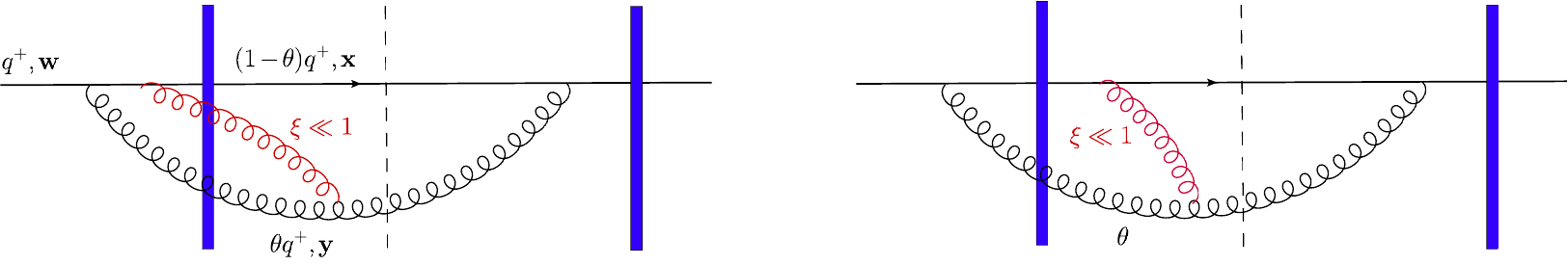}
  \caption{Graphs representing virtual corrections to the $qg$ dijet cross-section. Both shown
  graphs count for the BK evolution of the dipole  $\mathcal{S}(\bm{x},\,\bm{y})$ which enters
  the second term in the r.h.s. of  \eqn{pALOqchannel}.
     \label{qgg-virtual}}
\end{figure}

\section{Collinear divergences: recovering the DGLAP evolution}
\label{sec:DGLAP}

Besides the soft divergences that we have just discussed, the NLO corrections to the 
dijet cross-section are expected to also contain {\it collinear} divergences.
In momentum space, they correspond to the limit where the daughter partons emerging
from a splitting have a very small relative transverse momentum (i.e. they make a
very small angle $\theta\to 0$). In the transverse coordinate space, this is the limit where the 
transverse separation between the two daughter partons becomes arbitrarily large. 

Before turning to explicit calculations,
let us first recall some general features about the origin and the treatment of
the collinear divergences in the framework of the hybrid factorisation (see e.g.
\cite{Chirilli:2012jd}):

\text{(i)} The collinear divergences are associated with (unmeasured) partons which are
emitted long before the hard process, or longtime after it. By the ``hard process'', we mean
the ensemble of the other interactions, that is, the scattering off the nuclear target (the shockwave)
and the emission of additional, relatively hard (non-collinear), partons.

\text{(ii)} The ``collinear'' parton may scatter with the shockwave, but its scattering does not matter for
 the inclusive dijet cross-section at NLO.

\text{(iii)}  If the collinear emission occurs prior to the hard process --- in the present case, that
should be a gluon emission by the incoming quark ---, then the associated divergence
can be reabsorbed into the DGLAP evolution of the parton distributions in the proton projectile
(here, the quark distribution $q_f(x_{p},\mu^{2})$, as visible in Eqs.~\eqref{pALO} and \eqref{pANLO-qq}).

\text{(iv)} If the collinear emission originates from one of the final partons --- in this case, that could be
either a gluon emission from the final quark, or the splitting of the final gluon into two gluons, or into
a quark-antiquark pair ---, then its treatment depends upon our definition of the measurement process.
If we measure two hadrons in the final state, then the collinear divergences should be absorbed into the
DGLAP evolution of the fragmentation functions for the quark, or the gluon, emerging from the hard 
process. If instead we measure two jets, whose definition involves an opening angle $R$, then
any emission at small angles $\theta < R$ must be viewed as a part of the final-state jets, whereas
emissions at larger angles $\theta > R$ are interpreted as NLO corrections to the hard process.
In this argument, the jet angle $R$ effectively acts as a collinear cutoff for the NLO corrections.
In this remaining part of this section, we shall assume a hadronic description for the
final state, for definiteness.

\text{(v)} The collinear divergences can be unambiguously separated 
from the {\it infrared} (or ``soft'') divergences associated with soft gluons. A given graph
can develop both soft and collinear divergences, but the overlapping divergences cancel
--- as  a consequence of probability conservation ---
 when adding together real and virtual corrections. This makes it possible to
disentangle soft from collinear divergences in practice and to ascribe them to
the B-JIMWLK evolution of the hard process and to the DGLAP evolutions of the initial and final 
states, respectively.

\subsection{DGLAP evolution for the initial quark}

Our first example refers the NLO contribution denoted with a subscript $(1)$ 
in \eqn{qgrNLO}. We recall that this contribution is obtained from the trijet cross-section \eqref{trijetqgg1}
by integrating out a gluon with momentum $k_3$, which can be any of the two gluons in the final
state, cf. Fig.~\ref{fig:qgg1}.

 \begin{figure}[!t]\center
 \includegraphics[scale=0.95]{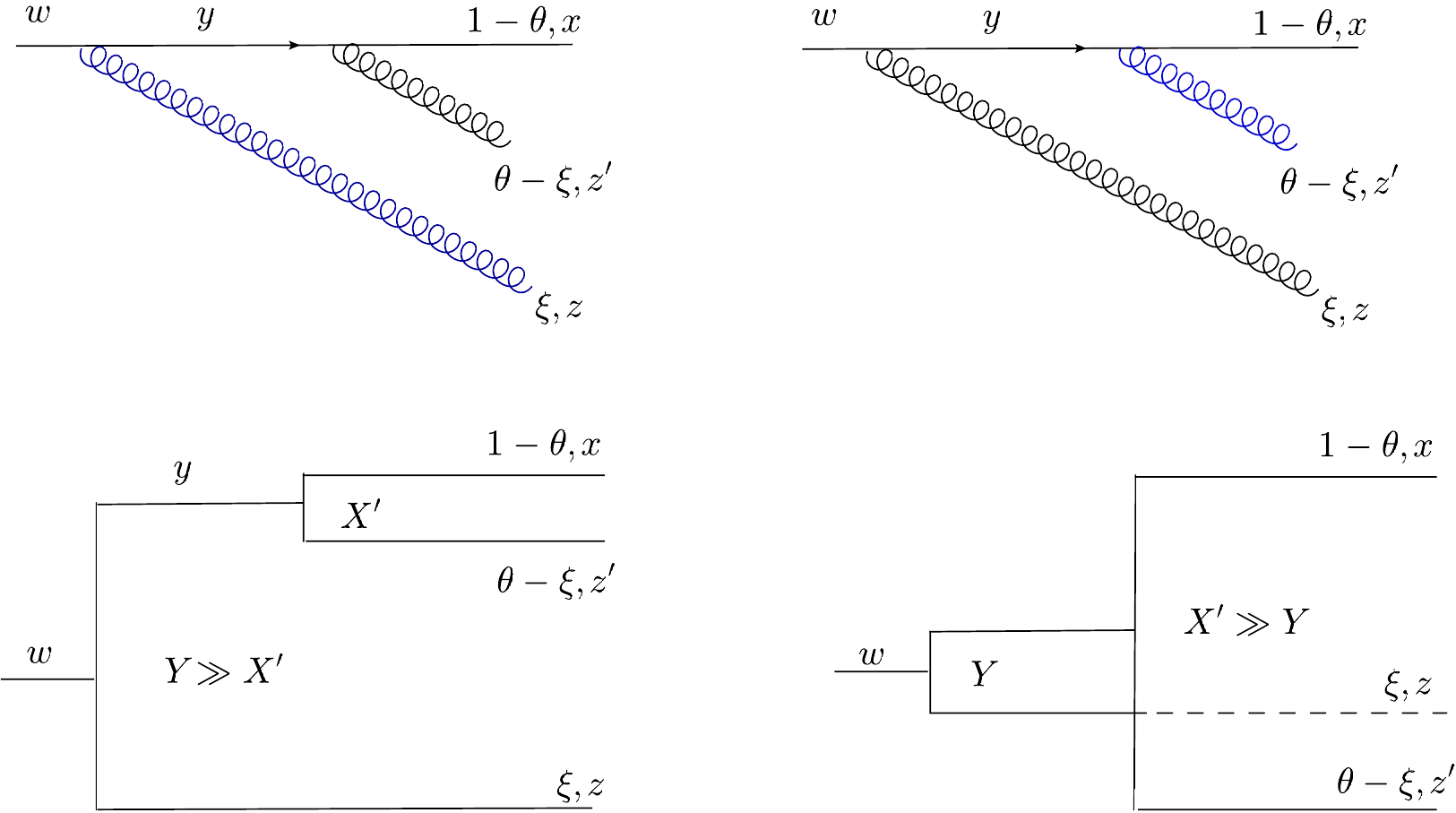}
\caption{Graphical illustration of the collinear limits corresponding to the two
gluon  emissions by the incoming quark.  The Feynman graphs in the first line show the topology
of the emission: the gluon emerging from a collinear splitting is represented in blue. The schematic
diagrams in the second line illustrate the geometry of the splittings in the transvserse
coordinate space.}
\label{fig:collin1}
\end{figure}

As already mentioned, the collinear regime corresponds to the situation where the transverse 
 separation between the two daughter partons
can be arbitrarily large. For the diagram in  Fig.~\ref{fig:qgg1}, there are two such limits,
one for each of the two gluon emissions, as illustrated in Fig.~\ref{fig:collin1}.
In the first case, depicted in the l.h.s. of Fig.~\ref{fig:collin1}, the first gluon emission is the
collinear one; one then has $|\bm{Y}|\gg |\bm{X}^{\prime}|$, where we recall that 
$\bm{X}'=\bm{x}-\bm{z}'$ and $\bm{Y}=\bm{y}-\bm{z}$. This covers the case of the
DGLAP evolution of the incoming quark distribution, to be discussed in this section. 
 In the other case (the r.h.s. of Fig.~\ref{fig:collin1}), the opposite inequality holds: $ |\bm{X}^{\prime}|\gg |\bm{Y}|$. This is the case of the final state DGLAP evolution, to be discussed in
the next subsection.

 \begin{figure}[!t]\center
 \includegraphics[scale=0.95]{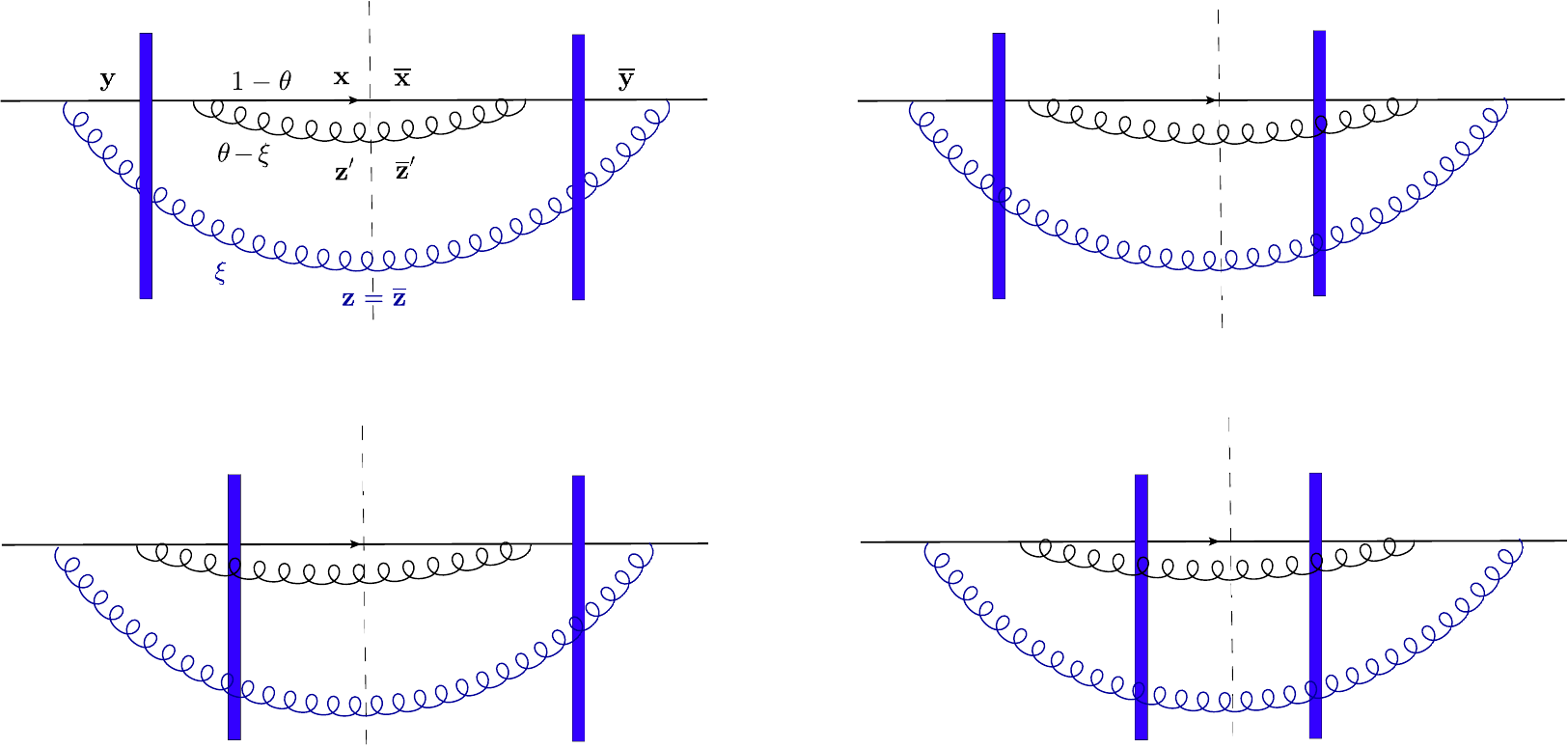}
\caption{The four graphs contributing to the limit where the first gluon emission (depicted in
blue) is collinear. These graphs are free of initial-state interactions.}
\label{fig:qgg1-coll1}
\end{figure}

Consider the  case  depicted in the l.h.s. of Fig.~\ref{fig:collin1}: the first gluon emission,
with longitudinal fraction $\xi$, is not measured ($\overline{\bm{z}}={\bm{z}}$) and is in
the collinear regime. Using  $\bm{Y}^{2}\gg (\bm{X}^{\prime})^{2}$
and similarly $\overline{\bm{Y}}^{2}\gg (\overline{\bm{X}}^{\prime})^{2}$, one finds that
the kernel \eqref{defK1}  simplifies to (recall \eqn{XIXI})
 \begin{align}\label{K1coll1}
\mathcal{K}_{1}^{mnpq}
\simeq 
\frac{32\delta^{mp}\delta^{nq}}
{(1-\xi)\bm{Y}^{2}\,\overline{\bm{Y}}^{2}}\,P_{q\to g}(\xi)\,P_{q\to g}(\xi')\,,
\end{align}
where $\xi'\equiv ({\vartheta-\xi})/({1-\xi})$ denotes the splitting fraction for the second gluon emission and
\beq
P_{q\to g}(z)\equiv \frac{1+(1-z)^2}{2z}\,,
\eeq
is the DGLAP splitting function for the process $q\to qg$.
The approximation in \eqn{K1coll1} holds so long as
\beq
\xi(1-\xi)^{2}\bm{Y}^{2} \gg (1-\vartheta)(\vartheta-\xi)(\bm{X}^{\prime})^{2}\,,
\eeq
together with a similar condition involving $\overline{\bm{Y}}$ and $\overline{\bm{X}}^\prime$.
In writing  \eqn{K1coll1}, we have ignored the instantaneous pieces of the effective vertices
which enter the numerator of $\mathcal{K}_{1}$ (that is, we have used \eqn{XIXI}). 
On physical grounds, it is quite obvious
that these pieces cannot yield collinear singularities: they correspond to effective graphs in which
the two emissions occur simultaneously. We shall shortly give a more mathematical argument in that sense.

\eqn{K1coll1} shows that, in the collinear limit\footnote{In general, i.e. beyond the collinear
limit, this factorisation would be spoilt by the dependence of the energy denominator
in \eqn{defK1} upon the  longitudinal fractions $\vartheta$ and $\xi$.}, 
the two gluon emissions  factorise from each other at the level of the kernel $\mathcal{K}_{1}$.
Let us now check that a similar factorisation also holds at the level of the colour structure,
i.e. for the $S$-matrices.

Since all the four terms within the squared brackets in 
 \eqn{trijetqgg1} are now multiplied by a same kernel, it follows that only four topologies (in terms
 of shockwave insertions) survive in this limit. These are the topologies shown in Fig.~\ref{fig:qgg1-coll1},
which do not involve initial-state interactions, as expected: 
the ``collinear'' gluon is emitted before the hard process,  both in the DA and in the
 CCA.  This gluon crosses the shockwave in all these graphs, but the Wilson lines describing its scattering
mutually cancel by unitarity: $V(\bm{z})V^{\dagger}(\overline{\bm{z}})=1$ when $\overline{\bm{z}}={\bm{z}}$.
Because of this cancellation, the colour structure of the four graphs in Fig.~\ref{fig:qgg1-coll1} is precisely
the same as for the LO $qg$ final state in \eqn{LOfinal}. 


Using Eqs.~\eqref{trijetqgg1} and \eqref{K1coll1}, one deduces
the following approximation for this particular dijet cross-section in the collinear limit for the unmeasured
gluon:
 \begin{align}\label{qgg1-coll1}
\frac{d\sigma_{(1)\,\rnlo, 1}^{qA\rightarrow qg+X}}{dp^{+}\,d^{2}\bm{p}\,dk^{+}\,d^{2}\bm{k}}
&\simeq  \frac{4\alpha_s C_F}{(2\pi)^6 (q^+)^2(1-\xi)}\,P_{q\to g}\Big(\frac{x_2}{x_1+x_2}\Big)\nonumber\\*[0.2cm] 
&\times \int_{\overline{\bm{x}},\,\bm{\overline{z}}^{\prime},\,\bm{x},\,\bm{z}^{\prime}}\,e^{-i\bm{p}\cdot(\bm{x}-\overline{\bm{x}})-i\bm{k}\cdot(\bm{z}^{\prime}-\overline{\bm{z}}^{\prime})}\,
\frac{(\bm{x}-\bm{z}')\cdot (\overline{\bm{x}}-\overline{\bm{z}}')}
{(\bm{x}-\bm{z}')^{2} (\overline{\bm{x}}-\overline{\bm{z}}')^{2}}\,
\nonumber\\*[0.2cm]
&\times\Big[
\mathcal{Q}(\bm{x},\,\bm{z}',\,\overline{\bm{z}}',\,\overline{\bm{x}})\,\mathcal{S}(\bm{z}',\,\overline{\bm{z}}') -\mathcal{S}(\bm{x},\,\bm{z}')\,\mathcal{S}(\bm{z}',\,\overline{\bm{y}})
 -\mathcal{S}(\bm{y},\,\overline{\bm{z}}')\,\mathcal{S}(\overline{\bm{z}}', \overline{\bm{x}})
 +\mathcal{S}\left(\bm{y},\,\bm{\overline{y}}\right)\Big]
 \nonumber\\*[0.2cm] &\times  \frac{4\alpha_s C_F}{(2\pi)^2}\,
P_{q\to g}(\xi) \int_{\bm{z}}
\frac{(\bm{y}-\bm{z})\cdot (\overline{\bm{y}}-{\bm{z}})}
{(\bm{y}-\bm{z})^{2} (\overline{\bm{y}}-{\bm{z}})^{2}}\,.
 \end{align}
 For more clarity, we have used the same notations for the momenta of the two produced partons
as in \eqn{LOfinal}:  $(p^{+},\,\bm{p})$ and $(k^{+},\,\bm{k})$ refer to the produced quark and gluon, respectively.
Furthermore, $x_1=p^+/Q^+$ and $x_2=k^+/Q^+$, with $Q^+$ the longitudinal momentum
of the incoming proton.  The original longitudinal fractions $\vartheta$ and $\xi$ should now be evaluated as
\begin{equation}\label{longit32}
 \vartheta 
 \,=1-\frac{x_1}{x_p}\,,
 \qquad\xi 
 \,=1-\frac{x_1+x_2}{x_p}\ \Longrightarrow \ \xi'
 \,\equiv \,\frac{\vartheta-\xi}{1-\xi} 
 \,=\,\frac{x_2}{x_1+x_2}\,.
 \end{equation}
 The transverse coordinate $\bm{y}$ of the intermediate quark is related to the coordinates
 $\bm{x}$ and $\bm{z}'$ of the produced partons via  \eqn{defyw2}, which becomes
 \beq
 \bm{y}\,=\,\frac{x_1\bm{x}+x_2\bm{z}^{\prime}}{x_1+x_2}\,\qquad\mbox{(and similarly
 for $\overline{\bm{y}}$)}\,.
 \eeq
 
\eqn{qgg1-coll1} exhibits the expected factorisation of the (unmeasured) collinear emission from
the LO dijet cross-section  \eqref{LOfinal} --- here initiated by a quark with original longitudinal 
momentum $q^+(1-\xi)$.  The integral over $\bm{z}$ in the last line, which physically represents 
the integral over the transverse phase space for the collinear emission, is logarithmically divergent
 at large  $|\bm{z}|$. To exhibit this singularity, 
 let us introduce a low-momentum cutoff $\Lambda$ on the transverse momentum of the
 unmeasured gluon, meaning an upper cutoff $\sim 1/\Lambda$ on the transverse separations
 $|\bm{y}-\bm{z}|$ and $|\overline{\bm{y}}-{\bm{z}}|$.
  With this regularisation,  the integral over $\bm{z}$ in \eqn{qgg1-coll1} can be 
 estimated as (below, $r\equiv |\bm{y}-\overline{\bm{y}}|$)
 \beq\label{colldiv}
  \int {d^2\bm{z}}
\frac{(\bm{y}-\bm{z})\cdot (\overline{\bm{y}}-{\bm{z}})}
{(\bm{y}-\bm{z})^{2} (\overline{\bm{y}}-{\bm{z}})^{2}}\,\to
 \int\frac{d^2\bm{q}}{\bm{q}^2}\,e^{-i\bm{q}\cdot(\bm{y}-\overline{\bm{y}})}\,
 \Theta(q-\Lambda)\,=\,\pi\ln\frac{1}{r^2\Lambda^2}
 \,=\,\pi\ln\frac{1}{r^2\mu^2} + \pi\ln\frac{\mu^2}{\Lambda^2}
 \,.\eeq
In the last equality, $\mu$ is a generic scale obeying $\Lambda \ll \mu < 1/r$.  In what follows,
it will be used as a renormalisation scale to subtract the collinear divergence at $\Lambda\to 0$
from the genuine NLO correction. The divergence will be then absorbed into the DGLAP evolution of the
quark distribution for the incoming quark.

To that aim, one should recall that the physical cross-section also includes a convolution with
$q_f(x_{p},\mu^{2})$, as visible in \eqn{pALO} at LO and in \eqn{pANLO-qq} for a 
particular ``real'' NLO correction. For the NLO contribution at hand, the relevant convolution
reads
\begin{equation}\label{qPDF}
\int_0^1 dx_{p}\,q_f(x_{p},\mu^{2})\,\Theta(x_p-x_1-x_2)\,\big(\cdots\big)
\,=\,\int_0^1\frac{d\xi}{1-\xi}\,\frac{x_1+x_2}{1-\xi}\,q_f\Big(\frac{x_1+x_2}{1-\xi},\mu^{2}\Big)
\,\Theta(1-\xi -x_1-x_2)\,\big(\cdots\big)\,,
 \end{equation}
where the dots within the integrand stay for the partonic cross-section in \eqn{qgg1-coll1}. In the 
integral in the r.h.s. we changed variable $x_p\to \xi$ according to \eqref{longit32};
the step function comes from the condition $x_p\le 1$. When using this convolution 
together with \eqn{qgg1-coll1}, one should also notice that  $q^+(1-\xi)=p^++k^+=(x_1+x_2)Q^+$
is the total longitudinal momentum of the produced dijet and thus it is independent
of $\xi$. Hence, 
the whole $\xi$ dependence of the final result is encoded in the following integral
\beq\label{DeltaqPDF}
x\Delta q_f(x,\mu^{2})\,\equiv  \frac{\alpha_s C_F}{\pi}
\int_0^{1-x}{d\xi}\,\frac{x}{1-\xi}\,q_f\Big(\frac{x}{1-\xi},\mu^{2}\Big)
P_{q\to g}(\xi)\,\ln\frac{\mu^2}{\Lambda^2}\,,
\eeq
where it is understood that $x\equiv x_1+x_2$. As suggested by its notation, the r.h.s. in
\eqref{DeltaqPDF} is precisely the result of one ``real'' step in the DGLAP
evolution of the quark distribution inside the proton, where the step consists in integrating
out gluon radiation with virtualities $q^2$ within the range $\Lambda^2 < q^2 < \mu^2$.
The corresponding virtual graphs are expected to add the ``plus'' prescription to the splitting function
$P_{q\to g}(\xi)$ and thus remove the apparent divergence of \eqref{DeltaqPDF} at $\xi\to 0$.
This will be checked in our subsequent paper (see also the discussion at the end of this subsection).

By inspection of Eqs.~\eqref{qgg1-coll1}--\eqref{DeltaqPDF}, one can see that the original
collinear divergence from the NLO correction has been transferred to the DGLAP evolution 
of the incoming quark distribution in the LO result.   The finite reminder, as obtained by
keeping only the second term, $ \pi\ln(1/r^2\mu^2)$, in the r.h.s. of 
\eqn{colldiv}, represents a genuine NLO correction to the hard process. However, our above
calculation for this correction is only correct to leading logarithmic accuracy w.r.t. the
logarithm $\ln(p_s^2/\mu^2)$, with $p_s^2 \equiv {\rm min}(\bm{p}^2, \,\bm{k}^2)$. (The
typical value of $r\equiv |\bm{y}-\overline{\bm{y}}|$ is fixed
by the Fourier transforms in \eqref{qgg1-coll1} as $r^2\sim 1/p_s^2$.)
 The $\mu$--dependence of the cross-section introduced by this and other similar corrections
compensates the respective
dependence of the quark distribution. This ensures that the result for the cross-section
is independent of  the arbitrary renormalisation scale $\mu$ at NLO. 

In order to have a complete NLO result, which also includes the numerical
coefficient under the log, one must go beyond our previous approximations in
Eqs.~\eqref{K1coll1} or \eqref{colldiv}. That is, one should exactly compute the contributions
from graphs which contain collinear divergences, like the 4 graphs Fig.~\ref{fig:qgg1-coll1},
by using dimensional regularisation together
with a suitable subtraction scheme, like $\overline{MS}$.

At this level, it is easy to understand why the graphs involving instantaneous propagators
cannot generate collinear divergences. As clear by inspection of \eqref{qgg1-coll1}, such
divergences occur via the integration over the Weisz{\"a}cker-Williams kernel $\bm{Y}^{m}/
\bm{Y}^2$ which describes a bremsstrahlung process in transverse space (recall e.g.
\eqn{trijetqgg1}). This kernel is inherent in the graphs 
involving the regular part of the (quark or gluon) propagator, but it is absent from the
instantaneous graphs. Said differently, the contribution from the instantaneous vertices
do not involve enough powers of $\bm{Y}^{m}$ and/or $\overline{\bm{Y}}^{p}$ to
generate a divergence when integrating over $\bm{z}$.

It is finally interesting to compare the graphs in Fig.~\ref{fig:qgg1-coll1}, in which the
first emitted gluon is {\it collinear}, to those  in Fig.~\ref{qgg1-first-soft}, where the same gluon is
{\it soft}. In Fig.~\ref{fig:qgg1-coll1} there are no initial-state interactions, whereas in 
Fig.~\ref{qgg1-first-soft} the final-state interactions are missing. This is in agreement with the 
physical expectations that a collinear gluon must be emitted very early, well before the hard
process, unlike a soft gluon, which must be the closest emission with respect to 
the collision. This physical distinction should allow us to unambiguously separate
between soft and collinear divergences.

Yet, by inspection of these two figures, one sees that one particular
graph contributes in both cases: this is the first graph in both Fig.~\ref{fig:qgg1-coll1} and 
Fig.~\ref{qgg1-first-soft}. This graph has generated the term proportional to $\mathcal{S}\left(\bm{y},\,\bm{\overline{y}}\right)$ in \eqn{qgg1-coll1}, which is in fact an {\it exact} evaluation for that
graph\footnote{Remember that this particular graph originates from
the last term within the square brackets in \eqref{trijetqgg1}, for which \eqn{K1coll1} becomes
exact.} --- up to regularisation issues, of course.
 And indeed, if one takes the soft limit $\xi\to 0$ in \eqn{qgg1-coll1}, one recovers one of 
the two terms proportional to $\mathcal{S}\left(\bm{w},\,\bm{\overline{w}}\right) $ in \eqn{qgevolu1}.
(The other such a term comes the fourth graph in Fig.~\ref{qgg1-first-soft}, in which the
soft gluon is never intersecting the shockwave. We recall that, in writing  \eqn{qgevolu1}, 
we have identified $\bm{y}=\bm{w}$ and $\bm{\overline{y}}=\bm{\overline{w}}$.)

This discussion makes clear that the first graph in  Fig.~\ref{fig:qgg1-coll1} (or in
Fig.~\ref{qgg1-first-soft}) contributes to both the collinear (DGLAP) and the soft  (B-JIMWLK)
evolutions. In particular, it contains an overlapping, soft and collinear, divergence. This might seem
to contradict our previous argument, that these two types of divergences can be separated from
each other. In fact, there is no contradiction: such overlapping divergence are eventually
cancelled after adding the virtual corrections.

 \begin{figure}[!t]\center
 \includegraphics[scale=1.]{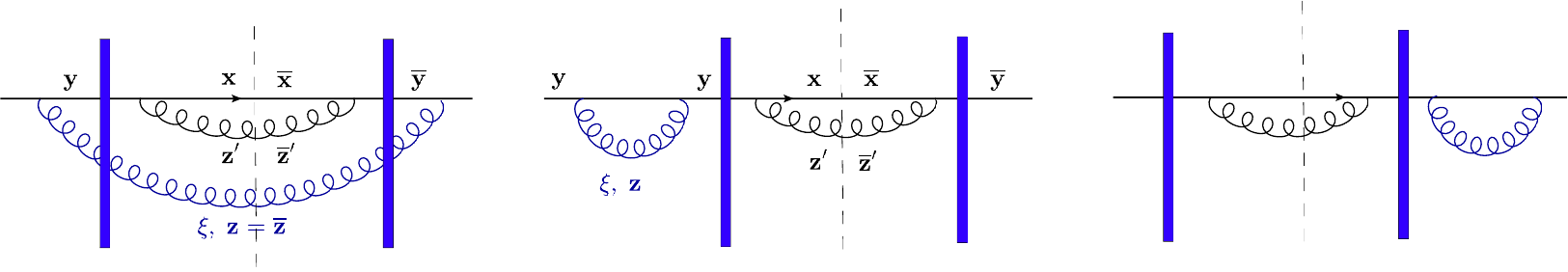}
\caption{The set of ``real'' and ``virtual'' graphs which count for the cancellation of
the overlapping, soft and collinear, divergences associated with the first gluon emission 
by the quark. }
\label{fig:qgg1-virt1}
\end{figure}

We shall systematically study the virtual corrections in a subsequent paper. Here we merely
show in Fig.~\ref{fig:qgg1-virt1} the graphs relevant to the above discussion: the real
graph that has already appeared as the first diagram  Fig.~\ref{fig:qgg1-coll1} together with the
virtual graphs which remove its collinear divergences in the limit where the first emission is {\it both}
collinear and soft ($\xi\ll 1$). 
The soft limit is indeed important for this argument: in general, the real NLO corrections and the
virtual ones are weighted by quark distribution functions with different arguments:
$x_p=(x_1+x_2)/(1-\xi)$ for the real corrections, cf. \eqn{qPDF}, but
$x_p=x_1+x_2$ for the virtual graphs, whose final state involves only two
partons:  the measured quark and gluon. However, when $\xi\ll 1$, these two
weighting factors become identical with each other 
and then it becomes possible to observe the cancellation of
the overlapping, soft and collinear, divergences. 

Let us show how this works for the three graphs in Fig.~\ref{fig:qgg1-virt1}.  Each of the two virtual graphs
gives a contribution which is simply the product between the LO dijet cross-section \eqref{LOfinal} 
and a ``tadpole'' describing the virtual gluon emission.  The effect of adding these contributions to the 
real term in \eqn{qgg1-coll1} is to replace the emission kernel for the first emission,  cf. \eqn{colldiv}, 
by the {\it dipole kernel},
\beq\label{Mdipole}
   \mathcal{M}(\bm{y},\overline{\bm{y}},\bm{z})\equiv
  \frac{1}{(\bm{y}-\bm{z})^{2}} + \frac{1}{(\overline{\bm{y}}-{\bm{z}})^{2}}
  -2\frac{(\bm{y}-\bm{z})\cdot (\overline{\bm{y}}-{\bm{z}})}
{(\bm{y}-\bm{z})^{2} (\overline{\bm{y}}-{\bm{z}})^{2}}\,=\,\frac{(\bm{y}-\overline{\bm{y}})^2}
{(\bm{y}-\bm{z})^{2}(\overline{\bm{y}}-{\bm{z}})^{2}}
\,,
\eeq
which decreases much faster at large $|\bm{z}|$ then the original kernel in  \eqn{colldiv}: as 
$1/\bm{z}^4$ instead of $1/\bm{z}^2$. So, the would-be logarithmic singularity at large $|\bm{z}|$
disappears. The dipole kernel exhibits instead short-distance (ultraviolet) poles at
$\bm{z}=\bm{y}$ and $\bm{z}=\overline{\bm{y}}$, but they are ultimately inocuous, as they cancel
against other virtual corrections, not shown  in Fig.~\ref{fig:qgg1-virt1}. 


\subsection{DGLAP evolution for the final quark}
\label{sec:coll2}

Still for the topology illustrated in Fig.~\ref{fig:qgg1}, we now consider the case where the second 
gluon emission --- the one where the gluons carries a longitudinal momentum fraction $\vartheta-\xi$ in Fig.~\ref{fig:qgg1} --- is not measured ($\overline{\bm{z}}'={\bm{z}}'$)
and is collinear. As already explained, this means that the transverse
separation $\bm{X}'=\bm{x}-\bm{z}'$ between the two daughter partons emerging from the
second splitting is
much larger than the corresponding separation $\bm{Y}=\bm{y}-\bm{z}$ for the first emission:
$(\bm{X}')^2 \gg (\bm{Y})^2$. Under this assumption, the dominant contribution to the  trijet cross-section in \eqn{trijetqgg1} comes from the last term there --- the one obtained after the double replacement $\big(\bm{x},\bm{z}^{\prime}\rightarrow\bm{y}\,
\ \&\ \,\bm{\overline{x}},\bm{\overline{z}}^{\prime}\rightarrow\overline{\bm{y}}\big)$ --- since the 
corresponding kernel is the only one not to be suppressed at large $(\bm{X}')^2$. Not surprisingly,
the graphs which survive in this limit are those which are void of final state interactions, as illustrated in
Fig.~\ref{fig:qgg1-coll2}. We therefore expect the collinear divergence generated by these graphs
to express the DGLAP evolution of the final-state gluon in the leading-order dijet cross-section \eqref{LOfinal}.
Let us verify that this is indeed the case.

 \begin{figure}[!t]\center
 \includegraphics[scale=1.]{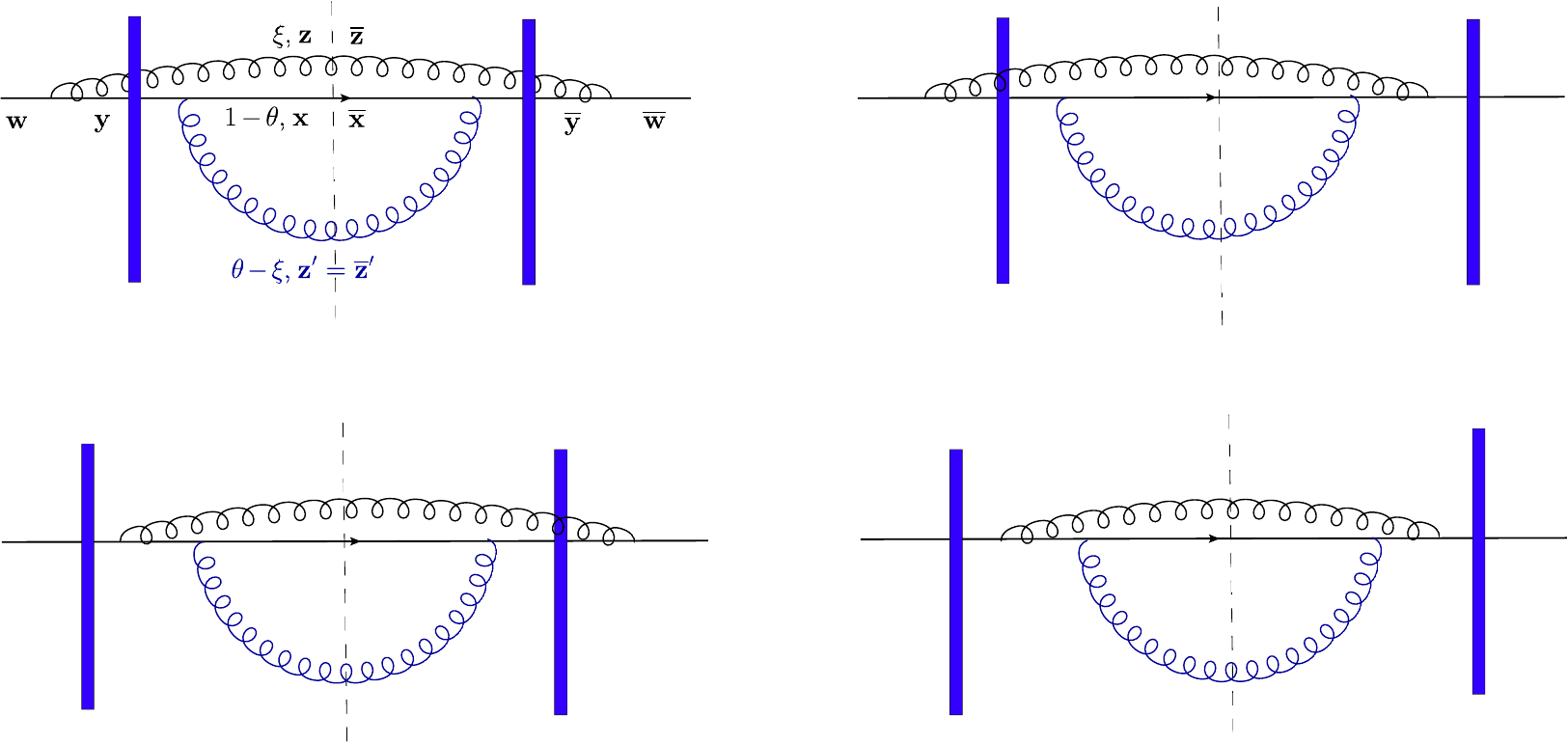}
\caption{The four graphs surviving in the collinear limit for the second gluon emission (depicted in
blue). These graphs are void of final-state interactions.}
\label{fig:qgg1-coll2}
\end{figure}

The kernel for the surviving terms has, clearly, the same expression as displayed in \eqn{K1coll1},
where the longitudinal momentum fractions must now be evaluated as
\begin{equation}\label{longit4}
 \vartheta 
 \,=1-\frac{x_1}{x_p}\,,
 \qquad\xi 
 \,=\,\frac{x_2}{x_p}\ \Longrightarrow \ \xi'
\equiv\,\frac{\vartheta-\xi}{1-\xi} 
 \,=\,\frac{x_p-(x_1+x_2)}{x_p-x_2}\,.
 \end{equation}
The main difference w.r.t. the previous subsection is that the kernel in \eqn{K1coll1} now
applies only to the last term within the square brackets in \eqn{trijetqgg1}. After also using $\overline{\bm{z}}'={\bm{z}}'$, we find that \eqn{qgg1-coll1} gets replaced by
 \begin{align}\label{qgg1-coll2}
\frac{d\sigma_{(1)\,\rnlo, 2}^{qA\rightarrow qg+X}}{dp^{+}d^{2}\bm{p}\,dk^{+}d^{2}\bm{k}}
&\simeq  \frac{4\alpha_s C_F}{(2\pi)^6 (q^+)^2(1-\xi)}\,P_{q\to g}(\xi)\nonumber\\*[0.2cm] 
&\times \int_{\overline{\bm{x}},\,\bm{\overline{z}},\,\bm{x},\,\bm{z}}\,e^{-i\bm{p}\cdot(\bm{x}-\overline{\bm{x}})-i\bm{k}\cdot(\bm{z}-\overline{\bm{z}})}\,
\frac{(\bm{y}-\bm{z})\cdot (\overline{\bm{y}}-\overline{\bm{z}})}
{(\bm{y}-\bm{z})^{2} (\overline{\bm{y}}-\overline{\bm{z}})^{2}}
\nonumber\\*[0.2cm]
&\times\Big[
\mathcal{Q}(\bm{y},\,\bm{z},\,\overline{\bm{z}},\,\overline{\bm{y}})\,\mathcal{S}(\bm{z},\,\overline{\bm{z}}) -\mathcal{S}(\bm{y},\,\bm{z})\,\mathcal{S}(\bm{z},\,\overline{\bm{w}})
 -\mathcal{S}(\bm{w},\,\overline{\bm{z}})\,\mathcal{S}(\overline{\bm{z}}, \overline{\bm{y}})
 +\mathcal{S}\left(\bm{w},\,\bm{\overline{w}}\right)\Big]
 \nonumber\\*[0.2cm] &\times  \frac{4\alpha_s C_F}{(2\pi)^2}\,
P_{q\to g}(\xi') \int_{\bm{z}'}
\frac{(\bm{x}-\bm{z}')\cdot (\overline{\bm{x}}-{\bm{z}}')}
{(\bm{x}-\bm{z}')^{2} (\overline{\bm{x}}-{\bm{z}}')^{2}}\,.
 \end{align}
For reasons to shortly become clear, it is convenient to consider $z_1\equiv 1-\xi'$ --- the splitting fraction
of the measured quark at the second emission vertex --- as an independent variable,
on the same footing as the external variables $x_1$ and $x_2$. Then the longitudinal fractions $\xi$
and $x_p$ and the transverse coordinate $\bm{y}$ should be understood as (recall 
Eqs.~\eqref{defyw2} and \eqref{longit4})
 \beq\label{xpz1}
 x_p=x_2 +\frac{x_1}{z_1}\,,\qquad
 \xi=\frac{z_1x_2}{x_1+z_1x_2}\,,\qquad  \bm{y}\,=\,z_1\bm{x}+(1-z_1)\bm{z}^{\prime}\,,
  \eeq
 together with a similar expression for $\overline{\bm{y}}$. 
 
  It is furthermore useful to change two of the integration variables, from $\bm{x}$ and
 $\overline{\bm{x}}$ to  $\bm{y}$ and $\overline{\bm{y}}$ (this will facilitate the comparison
 with the LO result in  \eqn{LOfinal}).   Recalling that  $\overline{\bm{z}}'={\bm{z}}'$,  one finds
 \beq
 \bm{p}\cdot(\bm{x}-\overline{\bm{x}}) = \frac{1}{z_1}\, \bm{p}\cdot(\bm{y}-\overline{\bm{y}}),\qquad
 \bm{x}-\bm{z}' = \frac{1}{z_1}\,( \bm{y}-\bm{z}')\,.\eeq
 When expressed in terms of these new variables,
 $\bm{w} = (1-\xi) \bm{y} + \xi \bm{z}$ becomes independent of  $\bm{z}'$, so the integral
 over $\bm{z}'$ factorizes.

After also convoluting with the initial quark distribution, as shown in the l.h.s. of \eqn{qPDF}, 
and changing the respective integration variable from $x_p$
to $z_1$ according to \eqn{xpz1}, one obtains the following expression for the collinear singularity encoded
in this particular NLO correction to the dijet cross-section
\begin{align}\label{qgg1-coll2fin}
\frac{d\sigma_{(1)\,\rnlo, 2}^{pA\rightarrow qg+X}}{dp^{+}d^{2}\bm{p}\,dk^{+}d^{2}\bm{k}}
&\simeq \int\frac{dz_1}{z_1^3}\,x_{p}\,q_f(x_{p},\mu^{2})\,
 \frac{4\alpha_s C_F}{(2\pi)^6 (q^+)^2}\,P_{q\to g}(\xi)\nonumber\\*[0.2cm] 
&\times \int_{\overline{\bm{y}},\,\bm{\overline{z}},\,\bm{y},\,\bm{z}}\,e^{-i\bm{p}\cdot(\bm{y}-\overline{\bm{y}})/z_1-i\bm{k}\cdot(\bm{z}-\overline{\bm{z}})}\,
\frac{(\bm{y}-\bm{z})\cdot (\overline{\bm{y}}-\overline{\bm{z}})}
{(\bm{y}-\bm{z})^{2} (\overline{\bm{y}}-\overline{\bm{z}})^{2}}
\nonumber\\*[0.2cm]
&\times\Big[
\mathcal{Q}(\bm{y},\,\bm{z},\,\overline{\bm{z}},\,\overline{\bm{y}})\,\mathcal{S}(\bm{z},\,\overline{\bm{z}}) -\mathcal{S}(\bm{y},\,\bm{z})\,\mathcal{S}(\bm{z},\,\overline{\bm{w}})
 -\mathcal{S}(\bm{w},\,\overline{\bm{z}})\,\mathcal{S}(\overline{\bm{z}}, \overline{\bm{y}})
 +\mathcal{S}\left(\bm{w},\,\bm{\overline{w}}\right)\Big]
 \nonumber\\*[0.2cm] &\times  \frac{4\alpha_s C_F}{(2\pi)^2}\,
P_{q\to q}(z_1) \int_{\bm{z}'}
\frac{(\bm{y}-\bm{z}')\cdot (\overline{\bm{y}}-{\bm{z}}')}
{(\bm{y}-\bm{z}')^{2} (\overline{\bm{y}}-{\bm{z}}')^{2}}\,,
 \end{align}
where we have also used  $x_1/(1-\xi)=x_p z_1$ and $P_{q\to g}(1-z_1) =P_{q\to q}(z_1) $. 
 
\eqn{qgg1-coll2} exhibits a factorised structure, as expected: it is the product of the LO dijet cross-section 
in \eqn{LOfinal} (but for final momenta $p_1=p/z_1$ and $k$)
 times the probability for an unobserved emission in the final state. This probability contains
a collinear singularity similar to that visible in \eqn{colldiv}, which here should be absorbed into the renormalisation of the fragmentation function of the final quark. 
Specifically, the singular piece of \eqn{qgg1-coll2}  can be written as
\begin{align}\label{qgg1-coll2div}
\frac{d\sigma_{(1)\,\rnlo, 2}^{pA\rightarrow qg+X}}{dp^{+}d^{2}\bm{p}\,dk^{+}d^{2}\bm{k}}\bigg |_{\rm coll}
=\int\frac{dz_1}{z_1^3}\,x_{p}\,q_f(x_{p},\mu^{2})\,
\frac{d\sigma_{\lo}^{pA\rightarrow qg+X}}{d^{3}p_1\,d^{3}k} \bigg |_{p_1=p/z_1}\,D_{g/q}(z_1,\mu^{2})\,,
 \end{align}
where
\beq\label{frag1}
D_{g/q}(z_1,\mu^{2})\equiv  \frac{\alpha_s C_F}{\pi}
P_{q\to q}(z_1)\,\ln\frac{\mu^2}{\Lambda^2}\,.
\eeq
is recognised as the contribution of the first step in the DGLAP evolution of the quark-to-gluon fragmentation function.

To summarise, Eqs.~\eqref{qgg1-coll2fin}--\eqref{frag1}  describe the beginning of the DGLAP evolution of the
quark fragmentation function in \eqn{pALOfrag}. The corresponding evolution of the gluon fragmentation
function will be discussed in the next two sections.

\subsection{DGLAP evolution for the final gluon: $g\to gg$ splitting}

We now consider the collinear limit for the second topology yielding real NLO corrections
to quark-gluon production, the one denoted by the subscript $(2)$ in \eqn{qgrNLO} and which
is illustrated in Fig.~\ref{final-qgg2}. In this case, a collinear divergence can be generated only
by the second emission (see the discussion towards the end of this section), in which case
it is associated with the DGLAP evolution of the fragmentation function for the final gluon.

 \begin{figure}[!t]\center
 \includegraphics[scale=1.]{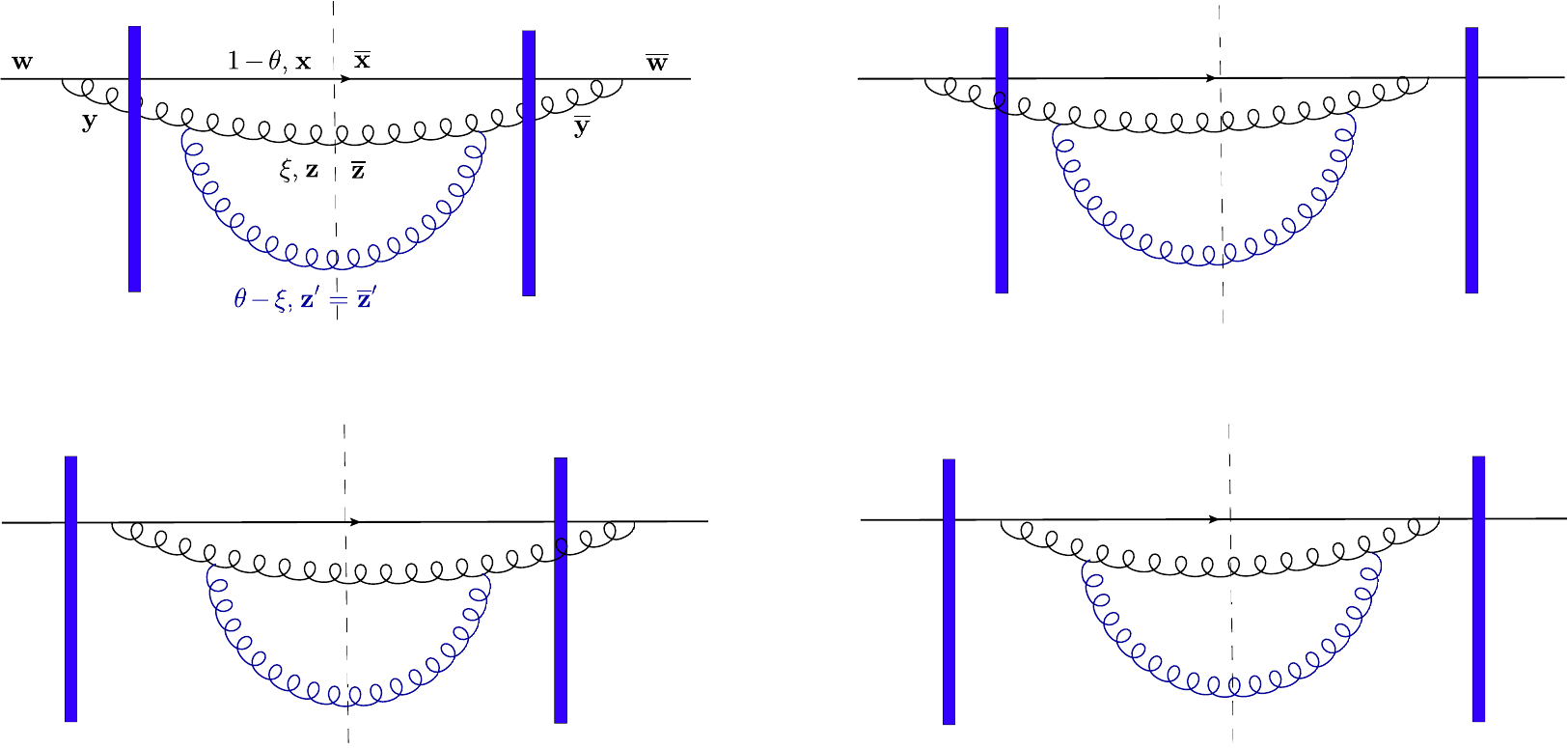}
\caption{The four graphs surviving in the collinear limit for the $g\to gg$ splitting
 (the unmeasured gluon is depicted in blue). These graphs are void of final-state interactions.}
\label{fig:qgg2-coll2}
\end{figure}

So let us consider the collinear limit  for the second emission in Fig.~\ref{final-qgg2}. This is
the limit where the transverse separation  $\bm{Z}=\bm{z}-\bm{z}'$ between the two final final 
gluons is much larger than the separation $\bm{R}=\bm{x}-\bm{y}$ between the final quark
and the intermediate gluon: $\bm{Z}^2\gg \bm{R}^2$. With reference to \eqn{trijetqgg2},
this hierarchy entails two important simplification. The first one refers to the energy denominator
and is fully similar to that discussed in the previous section: it implies that the only
surviving graphs are the four graphs without final-state interactions, shown in Fig.~\ref{fig:qgg2-coll2}.
These graphs correspond to the last term within the square brackets in \eqn{trijetqgg2} --- the only
one not to be suppressed at large $\bm{Z}^2$. The second simplification refers to the product 
\eqref{Pi2} of effective vertices: in  \eqn{trijetqgg2}, this product is multiplied by
$\bm{R}^{m}\,\bm{Z}^{n}\,\overline{\bm{R}}^{p}\,\overline{\bm{Z}}^{q}$. When computing the graphs 
in Fig.~\ref{fig:qgg2-coll2}, the integral over the coordinate $\bm{z}'$ of the unmeasured gluon
($\bm{z}'=\overline{\bm{z}}'$) can be factorised as
\beq\label{sing}
 \int_{\bm{z}'}\frac{\bm{Z}^{n}\,\overline{\bm{Z}}^{q}}{\bm{Z}^{2}\,
\overline{\bm{Z}}^{2}}\,\simeq\,\int_{\bm{z}'}\frac{\bm{Z}^{n}\,{\bm{Z}}^{q}}{\bm{Z}^{4}}
\,=\,\frac{\delta^{nq}}{2}
 \int_{\bm{z}'}
\frac{1}{\bm{Z}^{2}}\,\simeq\,\frac{\delta^{nq}}{2}
\,\pi\ln\frac{1}{(\bm{z}-\overline{\bm{z}})^2\Lambda^2}\,,
\eeq
where the approximate equality holds for the singular piece alone: in the collinear limit at hand,
the differences $\bm{Z}=\bm{z}-\bm{z}'$ and $\overline{\bm{Z}}=\overline{\bm{z}}-\bm{z}'$ can be
arbitrarily large, whereas $|\bm{z}-\overline{\bm{z}}|$ is fixed to a value $\sim 1/k_\perp$ by the
Fourier transform for the measured gluon; so, in evaluating the singular piece of \eqref{sing}, 
one can approximate $\bm{Z}\simeq \overline{\bm{Z}}$ and at the same time restrict 
the integral over $\bm{z}'$ to $\bm{Z}^2 > (\bm{z}-\overline{\bm{z}})^2$. The logarithmic divergence
at $\bm{Z}^2\to \infty$ has been regulated by a momentum cutoff $\Lambda^2$, as in \eqn{colldiv}.

\eqn{sing} shows that, for the purpose of extracting the collinear singularity, one can replace
$\bm{Z}^{n}\,\overline{\bm{Z}}^{q}\to (\bm{Z}\cdot \overline{\bm{Z}}){\delta^{nq}}/{2}$ within
the integrand  of  \eqn{trijetqgg2}. This allows us to simplify the tensorial
structure of the product of effective vertices (cf. \eqn{Pi2}):
\begin{align}
\Pi_{\lambda_{1}\lambda}^{ijmn}\,\Pi_{\lambda_{1}\lambda}^{ijpq\,*}
\,\bm{R}^{m}\,\bm{\overline{R}}^{p}\,(\bm{Z}\cdot \overline{\bm{Z}})\,\frac{\delta^{nq}}{2}
=&\,4\xi(\vartheta-\xi)\vartheta^{2}(1-\vartheta)^{2}\left[1+(1-\vartheta)^{2}\right]\nonumber\\*[0.2cm] 
&\times\left(\frac{1}{\vartheta^{2}}+\frac{1}{\xi^{2}}+\frac{1}{(\vartheta-\xi)^{2}}\right)(\bm{R}\cdot\bm{\overline{R}})\,(\bm{Z}\cdot\bm{\overline{Z}})\,.
\end{align}
As a check, we note that the eikonal limit $\xi\ll\vartheta$ of this result is consistent with \eqn{K2eik}.

The rest of the calculation is straightforward and yields (see also  Fig.~\ref{fig:qgg2-coll2}):
 \begin{align}\label{qgg2-coll2}
\frac{d\sigma_{(2)\,\rnlo, 2}^{qA\rightarrow qg+X}}{dp^{+}d^{2}\bm{p}\,dk^{+}d^{2}\bm{k}}
&\simeq  \frac{4\alpha_s C_F}{(2\pi)^6 (q^+)^2\vartheta}\,P_{q\to g}(\vartheta)\nonumber\\*[0.2cm] 
&\times \int_{\overline{\bm{x}},\,\bm{\overline{z}},\,\bm{x},\,\bm{z}}\,e^{-i\bm{p}\cdot(\bm{x}-\overline{\bm{x}})-i\bm{k}\cdot(\bm{z}-\overline{\bm{z}})}\,
\frac{(\bm{x}-\bm{y})\cdot (\overline{\bm{x}}-\overline{\bm{y}})}
{(\bm{x}-\bm{y})^{2} (\overline{\bm{x}}-\overline{\bm{y}})^{2}}
\nonumber\\*[0.2cm]
&\times\Big[
\mathcal{Q}(\bm{x},\,\bm{y},\,\overline{\bm{y}},\,\overline{\bm{x}})\,\mathcal{S}(\bm{y},\,\overline{\bm{y}}) -\mathcal{S}(\bm{x},\,\bm{y})\,\mathcal{S}(\bm{y},\,\overline{\bm{w}})
 -\mathcal{S}(\bm{w},\,\overline{\bm{y}})\,\mathcal{S}(\overline{\bm{y}}, \overline{\bm{x}})
 +\mathcal{S}\left(\bm{w},\,\bm{\overline{w}}\right)\Big]
 \nonumber\\*[0.2cm] &\times  \frac{4\alpha_s N_c}{(2\pi)^2}\,
P_{g\to g}(\xi/\vartheta) \int_{\bm{z}'}
\frac{(\bm{z}-\bm{z}')\cdot (\overline{\bm{z}}-{\bm{z}}')}
{(\bm{z}-\bm{z}')^{2} (\overline{\bm{z}}-{\bm{z}}')^{2}}\,,
 \end{align}
where $z_2\equiv \xi/\vartheta$ is the splitting fraction of the measured gluon and
\beq
P_{g\to g}(z)\equiv \frac{[1-z(1-z)]^2}{z(1-z)}\,,\eeq
is the gluon-to-gluon LO DGLAP splitting function.  It is understood that only the logarithmic collinear
singularity is properly encoded in \eqn{qgg2-coll2}, but not also the constant term under the logarithm.

At this stage, it is convenient to express all the longitudinal
fractions in terms of $x_1$, $x_2$ and $z_2$:
\beq\label{xpz2}
 x_p=x_1 +\frac{x_2}{z_2}\,,\qquad \vartheta=1-\frac{x_1}{x_p}= \frac{x_2}{x_2+z_1x_1}\,,
 \qquad  \xi=\vartheta z_2\,.
  \eeq
Also,  we make a change of integration variables, from $\bm{z}$ and
 $\overline{\bm{z}}$ to  $\bm{y}$ and $\overline{\bm{y}}$, by using
\beq
\bm{y}\,=\,z_2\bm{z}+(1-z_2)\bm{z}^{\prime} \ \Longrightarrow\  \
\bm{z}-\overline{\bm{z}} = \frac{1}{z_2}\, (\bm{y}-\overline{\bm{y}}),\qquad
 \bm{z}-\bm{z}' = \frac{1}{z_2}\,( \bm{y}-\bm{z}')\,,\eeq
together with  $\bm{w} = (1-\vartheta) \bm{x} + \vartheta\bm{y}$.

We finally add the convolution with the quark distribution function $q_f(x_{p},\mu^{2})$, but use
$z_2$ rather than $x_p$ as an integration variable. Putting all that together one finds
\begin{align}\label{qgg2-coll2fin}
\frac{d\sigma_{(2)\,\rnlo, 2}^{pA\rightarrow qg+X}}{dp^{+}d^{2}\bm{p}\,dk^{+}d^{2}\bm{k}}
&\simeq \int\frac{dz_2}{z_2^3}\,x_{p}\,q_f(x_{p},\mu^{2})\,
 \frac{4\alpha_s C_F}{(2\pi)^6 (q^+)^2}\,P_{q\to g}(\vartheta)\nonumber\\*[0.2cm] 
 &\times \int_{\overline{\bm{x}},\,\bm{\overline{y}},\,\bm{x},\,\bm{y}}\,e^{-i\bm{p}\cdot(\bm{x}-\overline{\bm{x}})-i\bm{k}\cdot(\bm{y}-\overline{\bm{y}})/z_2}\,
\frac{(\bm{x}-\bm{y})\cdot (\overline{\bm{x}}-\overline{\bm{y}})}
{(\bm{x}-\bm{y})^{2} (\overline{\bm{x}}-\overline{\bm{y}})^{2}}
\nonumber\\*[0.2cm]
&\times\Big[
\mathcal{Q}(\bm{x},\,\bm{y},\,\overline{\bm{y}},\,\overline{\bm{x}})\,\mathcal{S}(\bm{y},\,\overline{\bm{y}}) -\mathcal{S}(\bm{x},\,\bm{y})\,\mathcal{S}(\bm{y},\,\overline{\bm{w}})
 -\mathcal{S}(\bm{w},\,\overline{\bm{y}})\,\mathcal{S}(\overline{\bm{y}}, \overline{\bm{x}})
 +\mathcal{S}\left(\bm{w},\,\bm{\overline{w}}\right)\Big]
 \nonumber\\*[0.2cm] &\times  \frac{4\alpha_s N_c}{(2\pi)^2}\,
P_{g\to g}(z_2) \int_{\bm{z}'}
\frac{(\bm{y}-\bm{z}')\cdot (\overline{\bm{y}}-{\bm{z}}')}
{(\bm{y}-\bm{z}')^{2} (\overline{\bm{y}}-{\bm{z}}')^{2}}\,.
 \end{align}
This is analog to \eqn{qgg1-coll2fin} from the previous section and its subsequent discussion
is also similar:  the collinear divergence can be absorbed in one step of the DGLAP evolution
of the gluon-to-gluon fragmentation function:
\beq\label{frag2}
D_{g/g}(z_2,\mu^{2})\equiv  \frac{\alpha_s N_c}{\pi}
P_{g\to g}(z_2)\,\ln\frac{\mu^2}{\Lambda^2}\,.
\eeq

To conclude this section, let us explain why the first emission cannot contribute a collinear
divergence for this particular topology. The first emitted gluon is an intermediate gluon,
which decays into two other gluons prior the final state. One of these daughter gluons is measured
in the final state ($\overline{\bm{z}}\ne \bm{z}$), and the other one is not ($\overline{\bm{z}}'= \bm{z}'$).
Hence the coordinates, $\bm{y}$ and $\overline{\bm{y}}$, of the intermediate gluon are different in
the DA and the CCA, respectively (as also manifest in \eqn{qgg2-coll2}). 
Hence, the would-be collinear limit for the
first emission, that is,  $\bm{R}^2\gg \bm{Z}^2$, cannot generate a collinear divergence in the 
integral over the transverse coordinate $ \bm{z}'$ of the unmeasured daughter gluon.

One can similarly understand that the interference graphs responsible for 
the piece denoted by the subscript $(3)$ in \eqn{qgrNLO}  (see Fig.~\ref{qgg-interf})
do not generate  collinear
divergences either. This is consistent with the fact that the LO DGLAP evolution admits a probabilistic
picture and could not accommodate such interference effects.

\subsection{DGLAP evolution for the final gluon: $g\to q\bar q$ splitting}

As a final exemple for the DGLAP evolution of the quark-gluon cross-section, let us briefly 
consider the case
where the gluon undergoes a collinear splitting into a quark-antiquark pair, thus contributing to the
$qq\bar q$ final state. With reference to \eqn{trijetqq} for the respective cross-section, 
this situation corresponds to the case where $\bm{Z}^2\gg \bm{R}^2$ and one of the daughter fermions
of the gluon decay --- say the antiquark with longitudinal fraction  $\vartheta-\xi$ (see Fig.~\ref{S6q}) 
is not measured ($\overline{\bm{z}}'={\bm{z}}'$). Clearly, the treatment of the $g\to q\bar q$ collinear
splitting is entirely analog to that of the $g\to gg$ splitting discussed in the previous section. 
Once again, the surviving graphs are the four graphs corresponding to the last term in \eqn{trijetqq},
which contain no final-state interactions. (The respective $S$-matrix structure has been already exhibited
in \eqn{W1qg}.) And the product of effective vertices which enters the kernel
\eqref{defK0} can be simplified --- in so far as the collinear divergence is concerned --- by using the analog
of \eqn{sing}. This yields (recall also \eqn{Phi2})
\begin{align}
\hspace*{-0.6cm}
\Phi_{\lambda_{3}\lambda_{2}\lambda_{1}\lambda}^{ij}\,\Phi_{\lambda_{3}\lambda_{2}\lambda_{1}\lambda}^{mn}\,\bm{R}^{i}\,\bm{\overline{R}}^{\,m}(\bm{Z}\cdot\overline{\bm{Z}})\,\frac{\delta^{jn}}{2}
=\,8(1-\vartheta)^{2}\left(1+(1-\vartheta)^{2}\right)\left((\vartheta-\xi)^{2}+\xi^{2}\right)\left(\bm{R}\cdot\bm{\overline{R}}\right)\left(\bm{Z}\cdot\overline{\bm{Z}}\right).
\end{align}
Inserting this into \eqn{trijetqq} and integrating over $k^3$, one finds (we relabel  the final momenta
as $p$ and $k$ for the two measured quark, with $p$ referring to the leading quark)
\begin{align}\label{qqq-coll2}
\frac{d\sigma_{\rnlo, 2}^{qA\rightarrow qq+X}}{dp^{+}d^{2}\bm{p}\,dk^{+}d^{2}\bm{k}}
&\simeq  \frac{4\alpha_s C_F}{(2\pi)^6 (q^+)^2\vartheta}\,P_{q\to g}(\vartheta)\nonumber\\*[0.2cm] 
&\times \int_{\overline{\bm{x}},\,\bm{\overline{z}},\,\bm{x},\,\bm{z}}\,e^{-i\bm{p}\cdot(\bm{x}-\overline{\bm{x}})-i\bm{k}\cdot(\bm{z}-\overline{\bm{z}})}\,
\frac{(\bm{x}-\bm{y})\cdot (\overline{\bm{x}}-\overline{\bm{y}})}
{(\bm{x}-\bm{y})^{2} (\overline{\bm{x}}-\overline{\bm{y}})^{2}}
\nonumber\\*[0.2cm]
&\times\Big[
\mathcal{Q}(\bm{x},\,\bm{y},\,\overline{\bm{y}},\,\overline{\bm{x}})\,\mathcal{S}(\bm{y},\,\overline{\bm{y}}) -\mathcal{S}(\bm{x},\,\bm{y})\,\mathcal{S}(\bm{y},\,\overline{\bm{w}})
 -\mathcal{S}(\bm{w},\,\overline{\bm{y}})\,\mathcal{S}(\overline{\bm{y}}, \overline{\bm{x}})
 +\mathcal{S}\left(\bm{w},\,\bm{\overline{w}}\right)\Big]
 \nonumber\\*[0.2cm] &\times  \frac{4\alpha_s N_f}{(2\pi)^2}\,
P_{g\to q}(\xi/\vartheta) \int_{\bm{z}'}
\frac{(\bm{z}-\bm{z}')\cdot (\overline{\bm{z}}-{\bm{z}}')}
{(\bm{z}-\bm{z}')^{2} (\overline{\bm{z}}-{\bm{z}}')^{2}}\,,
 \end{align}
which also involves the LO DGLAP splitting function for the $g\to q\bar q$ splitting:
\beq
P_{g\to q}(z)\equiv \frac{z^2+(1-z)^2}{2}\,.\eeq
The similarity with \eqn{qgg2-coll2} is manifest. Once again, the collinear singularity encoded in
the above integral over $\bm{z}'$ is reabsorbed into a contribution to one-step in the DGLAP evolution
of the fragmentation function for gluon fragmenting into quarks:
\beq\label{frag3}
D_{q/g}(z,\mu^{2})\equiv  \frac{\alpha_s N_f}{\pi}
P_{g\to q}(z)\,\ln\frac{\mu^2}{\Lambda^2}\,.
\eeq

\bigskip
\noindent
\textbf{Acknowledgments}

We would like to thank G. Beuf, T. Lappi, M. Lublinsky, and A. H. Mueller for useful discussions. The work of E.I. is supported in part by the Agence Nationale de la Recherche project  ANR-16-CE31-0019-01.
The work of Y.M. was in part supported by the 2016-2017 Chateaubriand fellowship of the French embassy in Israel.  This work  is supported under the European Union's Horizon 2020 research and innovation programme by the European Research Council (ERC), grant agreement No. ERC-2015-CoG-681707 (Y.M. since 2018) and by the STRONG-2020 project (grant agreement No 824093).  The content  of  this article  does  not  reflect  the  official opinion  of  the  European  Union  and  responsibility  for  the information  and  views  expressed  therein  lies  entirely with  the  authors.

\appendix

\section{Definitions of states and field quantization \label{fieldef}}

In this Appendix, we summarise our conventions for the field quantisation and the
definition of the bare Fock states. More details on the light-cone wavefunction formalism, including
the complete expression of the QCD Hamiltonian in the light-cone gauge, can be found in
Appendix {\bf A} of Ref.~\cite{Iancu:2018hwa}.

Quantisation of the fields is performed in  usual manner introducing creation/annihilation operators and imposing commutation (anti-commutation) relations among them.
For the gauge fields we shall use the following expansion:
\begin{equation}\label{glufield}
A_{i}^{a}(x)=\int_{0}^{\infty}\frac{dk^{+}}{2\pi}\int\frac{d^{2}\bm{k}}{(2\pi)^{2}}\frac{1}{\sqrt{2k^{+}}}\left(a_{i}^{a}(k^{+},\bm{k})e^{-ik\cdot x}+a_{i}^{a\dagger}(k^{+},\bm{k})e^{ik\cdot x}\right).
  \end{equation}
The creation and annihilation operators obey the bosonic algebra:
\begin{equation}
\left[a_{i}^{a}(k^{+},\bm{k}),\, a_{j}^{b\dagger}(p^{+},\bm{p})\right]=(2\pi)^{3}\delta^{ab}\delta_{ij}\delta(k^{+}-p^{+})\delta^{(2)}(\bm{k}-\bm{p}).
 \end{equation}
Transforming to coordinate space,
     \begin{equation}\begin{split}\label{tran}
a_{i}^{a}(k^{+},\,\bm{k})\,=\,\int d^2\bm{z}\,\rme^{-i\bm{k}\cdot\bm{z}}\,a_{i}^{a}(k^{+},\,\bm{z})\;,
 \end{split}\end{equation}
the commutation relation becomes:
 \begin{equation}\label{composw}
 \left[a_{i}^{a}(k^{+},\bm{x}),\, a_{j}^{b\dagger}(p^{+},\bm{y})\right]=2\pi\delta^{ab}\delta_{ij}\delta(k^{+}-p^{+})\delta^{(2)}(\bm{x}-\bm{y}).
  \end{equation}
The quark fields can be expanded by the following expression:
\begin{equation}\label{quarfield}
\psi_{+}^{\alpha}(x)\,=\,\chi_{\lambda}\int_{0}^{\infty}\frac{dk^{+}}{2\pi}\int\frac{d^{2}\bm{k}}{(2\pi)^{2}}\frac{1}{\sqrt{2}}\left(b_{\lambda}^{\alpha}(k^{+},\bm{k})e^{-ik\cdot x}\,+\,d_{\lambda}^{\alpha\dagger}(k^{+},\bm{k})e^{ik\cdot x}\right).
 \end{equation}
The polarisation vectors are:
\begin{equation}\label{chis}
\chi_{+\frac{1}{2}}\,=\,\left(\begin{array}{c}
1\\
0
\end{array}\right),\qquad\quad\chi_{-\frac{1}{2}}\,=\,\left(\begin{array}{c}
0\\
1
\end{array}\right),
 \end{equation}
 \begin{equation}\label{lamiden}
\chi_{\lambda}\chi_{\lambda}^{\dagger}\,=\,I,\qquad\chi_{\lambda}^{\dagger}\chi_{\lambda}\,=\,2,\qquad\chi_{\lambda_{1}}^{\dagger}\,I\,\chi_{\lambda_{2}}\,=\,\delta_{\lambda_{1}\lambda_{2}},\qquad\chi_{\lambda_{1}}^{\dagger}\,\sigma^{3}\,\chi_{\lambda_{2}}\,=\,2\lambda_{1}\delta_{\lambda_{1}\lambda_{2}}.
 \end{equation}
The anti-commutation relations:
 \begin{equation}\begin{split}
\left\{ b_{\lambda_{1}}^{\alpha}(k^{+},\,\bm{k}),\, b_{\lambda_{2}}^{\beta\dagger}(p^{+},\,\bm{p})\right\} &=\,\left\{ d_{\lambda_{1}}^{\alpha}(k^{+},\,\bm{k}),\, d_{\lambda_{2}}^{\beta\dagger}(p^{+},\,\bm{p})\right\} \\
&=\,(2\pi)^{3}\,\delta_{\lambda_{1}\lambda_{2}}\,\delta^{\alpha\beta}\,\delta^{(2)}(\bm{k}-\bm{p})\,\delta(k^{+}-p^{+}).\\
 \end{split}\end{equation}
 Transforming the fields to coordinate space a la (\ref{tran}):
 \begin{equation}\begin{split}
\left\{ b_{\lambda_{1}}^{\alpha}(k^{+},\,\bm{x}),\, b_{\lambda_{2}}^{\beta\dagger}(p^{+},\,\bm{y})\right\} &=\,\left\{ d_{\lambda_{1}}^{\alpha}(k^{+},\,\bm{x}),\, d_{\lambda_{2}}^{\beta\dagger}(p^{+},\,\bm{y})\right\} \,\\
&=\,2\pi\,\delta_{\lambda_{1}\lambda_{2}}\,\delta^{\alpha\beta}\,\delta^{(2)}(\bm{x}-\bm{y})\,\delta(k^{+}-p^{+}).
 \end{split}\end{equation}

The multi-parton bare Fock states are obtained by acting with the relevant creation operators on the bare vacuum state. In this paper, we use both the 3-momentum representation  $k=(k^+,\bm{k})$ and  the mixed representation $(k^+,\bm{x})$, as obtained via the Fourier transform from transverse momenta to transverse coordinates. Let us present here a few representative examples.

\smallskip
$\bullet$ \textit{\textbf{The bare vacuum state  $\left|0\right\rangle $.}} This state obeys the following conditions:
  \begin{equation}\label{vacuum}
a_{i}^{a}(q^{+},\,\bm{q})\left|0\right\rangle \,=\,b_{\lambda}^{\alpha}(q^{+},\,\bm{q})\left|0\right\rangle \,=\,d_{\lambda}^{\alpha}(q^{+},\,\bm{q})\left|0\right\rangle \,=\,0.
  \end{equation}
  
  \smallskip
$\bullet$ \textit{\textbf{The bare quark state.}} In momentum space, this state is constructed as
  \begin{equation}\label{gsta}
\left|q_{\lambda}^{\alpha}(k^{+},\,\bm{k})\right\rangle \,\equiv\,{b_{\lambda}^{\alpha\dagger}(k^{+},\,\bm{k})}\left|0\right\rangle.
  \end{equation}
its scalar product is normalized as follows:
\begin{equation}\label{norcon1}
\left\langle q_{\lambda_{2}}^{\beta}(p^{+},\,\bm{p})\left|q_{\lambda_{1}}^{\alpha}(k^{+},\,\bm{k})\right.\right\rangle \,=\,(2\pi)^{3}\delta^{\alpha\beta}\,\delta_{\lambda_{1}\lambda_{2}}\,\delta^{(2)}(\bm{k}-\bm{p})\,\delta(k^{+}-p^{+}).
\end{equation}
The mixed representation of the bare quark state is obtained as 
\begin{align}\label{chanrep2}
\left|q_{\lambda}^{\alpha}(k^{+},\,\bm{x})\right\rangle \equiv\int 
\frac{d^{2}\bm{k}}{(2\pi)^{2}}\,\rme^{-i\bm{x}\cdot\bm{k}}\left|q_{\lambda}^{\alpha}(k^{+},\,\bm{k})\right\rangle=
 {b_{\lambda}^{\alpha\dagger}(k^{+},\,\bm{x})}\left|0\right\rangle,
\end{align}
and the dot product reads
\begin{equation}\label{norcon2}
\left\langle q_{\lambda_{2}}^{\beta}(p^{+},\,\bm{y})\left|q_{\lambda_{1}}^{\alpha}(k^{+},\,\bm{x})\right.\right\rangle \,=\,2\pi \,
\delta^{\alpha\beta}\,\delta_{\lambda_{1}\lambda_{2}}\,\delta^{(2)}(\bm{x}-\bm{y})\,\delta(k^{+}-p^{+}).
\end{equation}

\smallskip
$\bullet$ \textit{\textbf{The bare quark-gluon state.}} In momentum space, this state reads
  \begin{equation}\begin{split}\label{ggst}
&\left|q_{\lambda}^{\alpha}(p^{+},\,\bm{p})\,g_{i}^{a}(k^{+},\,\bm{k})\right\rangle \,\equiv\,{b_{\lambda}^{\alpha\dagger}(p^{+},\,\bm{p})\,a_{i}^{a\dagger}(k^{+},\,\bm{k})}\left|0\right\rangle\,,
\end{split}\end{equation}
The mixed representation of this state reads
\begin{align}\label{qgchanrep}
\left|q_{\lambda}^{\alpha}(p^{+},\,\bm{x})\,g_{i}^{a}(k^{+},\,\bm{y})\right\rangle & \equiv\int \frac{d^{2}\bm{p}}{(2\pi)^{2}}
\frac{d^{2}\bm{k}}{(2\pi)^{2}}\,\rme^{-i\bm{x}\cdot\bm{p}-i\bm{y}\cdot\bm{k}}
\left|q_{\lambda}^{\alpha}(p^{+},\,\bm{p})\,g_{i}^{a}(k^{+},\,\bm{k})\right\rangle=\nonumber\\*[.2cm]
&= {b_{\lambda}^{\alpha\dagger}(p^{+},\,\bm{x})\,a_{i}^{a\dagger}(k^{+},\,\bm{y})}\left|0\right\rangle.
\end{align}

\smallskip

 $\bullet$ \textit{\textbf{The bare 2 quarks plus an antiquark state:}}
\begin{equation}\begin{split}\label{qqst}
&\left|q_{\lambda_{1}}^{\alpha}(q^{+},\,\bm{q})\,q_{\lambda_{2}}^{\beta}(k^{+},\,\bm{k})\,\bar{q}_{\lambda_{3}}^{\gamma}(p^{+},\,\bm{p})\right\rangle \,\equiv\,b_{\lambda_{1}}^{\alpha\dagger}(q^{+},\,\bm{q})\,b_{\lambda_{2}}^{\beta\dagger}(k^{+},\,\bm{k})\,d_{\lambda_{3}}^{\gamma\dagger}(p^{+},\,\bm{p})\left|0\right\rangle ,
\end{split}\end{equation}

$\bullet$ \textit{\textbf{The bare quark plus 2 gluons state:}}
  \begin{equation}\begin{split}\label{qggst}
&\left|q_{\lambda}^{\alpha}(q^{+},\,\bm{q})\,g_{i}^{a}(k^{+},\,\bm{k})\,g_{j}^{b}(p^{+},\,\bm{p})\right\rangle \,\equiv\,b_{\lambda}^{\alpha\dagger}(q^{+},\,\bm{q})\,a_{i}^{a\dagger}(k^{+},\,\bm{k})\,a_{j}^{b\dagger}(p^{+},\,\bm{p})\left|0\right\rangle ,
\end{split}\end{equation}\\

 \section{Matrix elements\label{mateleapp}}

 In terms of the Fock space operators introduced in the previous Appendix, the  free QCD  Hamiltonian  takes the following form,
  \begin{equation}\begin{split}
 H_{0}=\int_{0}^{\infty}\frac{dk^{+}}{2\pi}&\int\frac{d^{2}\bm{k}}{(2\pi)^{2}}\frac{\bm{k}^{2}}{2k^{+}}\bigg(a_{i}^{a\dagger}(k^{+},\bm{k})\, a_{i}^{a}(k^{+},\bm{k})\\
&\left.+\,b_{\lambda}^{\alpha\dagger}(k^{+},\,\bm{k})\,b_{\lambda}^{\alpha}(k^{+},\,\bm{k})\,-\,d_{\lambda}^{\alpha}(k^{+},\,\bm{k})\,d_{\lambda}^{\alpha\dagger}(k^{+},\,\bm{k})\right),\\
  \end{split}\end{equation}
which in particular shows that the dispersion relation for free quarks and gluons is $E_{k}=\frac{\bm{k}^{2}}{2k^{+}}$. We shall not write the corresponding expression for the interaction piece of the
Hamiltonian, as this is not needed in full generality for the purposes of this paper. Rather, we show
only those matrix elements which enter the computation of the outgoing state in Sect.~\ref{sec:LCWF}:\\
  
 $\bullet$ \quad\textit{\textbf{Emission of a second gluon from the quark state}}
 \begin{equation}\begin{split}\label{qggqg}
&\left\langle q_{\lambda_{2}}^{\beta}(u)\,g_{j}^{b}(t)\,g_{l}^{c}(p)\left|\mathsf{H}_{q\rightarrow qg}\right|q_{\lambda_{1}}^{\alpha}(s)\,g_{i}^{a}(k)\right\rangle \\
&=\,\frac{gt_{\beta\alpha}^{c}\delta^{ab}\delta_{ij}}{2\sqrt{2p^{+}}}\chi_{\lambda_{2}}^{\dagger}\left[\frac{2\bm{p}^{l}}{p^{+}}-\frac{\sigma\cdot\bm{u}}{u^{+}}\sigma^{l}-\sigma^{l}\frac{\sigma\cdot\bm{s}}{s^{+}}\right]\chi_{\lambda_{1}}(2\pi)^{6}\delta^{(3)}(k-t)\delta^{(3)}(s-u-p)\\
&+\,\frac{gt_{\beta\alpha}^{b}\delta^{ac}\delta_{il}}{2\sqrt{2t^{+}}}\chi_{\lambda_{2}}^{\dagger}\left[\frac{2\bm{t}^{j}}{t^{+}}-\frac{\sigma\cdot\bm{u}}{u^{+}}\sigma^{j}-\sigma^{j}\frac{\sigma\cdot\bm{s}}{s^{+}}\right]\chi_{\lambda_{1}}(2\pi)^{6}\delta^{(3)}(k-p)\delta^{(3)}(s-u-t).
 \end{split}\end{equation}\\

  $\bullet$ \quad\textit{\textbf{Gluon splits into quark and antiquark pair}}
 \begin{equation}\begin{split}\label{qqqqg}
&\left\langle \bar{q}_{\lambda_{4}}^{\epsilon}(u)\,q_{\lambda_{3}}^{\delta}(t)\,q_{\lambda_{2}}^{\gamma}(p)\left|\mathsf{H}_{g\rightarrow qq}\right|q_{\lambda_{1}}^{\beta}(s)\,g_{i}^{a}(k)\right\rangle \\
&=\frac{gt_{\gamma\epsilon}^{a}\delta^{\beta\delta}\delta_{\lambda_{1}\lambda_{3}}}{2\sqrt{2k^{+}}}\chi_{\lambda_{2}}^{\dagger}\left[\frac{2\bm{k}^{i}}{k^{+}}-\frac{\sigma\cdot\bm{p}}{p^{+}}\sigma^{i}-\sigma^{i}\frac{\sigma\cdot\bm{u}}{u^{+}}\right]\chi_{\lambda_{4}}(2\pi)^{6}\delta^{(3)}(s-t)\delta^{(3)}(k-p-u)\\
&+\frac{gt_{\delta\epsilon}^{a}\delta^{\beta\gamma}\delta_{\lambda_{1}\lambda_{2}}}{2\sqrt{2k^{+}}}\chi_{\lambda_{3}}^{\dagger}\left[\frac{2\bm{k}^{i}}{k^{+}}-\frac{\sigma\cdot\bm{t}}{t^{+}}\sigma^{i}-\sigma^{i}\frac{\sigma\cdot\bm{u}}{u^{+}}\right]\chi_{\lambda_{4}}(2\pi)^{6}\delta^{(3)}(s-p)\delta^{(3)}(k-t-u).
 \end{split}\end{equation}\\ 
 
  $\bullet$ \quad\textit{\textbf{Triple gluon interaction in presence of a quark}}
\begin{eqnarray}\label{qgqgg_split}
&&\left\langle q_{\lambda_{2}}^{\beta}(t)\,g_{l}^{c}(q)\,g_{n}^{b}(p)\left|\mathsf{H}_{g\rightarrow gg}\right|q_{\lambda_{1}}^{\alpha}(s)\,g_{m}^{a}(k)\right\rangle =\frac{igf^{abc}\delta^{\alpha\beta}\delta_{\lambda_{1}\lambda_{2}}}{2\sqrt{2k^{+}p^{+}q^{+}}}(2\pi)^{6}\delta^{(3)}(s-t)\delta^{(3)}(k-p-q)\nonumber\\
&&\times\left[\left(\bm{p}^{m}-\bm{q}^{m}+\frac{q^{+}-p^{+}}{k^{+}}\bm{k}^{m}\right)\delta_{nl}+\left(\bm{k}^{n}+\bm{q}^{n}-\frac{k^{+}+q^{+}}{p^{+}}\bm{p}^{n}\right)\delta_{ml}\right.\\
&&\left.+\left(\frac{k^{+}+p^{+}}{q^{+}}\bm{q}^{l}-\bm{p}^{l}-\bm{k}^{l}\right)\delta_{mn}\right].\nonumber\end{eqnarray}\\

      $\bullet$ \quad\textit{\textbf{Quark produces quark and antiquark pair instantaneously}}
 \begin{equation}\begin{split}\label{instqq}
\left\langle q_{\lambda_{2}}^{\beta}(k)\,\bar{q}_{\lambda_{3}}^{\gamma}(q)\,q_{\lambda_{4}}^{\delta}(s)\left|\mathsf{H}_{q\rightarrow qqq}\right|q_{\lambda_{1}}^{\alpha}(p)\right\rangle \,=\,\frac{g^{2}t_{\beta\alpha}^{a}t_{\delta\gamma}^{a}}{(q^{+}+s^{+})^{2}}\chi_{\lambda_{4}}^{\dagger}\,\chi_{\lambda_{3}}\,\chi_{\lambda_{2}}^{\dagger}\,\chi_{\lambda_{1}}(2\pi)^{3}\delta(k-p+q+s).
  \end{split}\end{equation}
  
   $\bullet$ \quad\textit{\textbf{Instantaneous emission of a two gluons from the quark state (inst. q)}}
  \begin{equation}\begin{split}\label{instq}
&\left\langle q_{\lambda_{2}}^{\gamma}(u)\,g_{j}^{b}(t)\,g_{l}^{c}(p)\left|\mathsf{H}_{q\rightarrow qgg}^{inst\:q}\right|q_{\lambda_{1}}^{\alpha}(s)\right\rangle \\
&=\,\frac{g^{2}}{\sqrt{t^{+}p^{+}}}\left(\frac{t_{\gamma\beta}^{b}t_{\beta\alpha}^{c}}{s^{+}-p^{+}}\chi_{\lambda_{2}}^{\dagger}\sigma_{j}\sigma_{l}\chi_{\lambda_{1}}+\frac{t_{\gamma\beta}^{c}t_{\beta\alpha}^{b}}{s^{+}-t^{+}}\chi_{\lambda_{2}}^{\dagger}\sigma_{l}\sigma_{j}\chi_{\lambda_{1}}\right)(2\pi)^{3}\delta(t+p+u-s).
  \end{split}\end{equation}

 $\bullet$ \quad\textit{\textbf{Instantaneous emission of two gluons from the quark state (inst. g)}}
 \begin{equation}\begin{split}\label{instg}
\left\langle q_{\lambda_{2}}^{\beta}(s)\,g_{j}^{c}(p)\,g_{i}^{b}(k)\left|\mathsf{H}_{q\rightarrow qgg}^{inst\:g}\right|q_{\lambda_{1}}^{\alpha}(q)\right\rangle \,=\,\frac{ig^{2}f^{abc}t_{\beta\alpha}^{a}(p^{+}-k^{+})}{2\sqrt{k^{+}p^{+}}(k^{+}+p^{+})^{2}}\chi_{\lambda_{2}}^{\dagger}\chi_{\lambda_{1}}(2\pi)^{3}\delta_{ij}\delta(k+p+s-q).
  \end{split}\end{equation}


\providecommand{\href}[2]{#2}\begingroup\raggedright\endgroup

\end{document}